\documentclass[a4paper, useAMS,usenatbib,usedcolumn]{mn2e}
\usepackage[figuresright]{rotating}
\usepackage{lscape} 

\usepackage{graphics}

\usepackage{bigdelim}
\usepackage{bigstrut}

\usepackage{setspace}

\usepackage{natbib}
\usepackage{mathtools}
\usepackage{amsmath}
\usepackage{amssymb}
\usepackage[subnum]{cases}

\newcommand{\kms}{\mbox{${\rm km\;s^{-1}}$}}

\newcommand{\sauron}{{{SAURON}}}
\newcommand{\atlas}{{\texttt {ATLAS$^{\rm{3D}}$}}}
\newcommand{\deimos}{{{DEIMOS}}}

\newcommand{\Hb}{H$\beta$}
\newcommand{\Hi}{{\sc H$\,$i}}

\newcommand{\placetabA}	
{
	\begin{table*}
	\begin{center}
	\begin{tabular}{ccccccccccc}
		\hline
		\\
		Galaxy 			&	R.A.	& Dec.		&  $\rm{R_{e}}$		& PA		& b/a	& $\rm{V_{sys}}$	& $\rm{\sigma_{0}}$ 	& Morph	& Distance	& $\rm{M_{K}}$	\\
							&	(hh mm ss)		& (dd mm ss)			&	(arcsec)		& (deg)		&	& ($\rm{km/s}$)	& ($\rm{km/s}$)	&	& ($\rm{Mpc}$)	& ($\rm{mag}$)		\\		
		(1)				& (2)		&	(3)	&	(4)			& (5)		& (6)	& (7)			& (8)			& (9)	& (10)		& (11)		\\		
		\\
		\hline
		NGC~720		& 01 53 00.50		& $-$13 44 19.2	& 33.9				& 142.3	& 0.57 	& 1745			& 241			& E5	& 26.9		& -24.60	\\
		NGC~821		& 02 08 21.14		& +10 59 41.7	& 39.8				& 31.2		& 0.65	& 1718			& 200			& E6	& 23.4		& -23.99	\\	
		NGC~1023	& 02 40 24.01		& +39 03 47.8 & 47.9				& 83.3		& 0.63	& 602			& 204			& S0	& 11.1		& -24.01	\\ 	
		NGC~1400	& 03 39 30.84		& $-$18 41 17.1	& 29.3				& 36.1		& 0.89	& 558			& 252			& E1/S0	& 26.8		& -24.30	\\	%
		NGC~1407	& 03 40 11.86		& $-$18 34 48.4	& 63.4				& 58.3		& 0.95	& 1779			& 271			& E0	& 26.8		& -25.40	\\  %
		NGC~2768	& 09 11 37.50		& +60 02 14.0	& 63.1				& 91.6		& 0.53	& 1353			& 181			& E6/S0	& 21.8		& -24.71	\\
		NGC~2974	& 09 42 33.28		& $-$03 41 56.9	& 38.0				& 44.2		& 0.59	& 1887			& 238			& E4	& 20.9		& -23.62	\\	
		NGC~3115	& 10 05 13.98		& $-$07 43 06.9 	& 32.1				& 43.5		& 0.51	& 663			& 267			& S0	& 9.4		& -24.00	\\	
		NGC~3377	& 10 47 42.33		& +13 59 09.3	& 35.5				& 46.3		& 0.50	& 690			& 139			& E5-6	& 10.9		& -22.76	\\	
		NGC~4111	& 12 07 03.13 		& +43 03 56.6 & 12.0				& 150.3		& 0.42	& 792			& 149			& S0	& 14.6		& -23.27	\\
		NGC~4278	& 12 20 06.82 		& +29 16 50.7	& 31.6				& 39.5			& 0.90	& 620			& 237			& E1-2	& 15.6		& -23.80	\\	
		NGC~4365	& 12 24 28.28		& +07 19 03.6	& 52.5				& 40.9		& 0.75	& 1243			& 256			& E3	& 23.3		& -25.21	\\	
		NGC~4374	& 12 25 03.74		& +12 53 13.1 & 52.5				& 128.8		& 0.85	& 1017			& 283			& E1	& 18.5		& -25.12	\\
		NGC~4473	& 12 29 48.87		& +13 25 45.7 & 26.9				& 92.2		& 0.58	& 2260			& 179			& E5	& 15.3		& -23.77	\\	
		NGC~4494	& 12 31 24.10		& +25 46 30.9 & 49.0				& 176.3		& 0.83	& 1342			& 150			& E1-2	& 16.6		& -24.11	\\	
		NGC~4526	& 12 34 03.09		& +07 41 58.3 & 44.7				& 113.7		& 0.64	& 617			& 251			& S0	& 16.4		& -24.62	\\	
		NGC~4649	& 12 43 39.98		& +11 33 09.7 & 66.1				& 91.3		& 0.84	& 1110			& 335			& E2/S0	& 17.3		& -25.46	\\	
		NGC~4697	& 12 48 35.88		& $-$05 48 02.7 	& 61.7				& 67.2		& 0.55	& 1252			& 171			& E6	& 11.4		& -23.93	\\
		NGC~5846	& 15 06 29.28		& +01 36 20.3 & 58.9				& 53.3		& 0.94	& 1712			& 239			& E0-1/S0	& 24.2		& -25.01	\\
		NGC~7457	& 23 00 59.93		& +30 08 41.8	& 36.3				& 124.8		& 0.53	& 844			& 69			& S0	& 12.9		& -22.38	\\	
		\hline
		NGC~3607	& 11 16 54.64		& +18 03 06.3 & 38.9				& 124.8		& 0.87	& 942			& 224			& S0	& 22.2		& -24.74	\\
		NGC~5866	& 15 06 29.50		& +55 45 47.6 & 36.3				& 125.0		& 0.43	& 755			& 159			& S0	& 14.9		& -24.00	\\
		\hline
	\end{tabular}
	\end{center}\caption{Galaxy parameters. 
		Notes: All the galaxies are part of the original SLUGGS survey, apart from NGC~3607 and NGC~5866 
		which have been added later to our sample. 
		The columns present: 
		(1) Galaxy name. 
		(2) Right ascension and (3) declination in J2000 coordinates taken from the NASA/IPAC Extragalactic database (NED\protect\footnotemark[2]). 
		(4) Effective radius from \citet{Arnold14}.
		(5) Photometric position angle and (6) axial ratio from \citet{Emsellem11}, except 
		NGC~720 \citep{Cappellari07}, NGC~821 \citep{Krajnovic11}, NGC~1400, NGC~1407 \citep{Spolaor08a} and 
		NGC~3115 \citep{Capaccioli87}. 
		(7) Systemic heliocentric velocity from \citet{Cappellari11a}, except  
		NGC~720, NGC~1400, NGC~1407, NGC~3115 and NGC~5846 
		(NED).
		(8) Central velocity dispersion from HyperLeda\protect\footnotemark[3]. 
		(9) Morphology from NED, combining the RSA and RC3 classifications.
		(10) Distance from \citet{Cappellari11a}. For NGC~720, NGC~1400, NGC~1407 and 
		NGC~3115 we adopt the values in \citet{Arnold14}. 
		(11) Extinction corrected total K-band absolute magnitude from the 2MASS extended 
		source catalog \citep{Jarrett00}, obtained following the same approach used in \citet{Emsellem11}.
		}			
	\label{tab:A}
	\end{table*}
}

\newcommand{\placetabB}
{
	\begin{table}
		\caption{CaT index definition passbands.}
	\label{tab:B}
	\begin{center}
	\begin{tabular}{cc}
		\hline
		Continuum passbands (\AA) &  CaT line passbands (\AA)	\\
		\hline
		C1: $8474-8483$	&	L1: $8483-8513$ 	\\
		C2: $8514-8526$ 	& L2: $8527-8557$ 	\\
		C3: $8563-8577$ 	& L3: $8647-8677$ 	\\
		C4: $8619-8642$ 	& 							\\
		C5: $8680-8705$	&  						\\	
		\hline
	\end{tabular}
	\end{center}
	\end{table}
}    

\newcommand{\placetabC}
{
	\begin{table}
	\begin{center}
	\begin{tabular}{lcc}
		\hline
		Galaxy &  Inner ($0.32-1~\rm{R_{e}}$) & Outer ($1-2.5~\rm{R_{e}}$) \\
					& (dex/dex)			& (dex/dex)			\\
		\hline
		NGC~1023		& -0.32 $\pm$ 0.03	& -0.33 $\pm$ 0.03\\
		NGC~1400		& -0.35 $\pm$ 0.03	& -1.34 $\pm$ 0.18\\
		NGC~2768		& -0.30 $\pm$ 0.02	& -1.35 $\pm$ 0.08\\
		NGC~3115		& -0.04 $\pm$ 0.02 	& -0.19 $\pm$ 0.03\\
		NGC~3377		& -0.69 $\pm$ 0.05	& -2.17 $\pm$ 0.25\\
		NGC~3607		& -0.49 $\pm$ 0.08	& -0.22 $\pm$ 0.31\\
		NGC~4111		& -0.66 $\pm$ 0.04	& -1.28 $\pm$ 0.05\\
		NGC~4278		& -0.06 $\pm$ 0.01	& +0.25 $\pm$ 0.09\\
		NGC~4365		& -0.20 $\pm$ 0.03	& +0.50 $\pm$ 0.16\\
		NGC~4374		& -0.03 $\pm$ 0.05	& -0.33 $\pm$ 0.23\\
		NGC~4473		& -0.22 $\pm$ 0.02	& -1.65 $\pm$ 0.25\\
		NGC~4526		& -0.40 $\pm$ 0.03	& -0.59 $\pm$ 0.09\\
		NGC~4649		& -0.22 $\pm$ 0.09	& -0.75 $\pm$ 0.06\\
		NGC~4697		& -0.52 $\pm$ 0.07	& -2.31 $\pm$ 0.08\\
		NGC~5846		& -0.46 $\pm$ 0.02	& ---	\\
		NGC~7457		& -1.10 $\pm$ 0.04	& -1.96 $\pm$ 0.16\\	
		\hline
	\end{tabular}
	\end{center}
		\caption{\textbf{Metallicity gradients measured on the radial profiles extracted from the 2D metallicity 
		maps. }
		The inner metallicity gradient values (second column) are measured 
		in the radial region $0.32 < R \leq 1~\rm{R_{e}}$ (corresponding 
		in logarithmic space to $-0.5 <\log (R/\rm{R_{e}}) \leq 0$). 
		The outer metallicity gradient values (third column) are measured 
		in the radial region $1 < R \leq 2.5~\rm{R_{e}}$ (corresponding
		in logarithmic space to $0 < \log (R/\rm{R_{e}}) \leq 0.4$). 
		In the cases where the maps do not extent out to such radii, we extrapolate the trend from the 
		available points outside $1~\rm{R_{e}}$. 
		In the case of NGC~5846, the lack of data points outside $1~\rm{R_{e}}$ prevents the measurement 
		of a reliable outer gradient. } \label{tab:C}
	\end{table}
}    

\newcommand{\placefigBdata}{
	\begin{figure}
    	\begin{center}
			\includegraphics[width=\columnwidth]{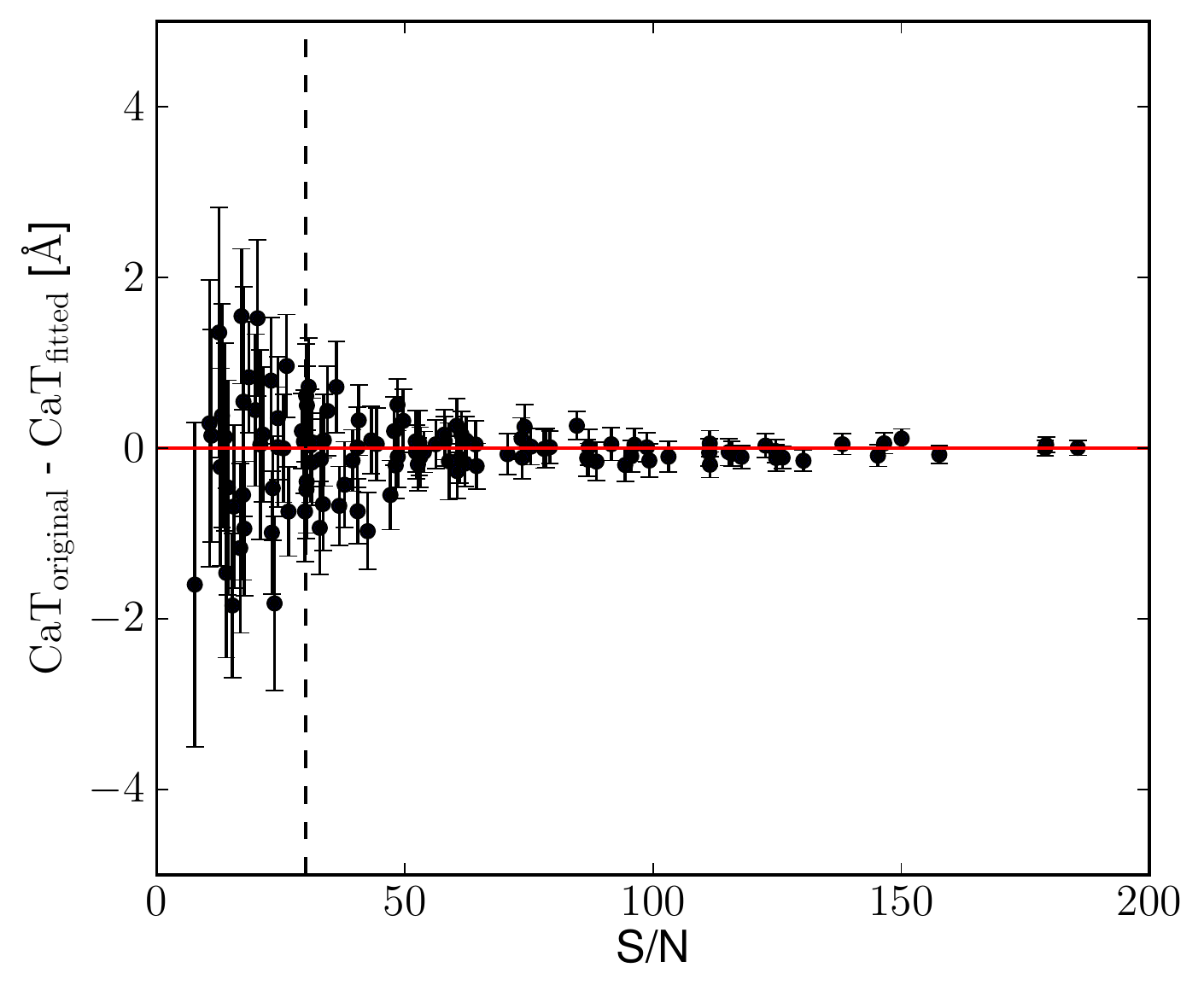}
		\end{center}
	    \caption[]{Differences between the NGC~5846 CaT indices measured 
	    on the original sky-subtracted spectra and on the fitted spectra versus 
	    the S/N.
	    The errorbars are obtained by summing in quadrature the uncertainties of each 
	    pair of CaT values. 
		The dashed vertical line indicates the cut in S/N we use for
		reliable measurements (i.e. $\rm{S/N} >30$).		 		
	    }
    \label{fig:B}
  \end{figure}
}

\newcommand{\placefigDdata}{
	\begin{figure}
    	\begin{center}
			\includegraphics[width=\columnwidth]{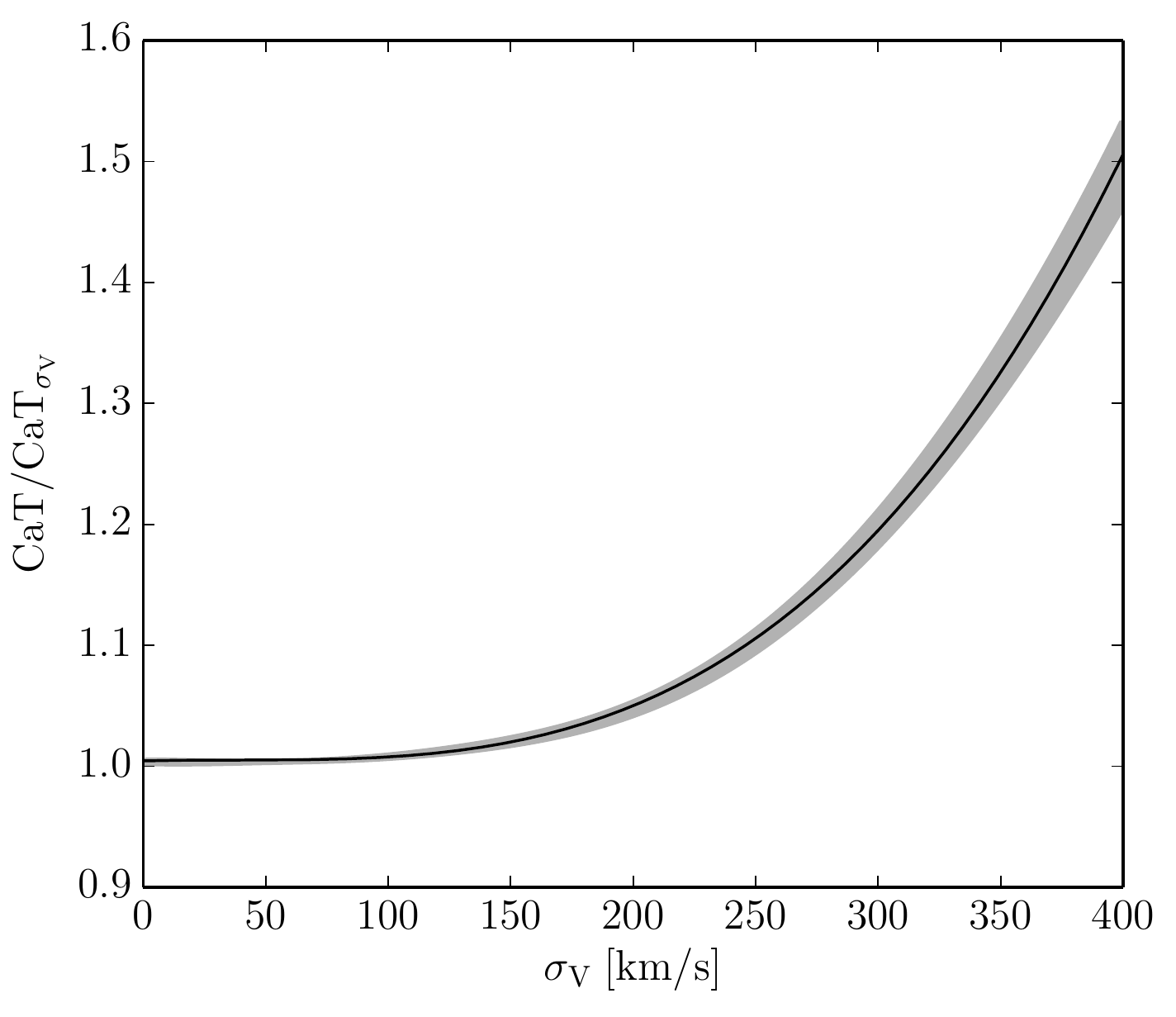}
		\end{center}
	    \caption[]{Applied velocity dispersion index correction.
	    The correction CaT/CaT$_{\sigma_{\rm{V}}}$ is shown against the velocity 
	    dispersion $\sigma_{\rm{V}}$. 
	    The relation has been derived from the measurement of the CaT index in the 
	    \citet{Vazdekis03} SSP model spectra convolved with Gaussians of corresponding 
	    width for a given velocity dispersion in the range $[0,400]$~\kms.
	    The values have been fitted with a third order polynomial curve in order to 
	    estimate a correction for each value of $\sigma_{\rm{V}}$. 
	    The shaded area shows the scatter in the correction due to models with 
	    different metallicity. 
	    The correction is less than 10\% for most of the galaxies in our sample.
	    }
    \label{fig:D}
  \end{figure}
}

\newcommand{\placefigEdata}{
	\begin{figure}
    	\begin{center}
			\includegraphics[width=\columnwidth]{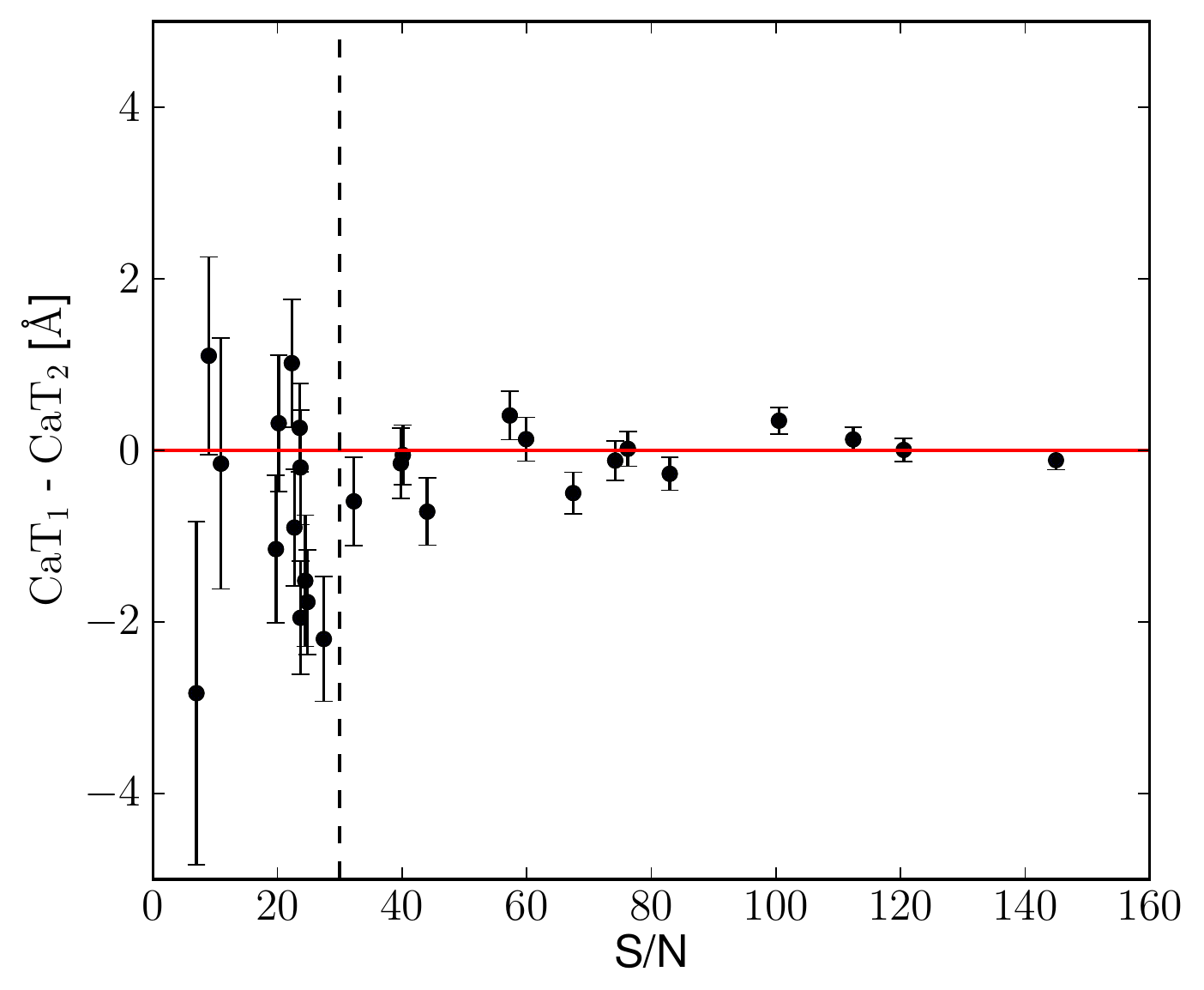}
		\end{center}
	    \caption[]{Differences between the CaT index values obtained on different nights 
	    from the same NGC~2768 mask, along with the S/N of the 
	    spectra.
	    The vertical errorbars are obtained averaging in quadrature the uncertainties of each couple of 
	    measurements, while the abscissa values are the mean of the measured signal-to-noise ratios of 
	    paired spectra.
	    The dashed vertical line indicates the limit in S/N we use (i.e. $\rm{S/N} >~30$).		 		
	    }
    \label{fig:E}
  \end{figure}
}

\newcommand{\placefigFdata}{
	\begin{figure}
    	\begin{center}
			\includegraphics[width=\columnwidth]{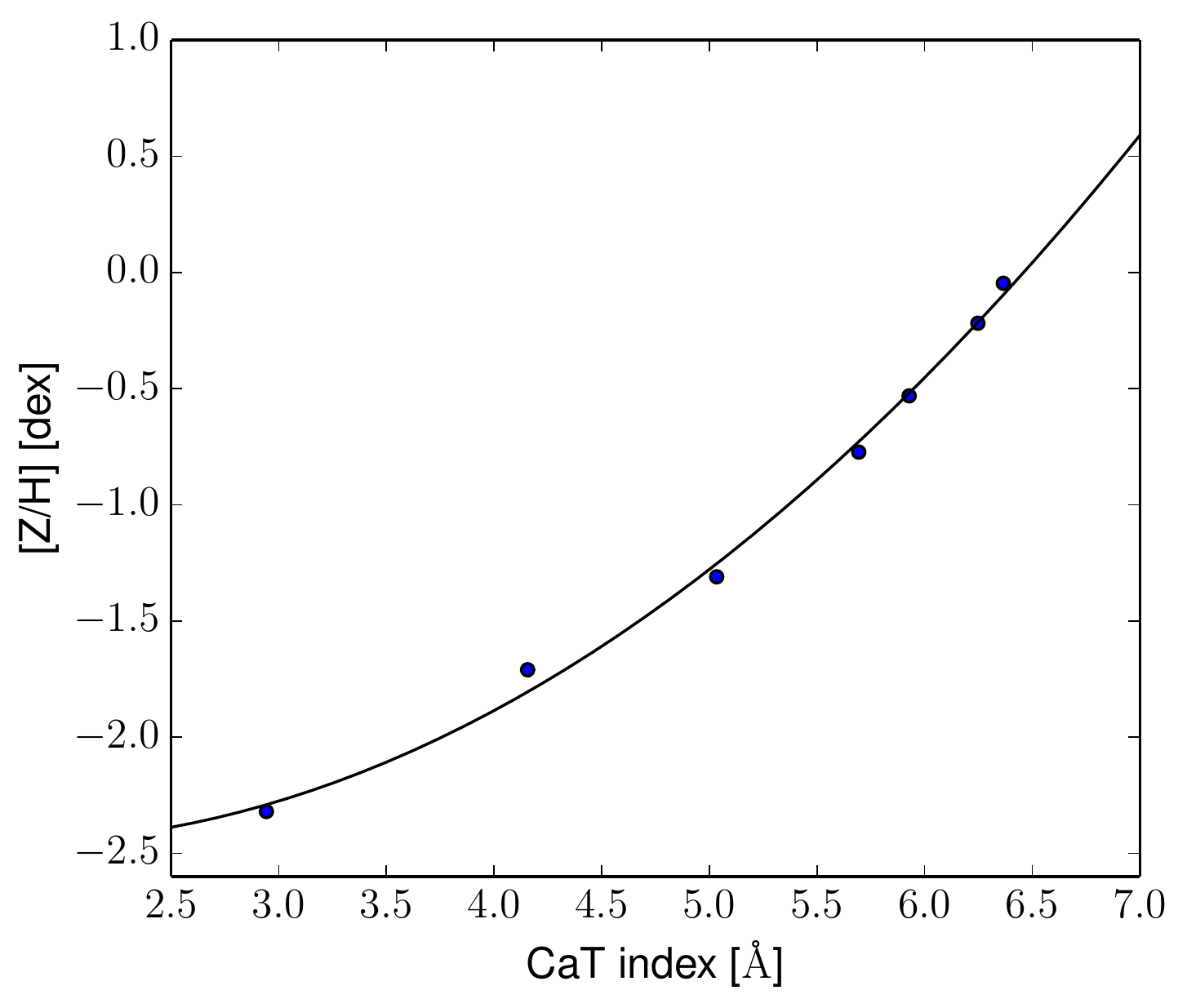}
		\end{center}
	    \caption[]{Relation between CaT index and metallicity.
	    The points are the CaT values measured on 7 spectra from the 
	    \citet{Vazdekis03} SSP library with a constant age of $12.6~\rm{Gyr}$ and 
	    a \citet{Salpeter55} IMF.
	    The line is a fitted second order polynomial.
	    }
    \label{fig:F}
  \end{figure}
}

\newcommand{\placefigGanalysis}{
	\begin{figure}
    	\begin{center}
			\includegraphics[width=\columnwidth]{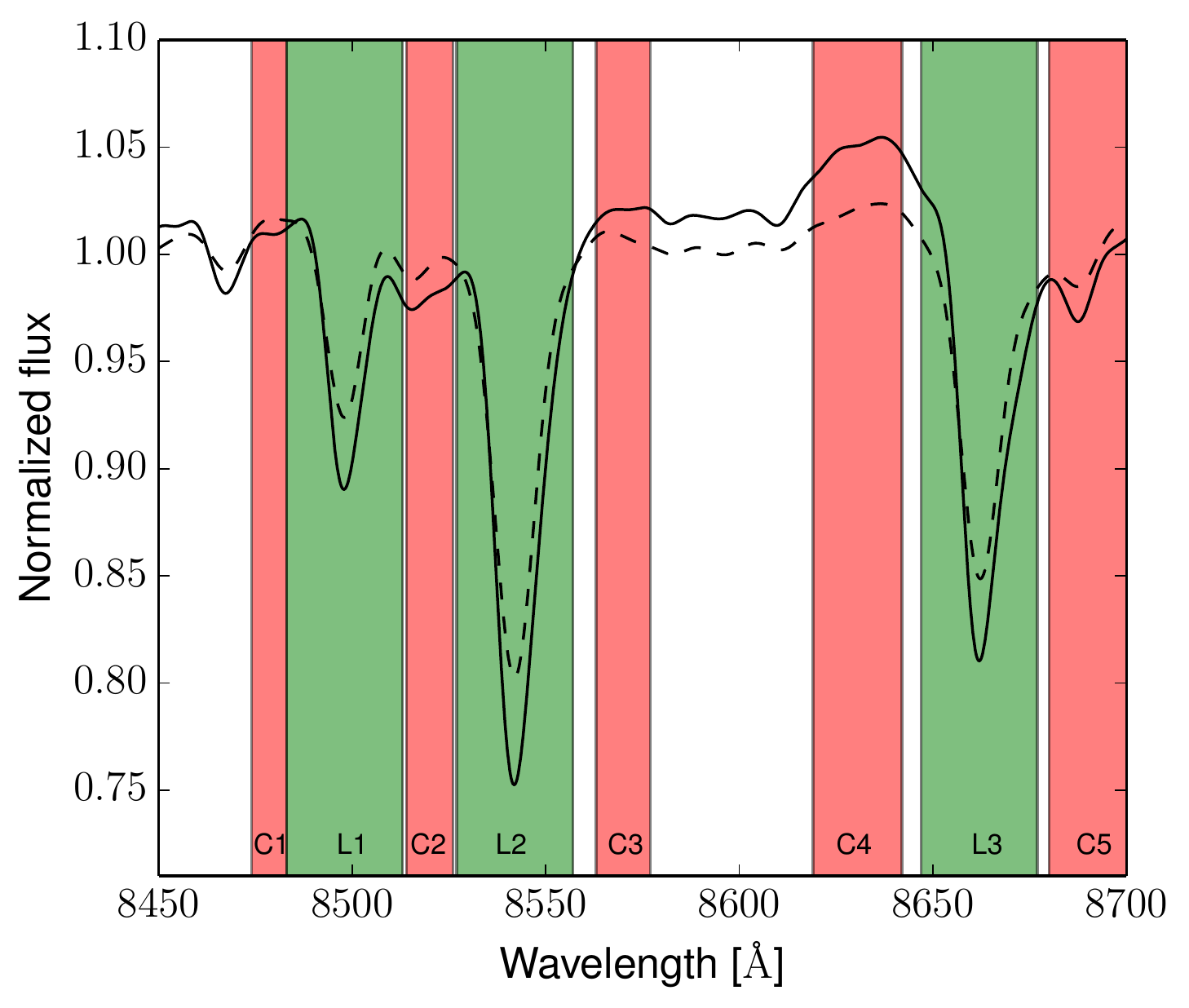}
		\end{center}
	    \caption[]{Comparison between two normalized stellar light spectra of 
	    NGC~2768 with similar velocity dispersion $\sigma_{\rm{V}}$ and S/N 
	    but different CaT indices. 
		The dashed line is the fit of the spectrum with CaT index $= 5.25~\rm{\AA}$, corresponding to 
		a metal-poor (sub-solar) stellar population, while the solid line is the fit of the spectrum with CaT 
		index $= 6.78~\rm{\AA}$, corresponding to a metal-rich (super-solar) stellar population.
	    The green regions (L1, L2 and L3) represent the absorption lines intervals, while the red regions 
	    (C1, C2, C3, C4 and C5) the continuum intervals, as defined in Table \ref{tab:A}.	
	    The difference between a metal-rich (dashed line) and a metal-poor stellar population is 
	    noticeable. 
	    }
    \label{fig:G}
  \end{figure}
}

\newcommand{\placefigHanalysis}{
	\begin{figure*}
    	\begin{center}
			\includegraphics[width=\textwidth]{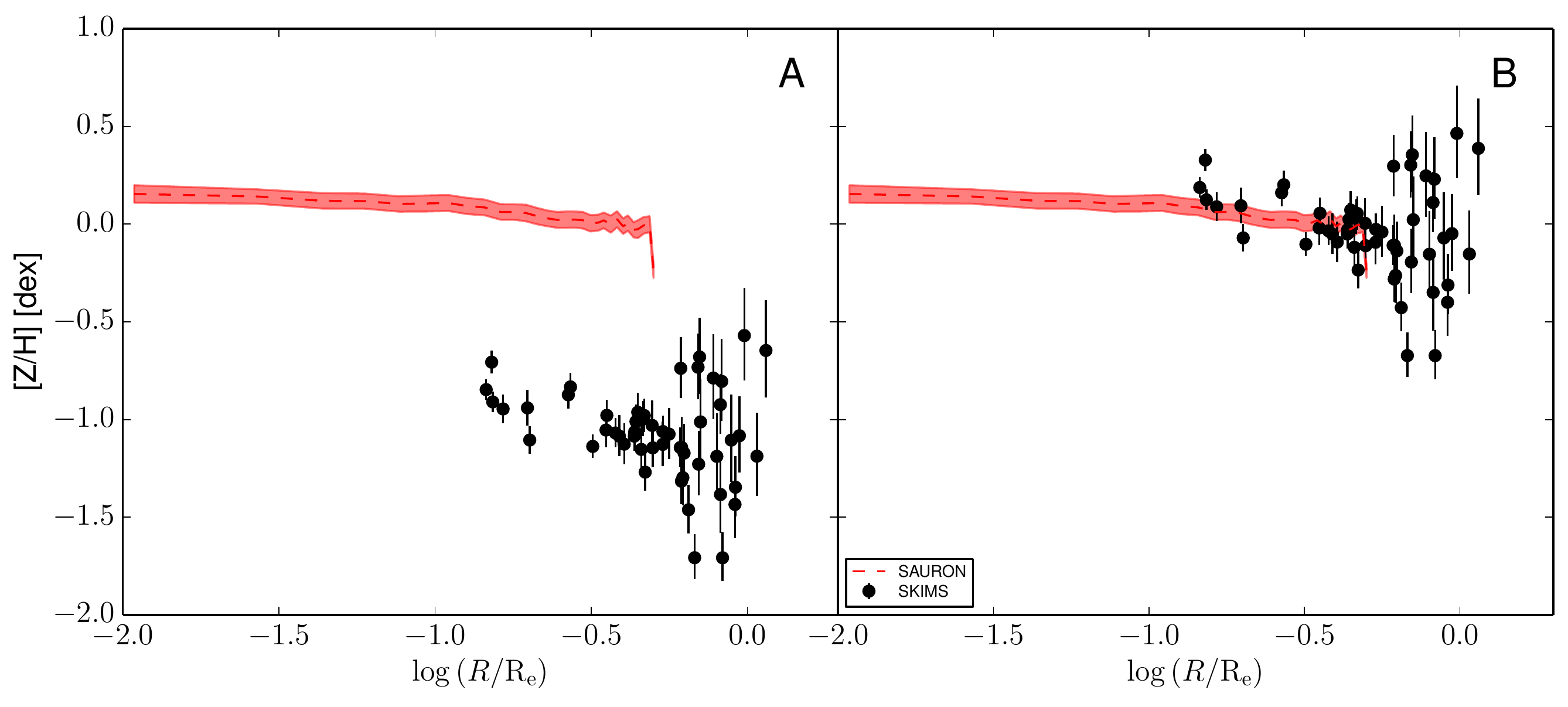}
		\end{center}
	    \caption[]{
	    NGC~5846 radial metallicity profile before and after the offset correction. 
	    Panels $A$ and $B$ show metallicity versus galactocentric radius scaled by $\rm{R_{e}}$. 
	    In both the panels the red dashed line shows the metallicity radial profile extracted from the 2D \sauron\ 
		metallicity map of NGC~5846. 
	    The black points are the values obtained from \deimos\ multislit observations (called SKiMS) before (panel $A$) and after 
	    (panel $B$) the empirical correction. 
	    }
    \label{fig:H}
  \end{figure*}
}

\newcommand{\placefigJanalysis}{			
	\begin{figure}										
    	\begin{center}
			\includegraphics[width=\columnwidth]{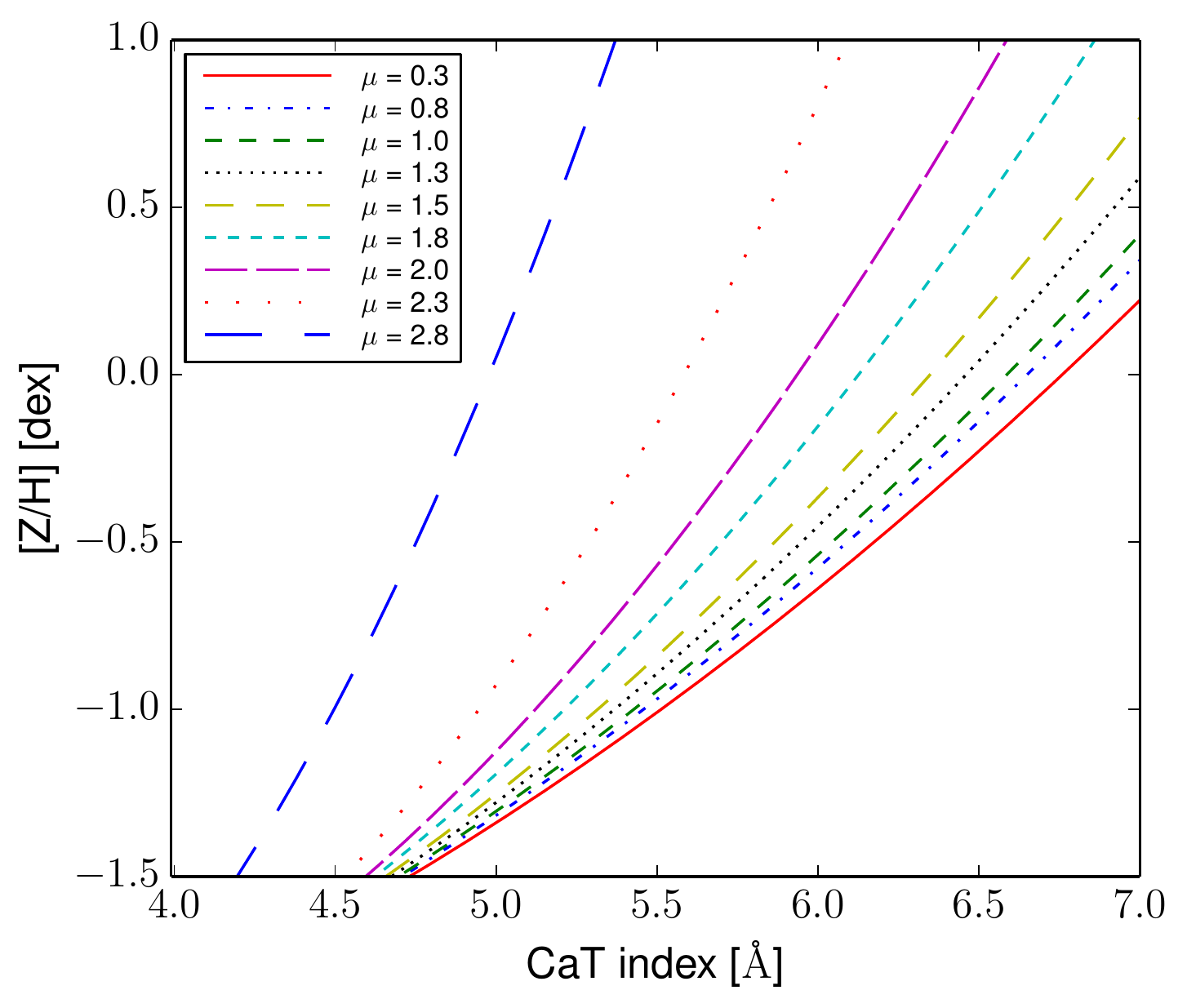}
		\end{center}
	    \caption[]{CaT to metallicity model relations for different IMF slopes. 
	    The different lines show the second order polynomials fit to CaT indices measured 
	    on the \citet{Vazdekis03} models for old ages (i.e. 12.6 Gyr). 
	    The lines are colour- and style- coded according to the assumed IMF slope 
	    ($\mu$), as in the legend. 
	    A \citet{Salpeter55} IMF has a slope $\mu = 1.3$.
	    For the same CaT index, steeper IMF slopes (higher $\mu$) give higher metallicities. 
	    From the CaT index we measure at $R=1~\rm{R_{e}}$ and the \sauron\ [Z/H] 
	    at the same galactocentric radius, we estimate the IMF slope which could compensate 
	    for the offset between our metallicities and those of \sauron. 
	    Because of the discrete number of available models, in order to obtain the IMF slope where the 
	    [Z/H]-CaT index point lies, we interpolate between the two closest curves. 
	     }
    \label{fig:J}
  \end{figure}
}

\newcommand{\placefigKanalysis}{
	\begin{figure}
    	\begin{center}
			\includegraphics[width=\columnwidth]{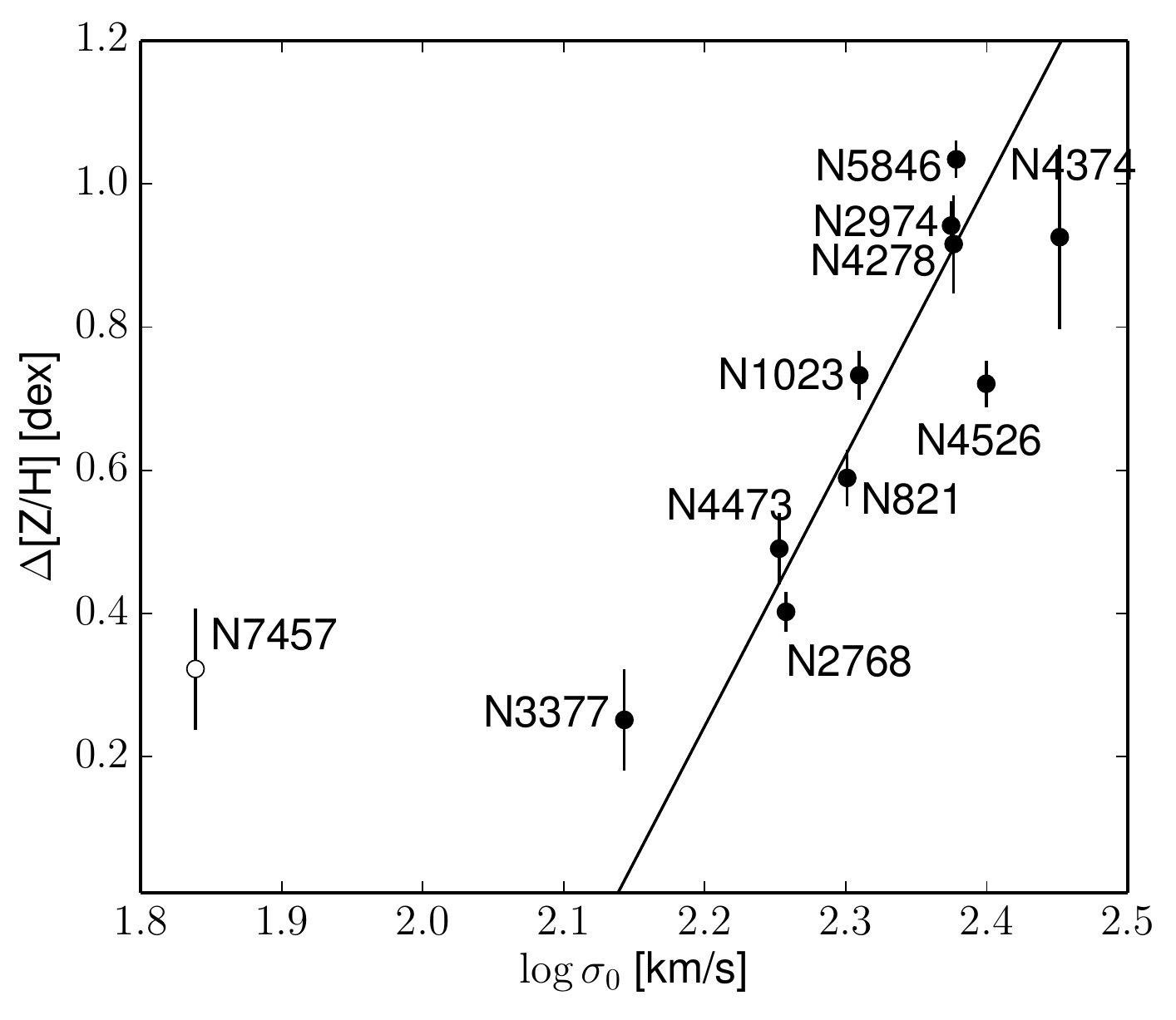}
		\end{center}
	    \caption[]{Metallicity empirical correction. 
	    The points represent the metallicity offsets between our values 
	    and the \sauron\ metallicities versus the central 
	    velocity dispersion $\sigma_{\rm{0}}$ obtained from HyperLeda. 
	    The straight line is the fit to the black data points, excluding the outlier NGC~7457 
	    (open circle). 
	    }
    \label{fig:K}
  \end{figure}
}

\newcommand{\placefigLanalysis}{
	\begin{figure*}
    	\begin{center}
			\includegraphics[width=\textwidth]{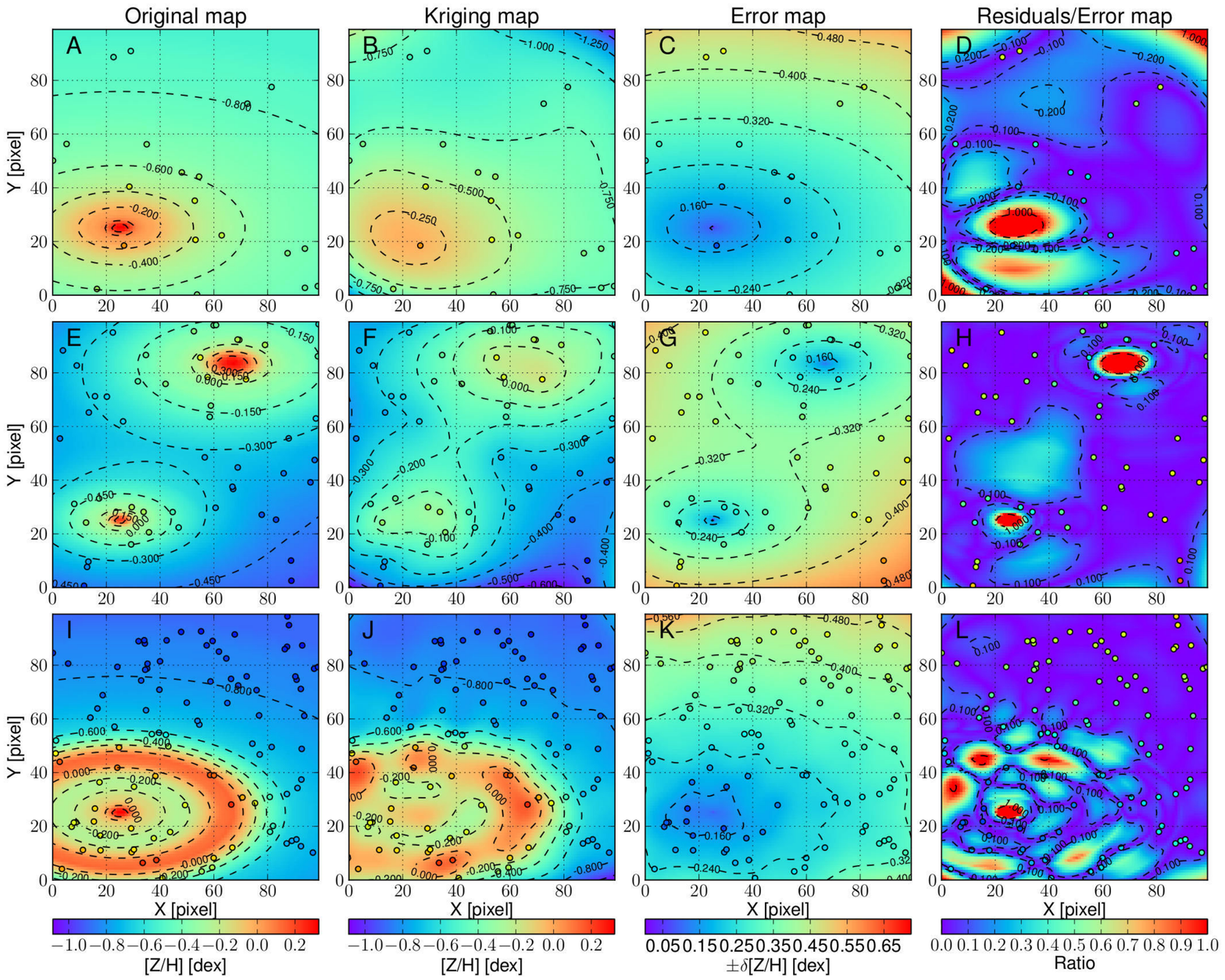}
     	\end{center}
	    \caption[]{Kriging test examples. 
	    The three rows of panels are three different simulations. 
	    The top row is an off-centre galaxy sampled with a low number of slits ($n = 20$), 
	    the middle row shows the double galaxy case and an average 
	    number of slits ($n = 50$) while the bottom row is a galaxy with a ring shaped  
	    substructure and a high number of sampling slits ($n = 100$). 
	    From left to right, the different columns show, respectively, the original mock metallicity field, 
	    the kriging fitted map, the error map and the map of the ratio between residuals and uncertainties. 
	    The metallicity uncertainties on each pixel are measured summing in quadrature the assigned intrinsic uncertainty 
	    and the kriging interpolation estimated uncertainty. 
	    The points show the position of the samplings, color-coded to show the metallicity in the first two 
	    columns, the metallicity associated uncertainty in the third and the ratio between residuals and uncertainties 
	    in the fourth column. 
	    The contours show regions with equal values. 
	    }
    \label{fig:L}
  \end{figure*}
}

\newcommand{\placefigManalysis}{
	\begin{figure}
    	\begin{center}
			\includegraphics[width=\columnwidth]{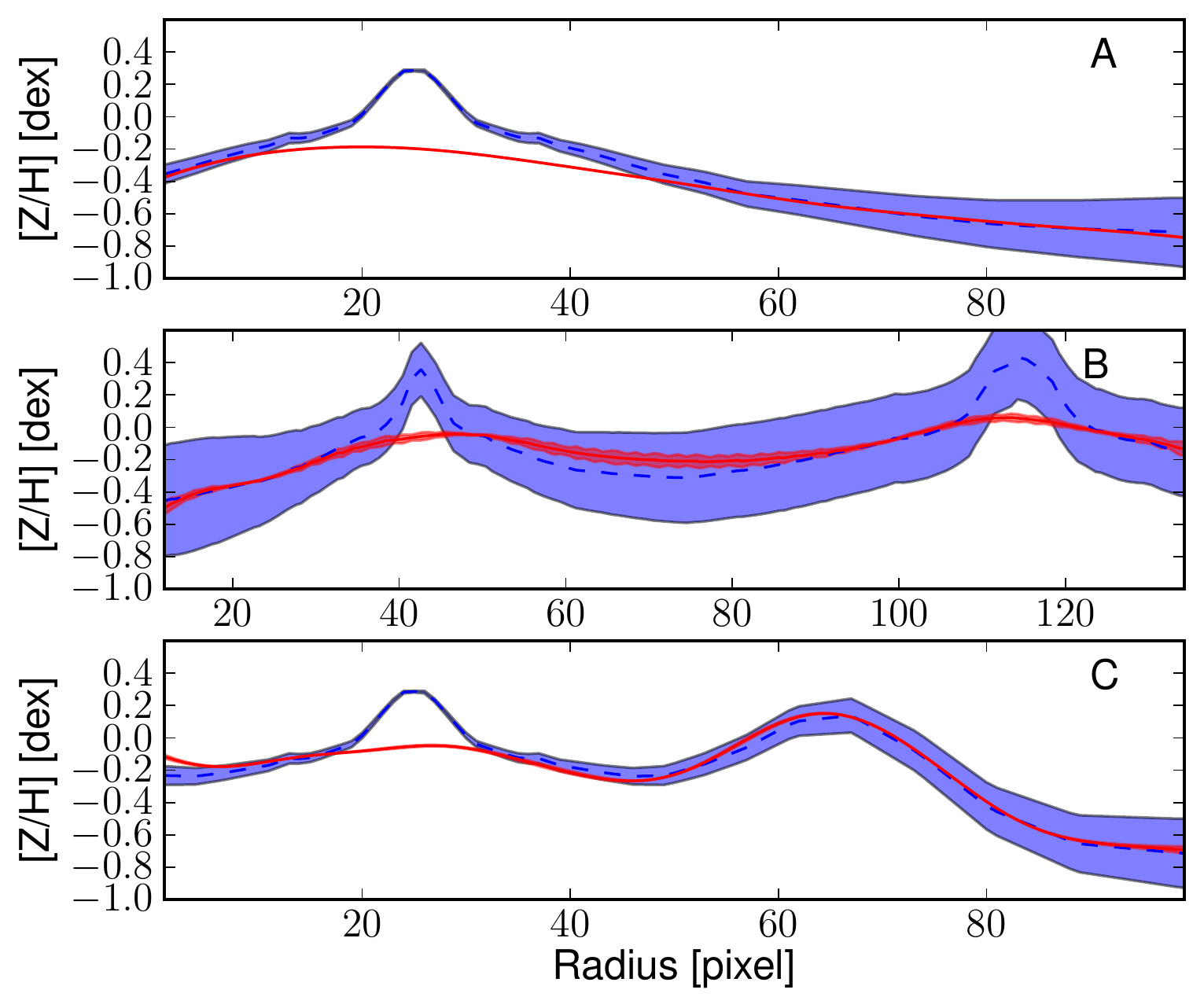}
	\end{center}
	    \caption[]{Examples of kriging-extracted virtual slit metallicity radial profiles. 
	    The blue dashed lines and the red solid lines show the metallicity profile extracted 
	    from a virtual slit in the original and in the kriging map, respectively.
	    The three panels are the three different cases presented in Figure \ref{fig:L}. 
	    In particular, panel A presents the not-centred galaxy case, the panel B the double galaxy 
	    case and panel C the ring substructure case. 
	    The profiles in A and C are horizontal slits, while the profile for the case B is extracted 
	    from a slit tilted to include the two galaxy centres. 
	    On the x-axis the distance of the pixels from the left edge of the grid is shown. 
	    The blue region shows the uncertainties associated with the original map points, while the red shadow presents 
	    the kriging interpolation estimated uncertainties. 
	    Exluding the regions where the original profile is steep (i.e. galaxy centres), the virtual and the real 
	    profiles match within the uncertainties. 
	    }
    \label{fig:M}
  \end{figure}
}

\newcommand{\placefigNanalysis}{
	\begin{figure}
    	\begin{center}
			\includegraphics[width=\columnwidth]{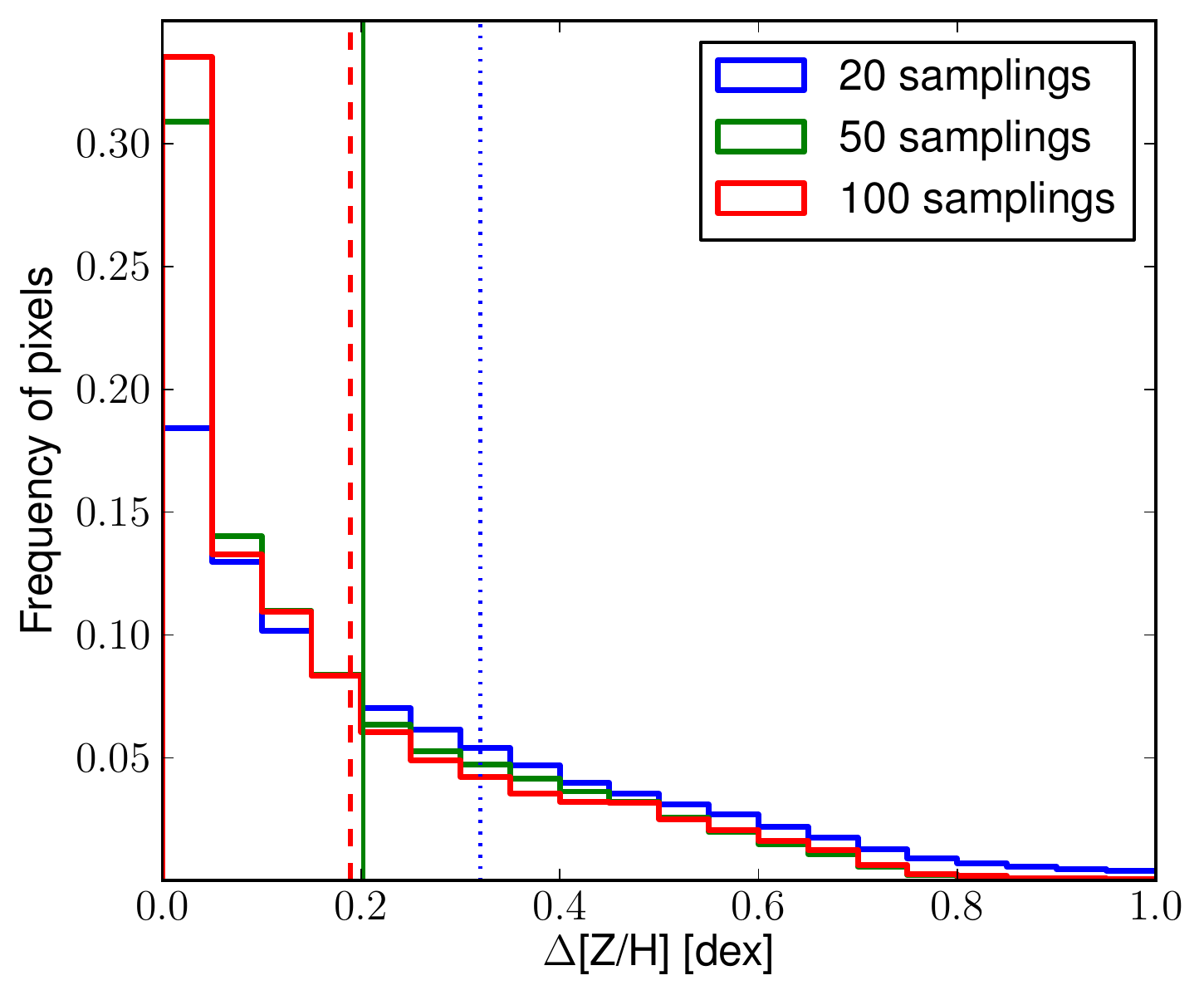}
	\end{center}
	    \caption[]{
		Frequency of residual over uncertainty ratio distributions with different samplings.
		For each pixel the residual is obtained as the absolute difference between the original and the kriging map values.
		The associated metallicity uncertainty is obtained as the sum in quadrature of the uncertainty in the original metallicity 
		with the kriging interpolation uncertainty, 
	    The horizontal axis shows ratio bins of width $0.05$ while the vertical axis presents the relative frequency of 
	    the pixels with a given ratio.
	    In the three different cases of low ($n=20$, in blue), average ($n=50$, in green) and high ($n=100$, 
	    in red) number of sampling points, the values are added for 100 different statistical realizations of point 
	    selection to reduce the systematic uncertainties.
	    The black dashed vertical line separates the pixels for which the residuals are less than the associated uncertainty from 
	    the ones in which kriging does not return a comparable value. 
	    The red dashed, green solid and blue dotted vertical lines show the limit within which 68\% of the pixels are enclosed, 
	    respectively in the high, average and low sampling cases (the values of which are presented in the legend). 
	    In all the different sampling cases, most of the kriging map points match with the original map ones within the uncertainty. 
	    }
    \label{fig:N}
  \end{figure}
}

\newcommand{\placefigOanalysis}{
	\begin{figure*}
    	\begin{center}
			\includegraphics[width=2\columnwidth]{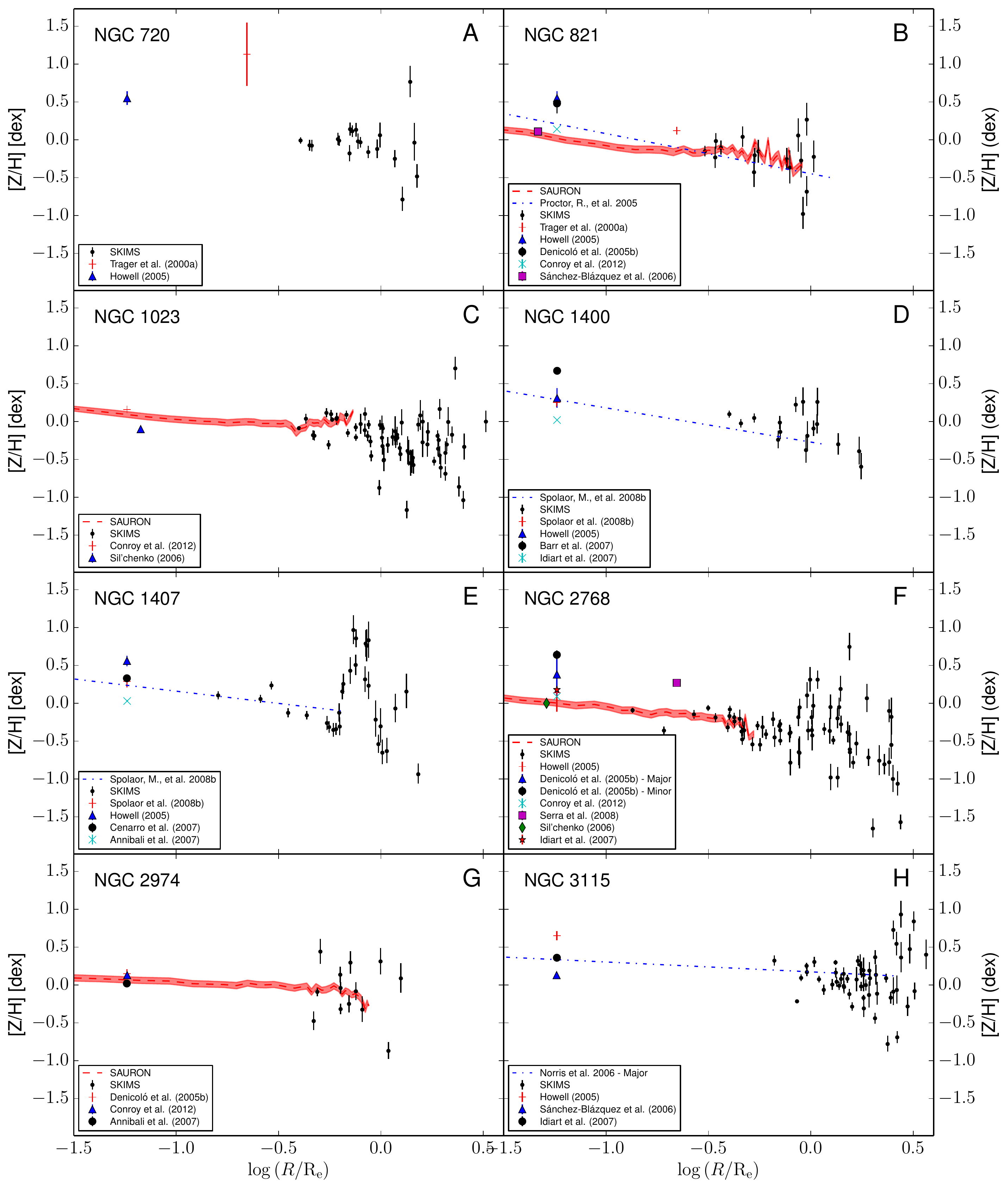}
	\end{center}
	    \caption[]{1D Metallicity profiles. 
	    The plots (A-V) show metallicity versus galactocentric radius scaled by $\rm{R_{e}}$. 
	    The CaT-derived metallicities, empirically corrected in all the cases except NGC~7457, are presented 
	    as black dots and labelled as `SKiMS'. 
		When available, the \sauron\ metallicity radial profiles are presented as a red solid line. 
		Other literature radial profiles are presented as blue dashed and dot-dashed lines. 
		The central metallicities from the literature have been plotted at a luminosity-scaled radius from 
		the centre, assuming a de Vaucouleurs profile for the galaxy surface brightness. 
		The sources for all the values and the radial profiles are presented in the legend panels on the bottom-left 
		corner of each plot. 
	    }
    \label{fig:O}
  \end{figure*}
  
  \addtocounter{figure}{-1}
	\begin{figure*}
    	\begin{center}
			\includegraphics[width=2\columnwidth]{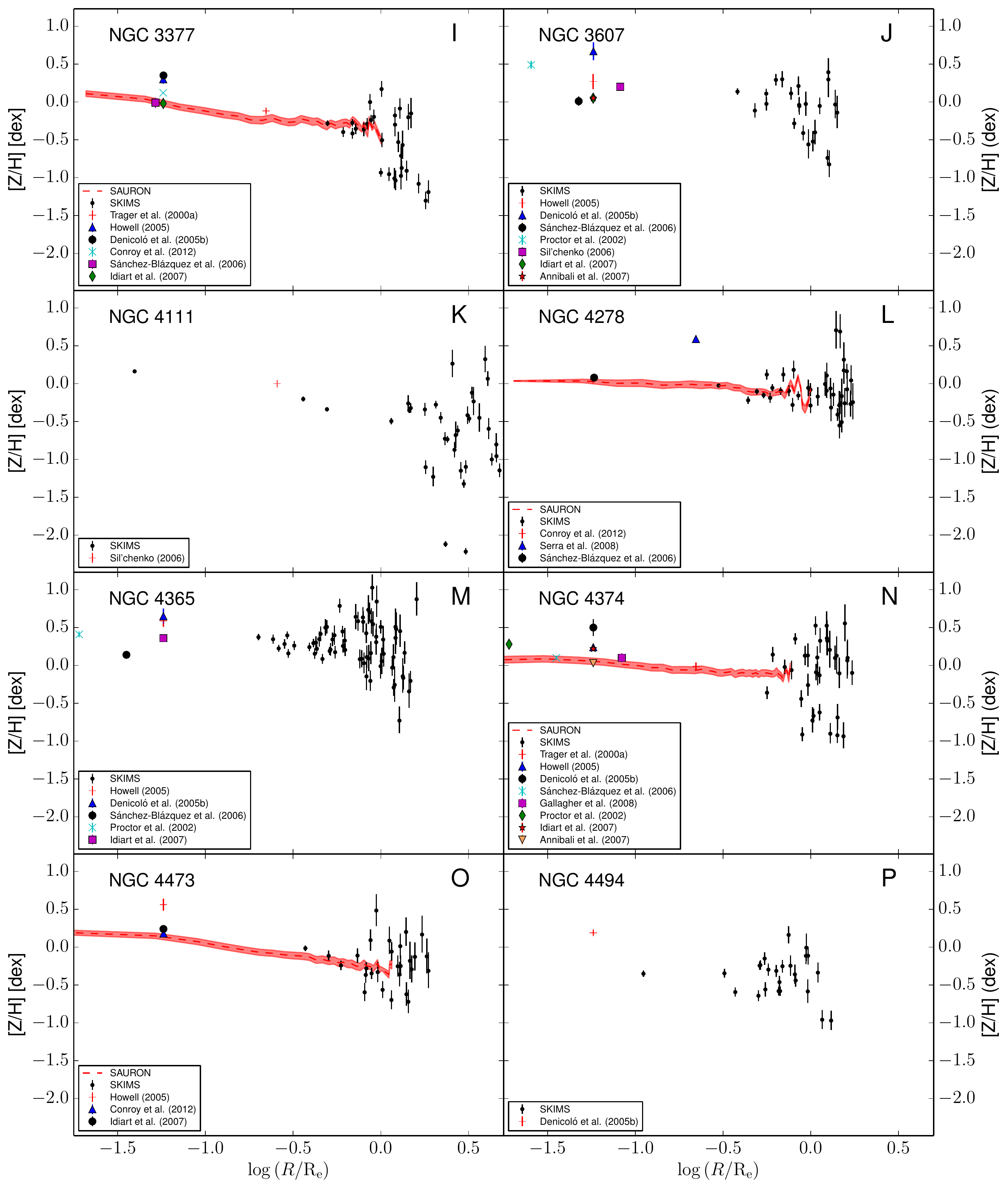}
	\end{center}
	    \caption[]{Continued. 
	    }
  \end{figure*}
  
  \addtocounter{figure}{-1}
	\begin{figure*}
    	\begin{center}
			\includegraphics[width=2\columnwidth]{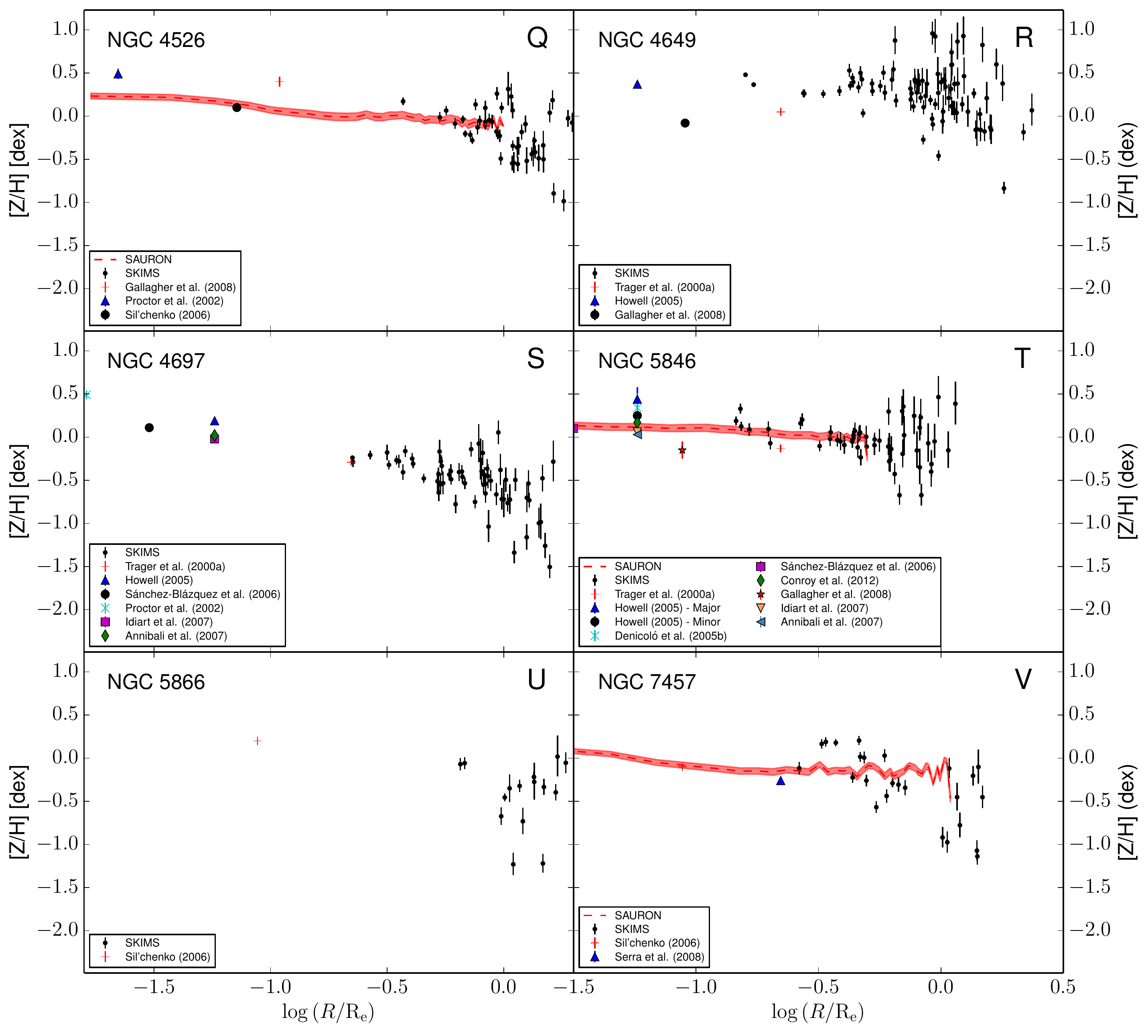}
	\end{center}
	    \caption[]{Continued. 
	    }
  \end{figure*}
}

\newcommand{\placefigPanalysis}{
	\begin{figure}
    	\begin{center}
			\includegraphics[width=\columnwidth]{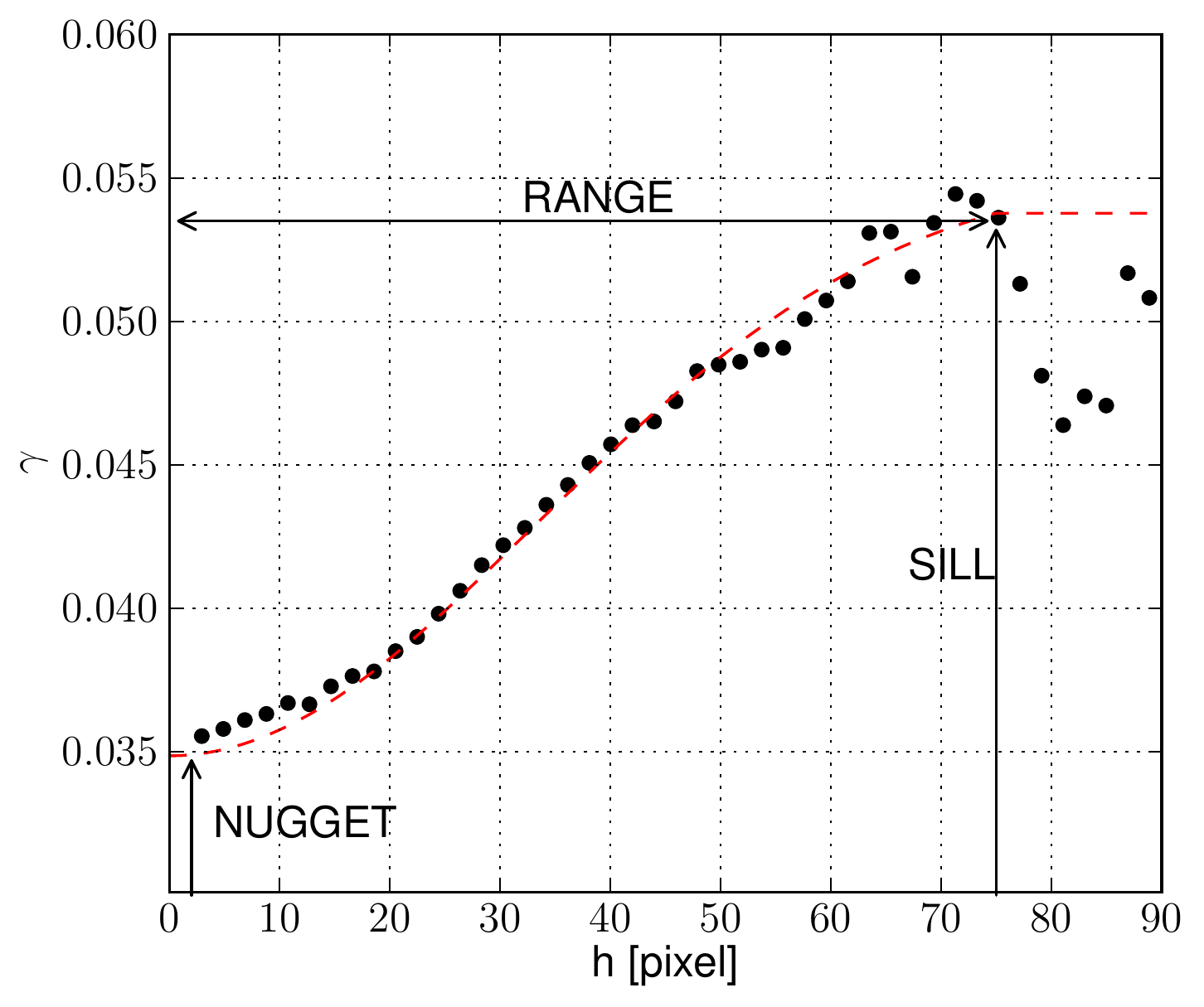}
	\end{center}
	    \caption[]{Semivariogram fitting example. 
	    The black points represent the mean value of the spatial autocorrelation $\gamma$ between pairs of points separated by a 
	    distance $h$.  
	    The red dashed line is the Gaussian function adopted to fit the distribution of points within the \textit{range} parameter, which is 
	    shown by the black horizontal arrow. 
	    The left vertical line shows the \textit{nugget} parameter, while the right vertical line shows the \textit{sill} parameter (see text).  
	    Pairs of points separated by a distance $h >$~\textit{range} are excluded from the fit. 
	    }
    \label{fig:P}
  \end{figure}
}

\newcommand{\placefigQanalysis}{
	\begin{figure*}
    	\begin{center}
				\includegraphics[width=2\columnwidth]{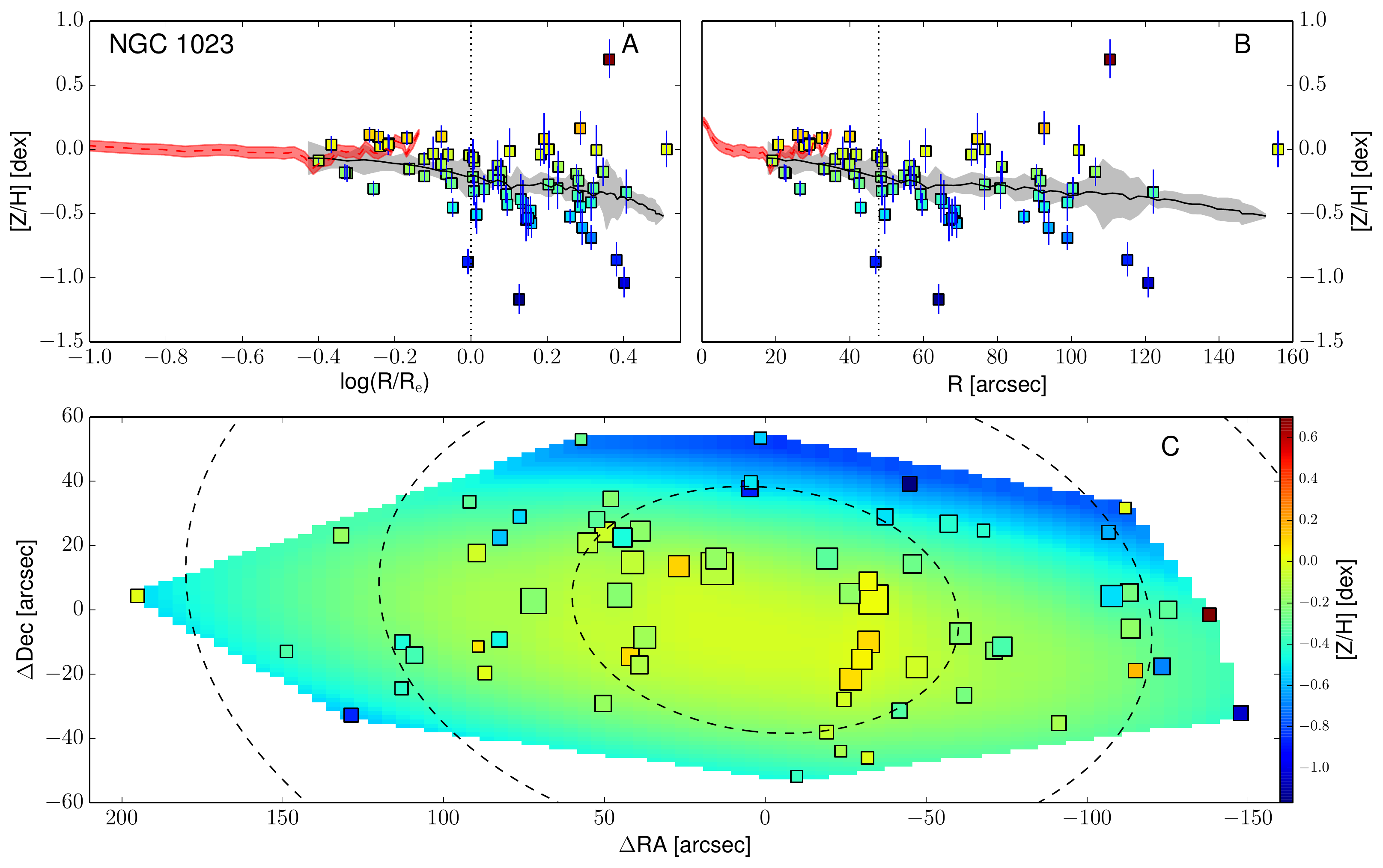}
		\end{center}
	    \caption[]{Empirically corrected 1D metallicity profiles and 2D metallicity maps. 
	    The \textit{A} panels show the metallicity profiles extracted from the kriging map (black lines) 
	    with the galactocentric radius in logarithmic space. 
	    The \textit{B} panels show the same profiles, but on a linear radial scale. 
	    In both these panels the measured metallicity data points are shown as squares 
	    colour coded according to their metallicity. 
	    The black dotted vertical line shows the radius corresponding to $1~\rm{R_{e}}$. 
	    When available, the literature profiles are overplotted as red dashed (if \sauron\ 
	    profiles) or blue dot-dashed lines (see Figure \ref{fig:O}). 
	 	In this case the red dashed line shows the metallicity radial profile extracted from the 2D \sauron\ 
		metallicity map.
	    The \textit{C} panels present the 2D metallicity maps from kriging and the measured 
	    data points colour coded according to their metallicity values. 
	    The size of each point is inversely proportional to its uncertainty. 
	    The black dashed lines show the isophotes at 1, 2 and $3~\rm{R_{e}}$, with ellipticity and PA from Table \ref{tab:A}. 
	    All of the data points, radial profiles and 2D metallicity maps have been corrected with the empirical relation presented 
	    in Equation \ref{eqn:correction}.
	  	 }
   \label{fig:Q}
  \end{figure*}
\clearpage
  \addtocounter{figure}{-1}
	\begin{figure}
    	\begin{center}
			\includegraphics[width=\columnwidth]{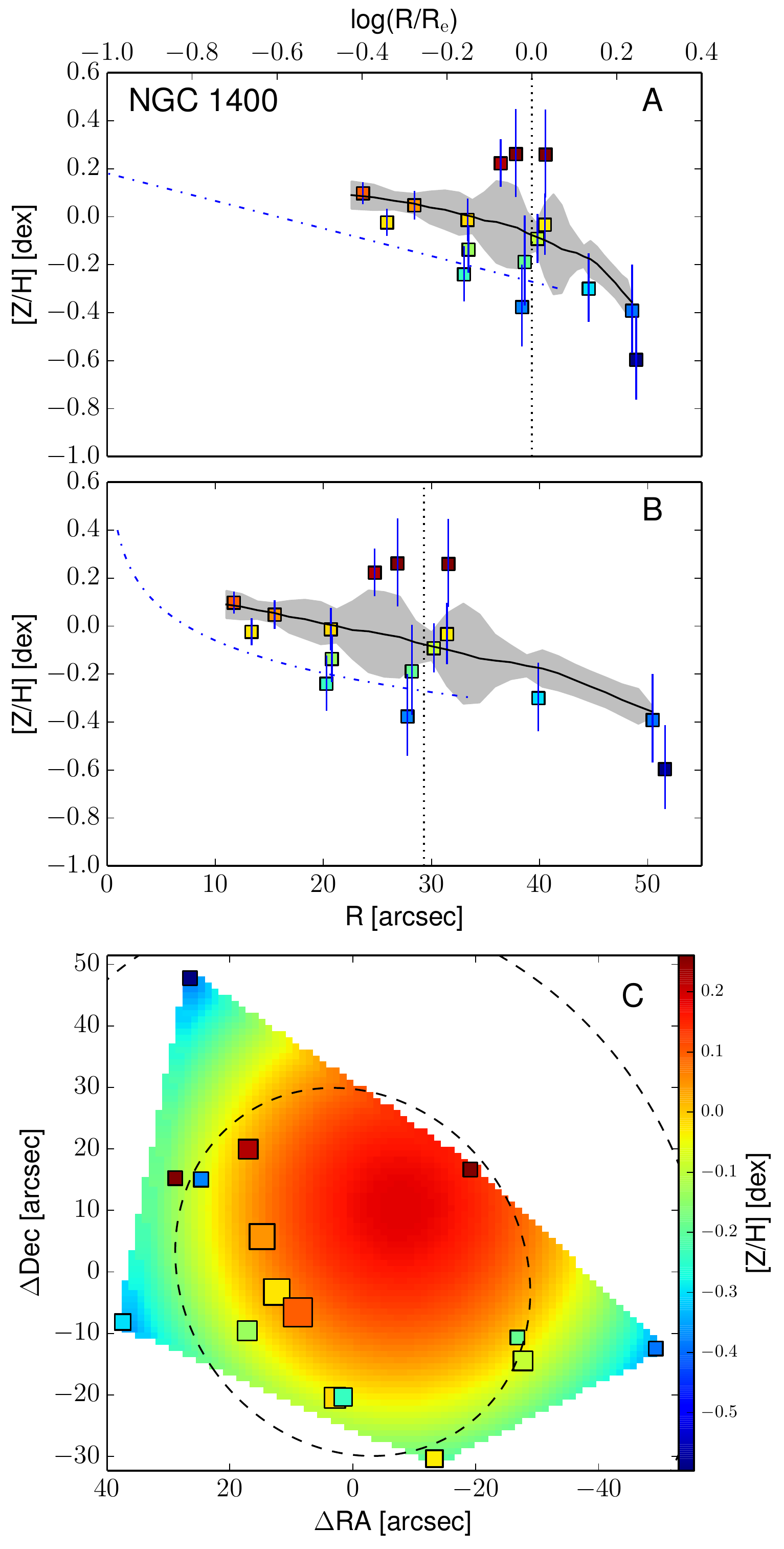}
	\end{center}
	    \caption[]{Continued. 
	    The dot-dashed blue line is the metallicity profile along the major axis 
	    as measured by \citet{Spolaor08b}. 
		 }
  \end{figure}

  \addtocounter{figure}{-1}
	\begin{figure}
    	\begin{center}
			\includegraphics[width=\columnwidth]{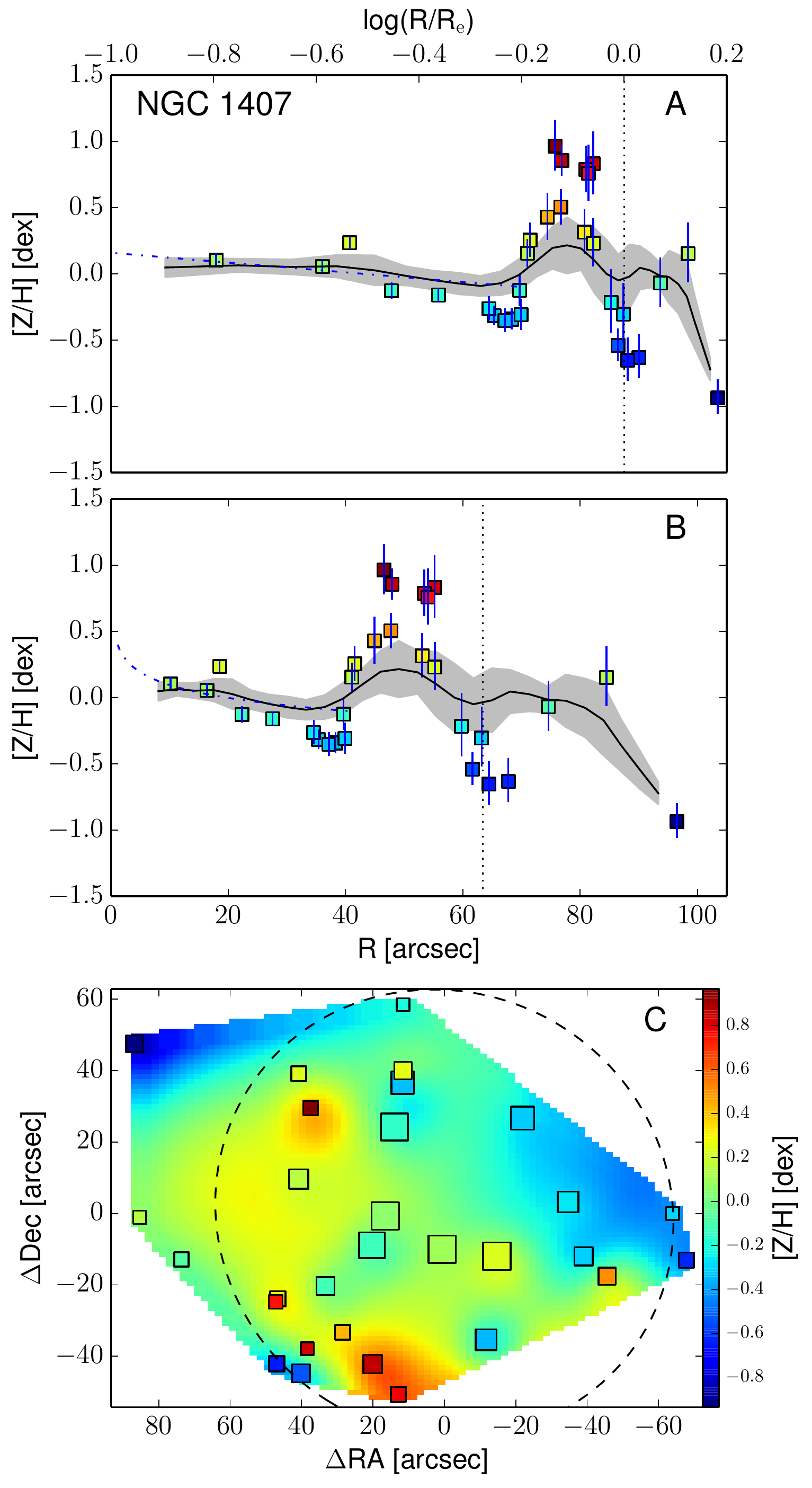}
	\end{center}
	    \caption[]{Continued. 
	    The dot-dashed blue line is the metallicity profile along the major axis 
	    as measured by \citet{Spolaor08b}. 
	    }
  \end{figure}

  \addtocounter{figure}{-1}
	\begin{figure*}
    	\begin{center}
			\includegraphics[width=2\columnwidth]{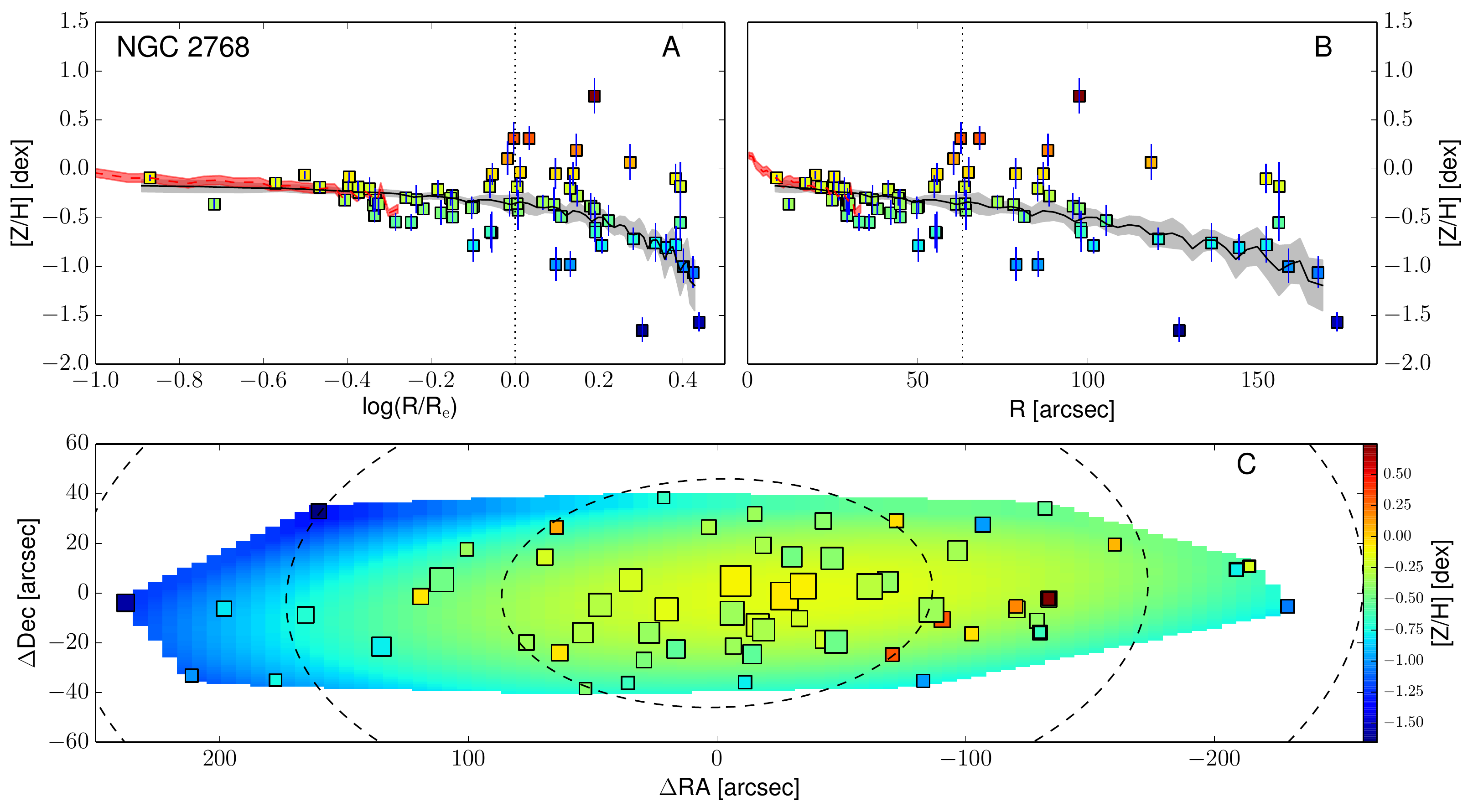}
	\end{center}
	    \caption[]{Continued. 
	    The red dashed line shows the metallicity radial profile extracted from the 2D \sauron\ 
		metallicity map. 
	    }
  \end{figure*}

  \addtocounter{figure}{-1}
	\begin{figure}
    	\begin{center}
			\includegraphics[width=\columnwidth]{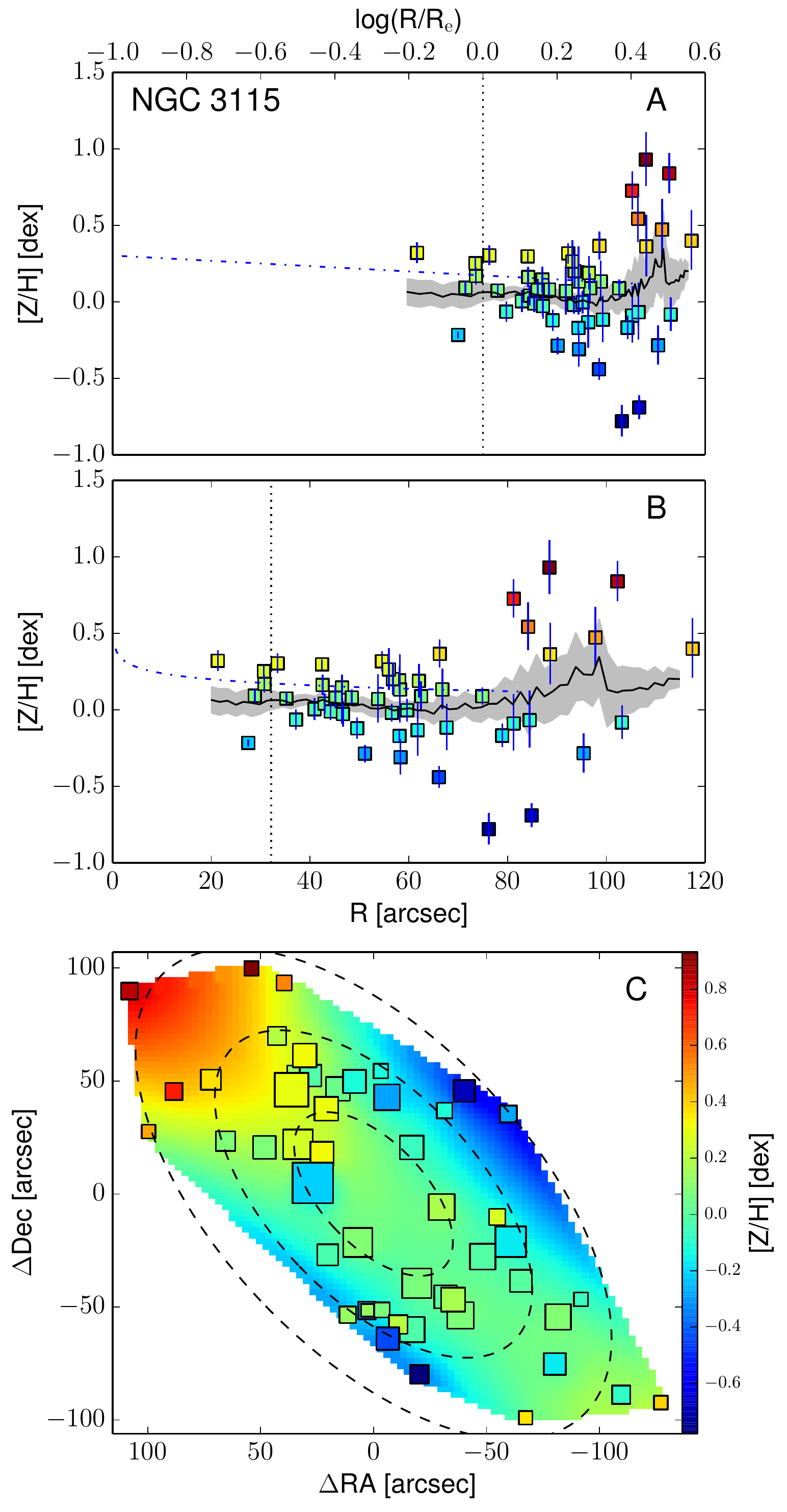}
	\end{center}
	    \caption[]{Continued. 
	    The dot-dashed line is the metallicity profile along the major axis 
	    as measured by \citet{Norris06}. 
	    }
  \end{figure}
\clearpage
  
  \addtocounter{figure}{-1}
	\begin{figure}
    	\begin{center}
			\includegraphics[width=\columnwidth]{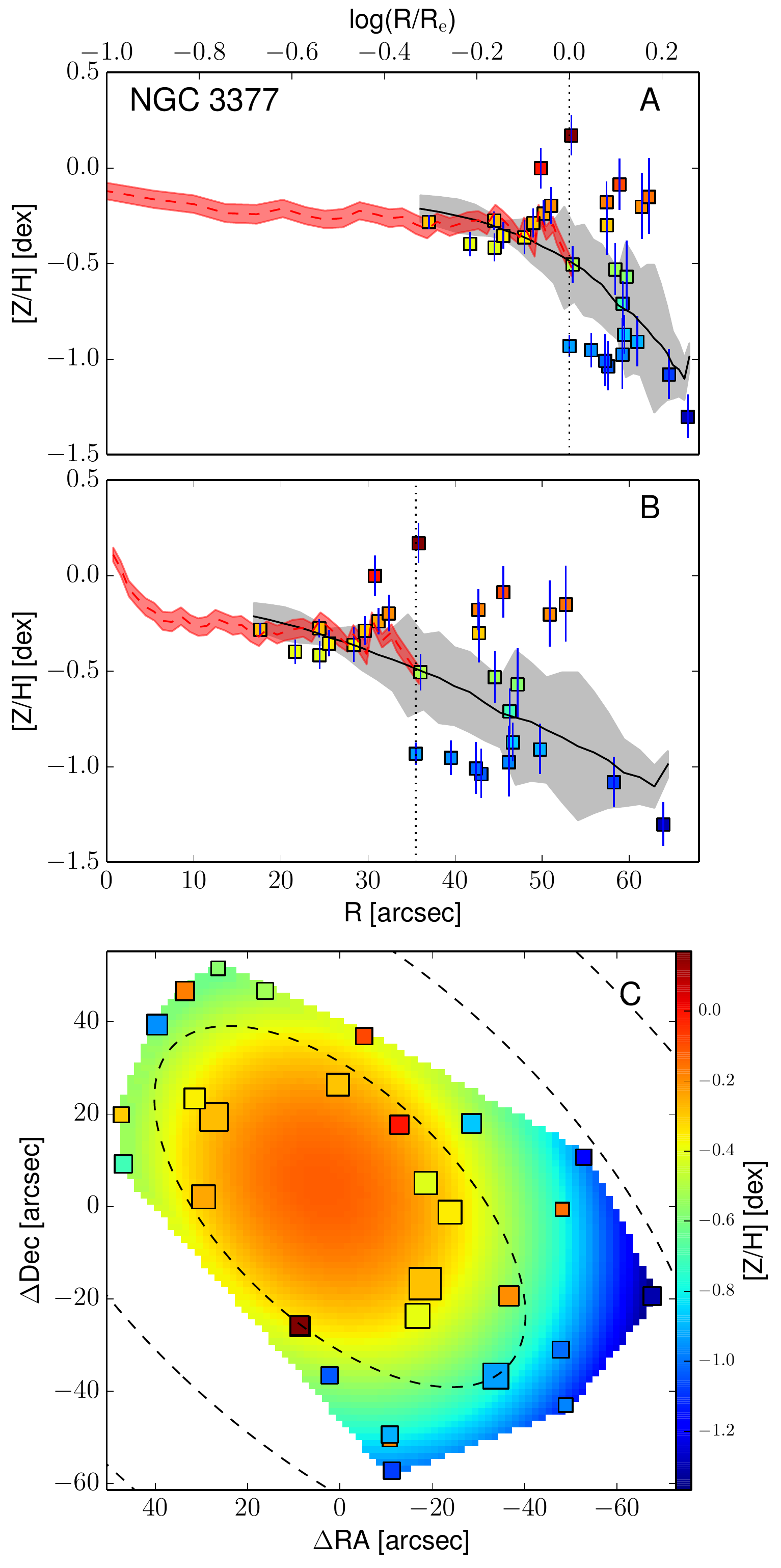}
	\end{center}
	    \caption[]{Continued. 
		The red dashed line shows the metallicity radial profile extracted from the 2D \sauron\ 
		metallicity map. 
	    }
  \end{figure}

  \addtocounter{figure}{-1}
	\begin{figure}
    	\begin{center}
			\includegraphics[width=\columnwidth]{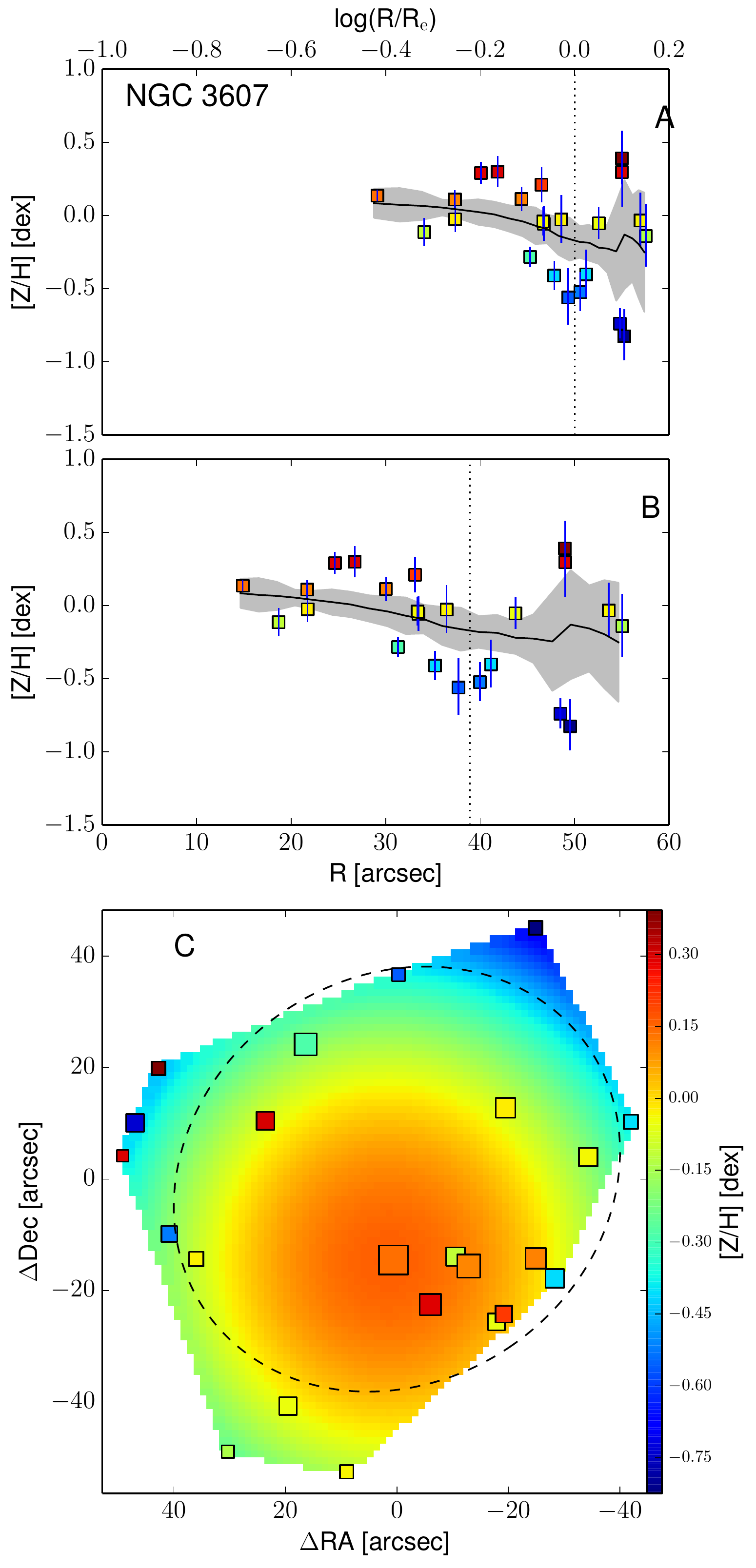}
	\end{center}
	    \caption[]{Continued. 
	    }
  \end{figure}
  
  \addtocounter{figure}{-1}
	\begin{figure}
    	\begin{center}
			\includegraphics[width=\columnwidth]{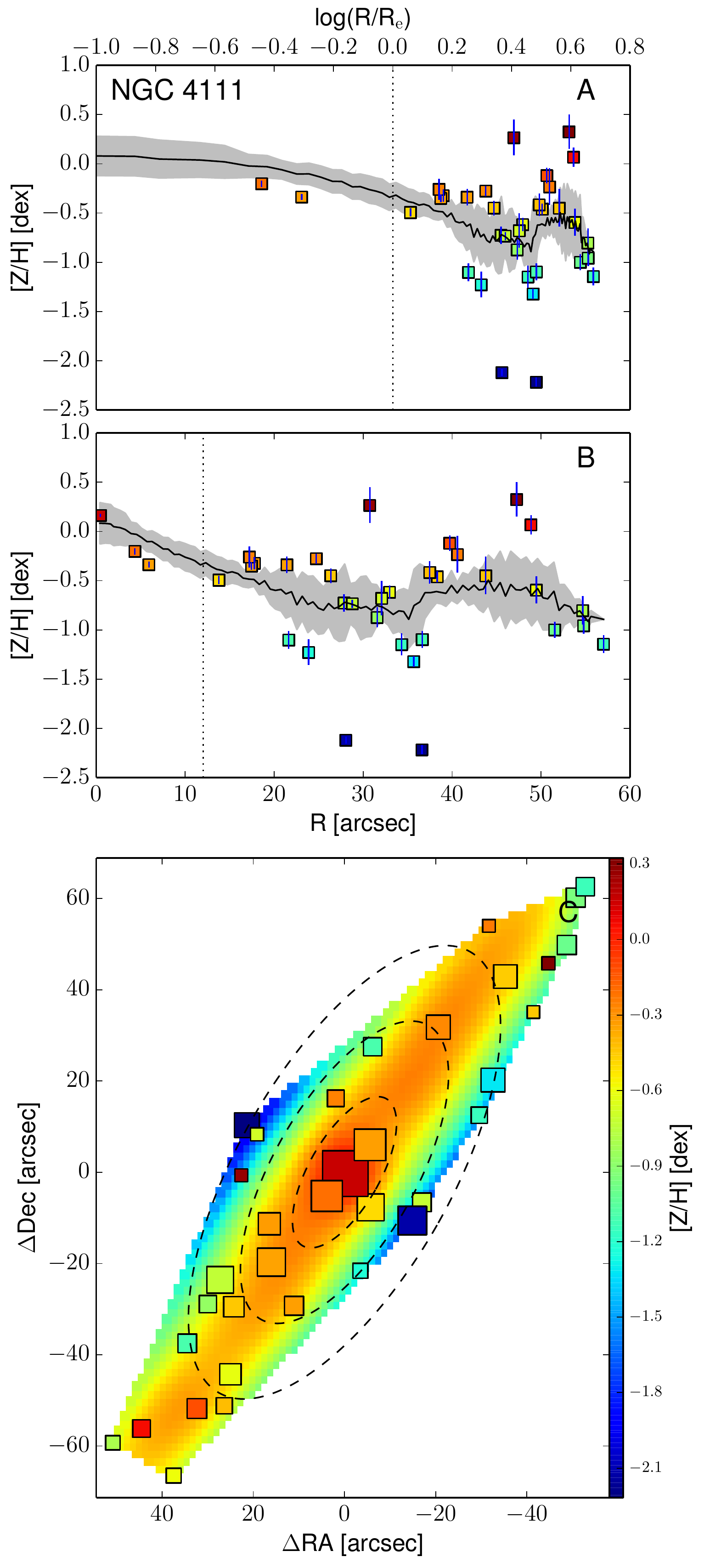}
	\end{center}
	    \caption[]{Continued. 
	    }
  \end{figure}
  
  \addtocounter{figure}{-1}
	\begin{figure}
    	\begin{center}
			\includegraphics[width=\columnwidth]{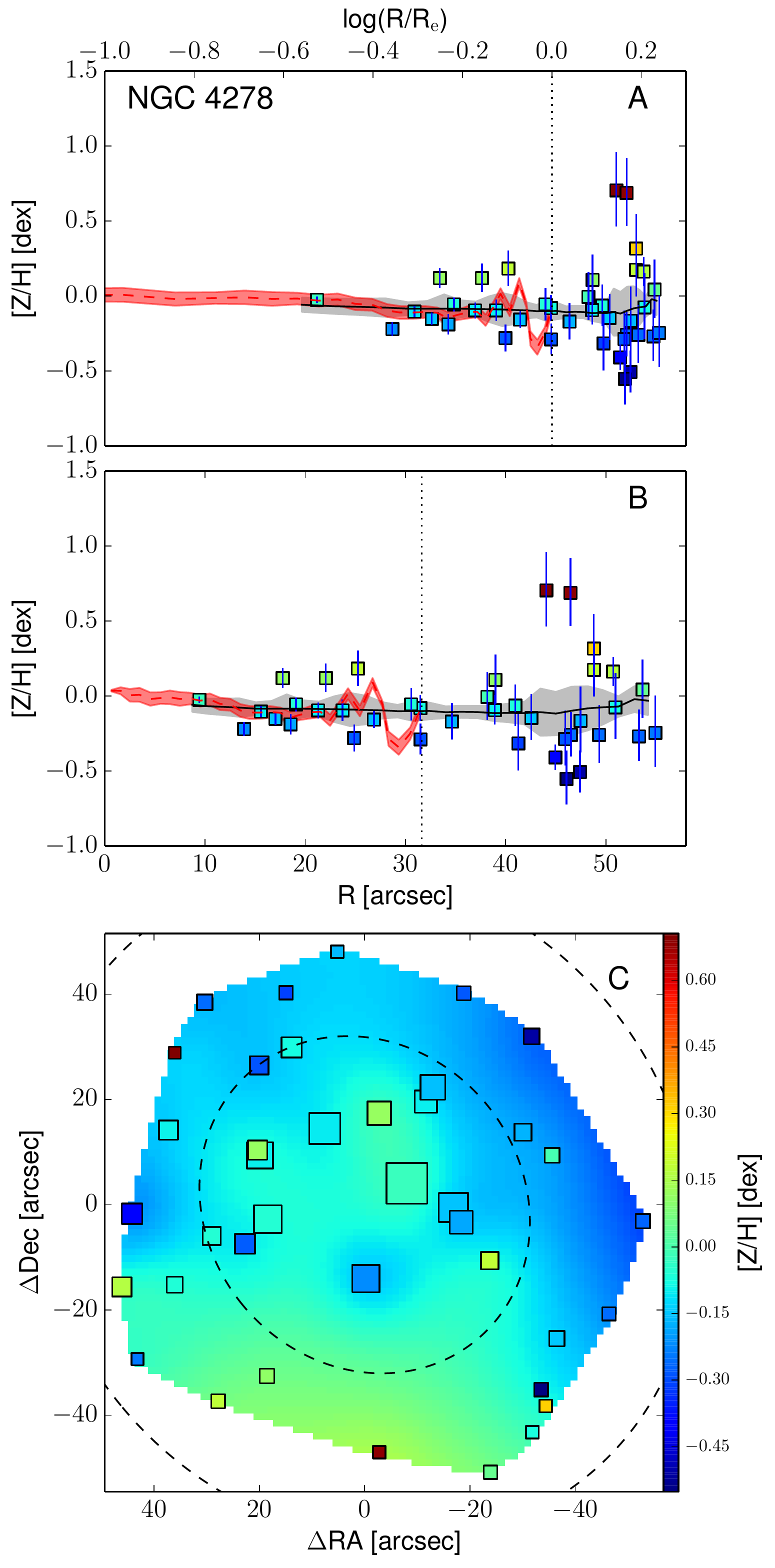}
	\end{center}
	    \caption[]{Continued. 
	    The red dashed line shows the metallicity radial profile extracted from the 2D \sauron\ 
		metallicity map. 
	    }
  \end{figure}

  \addtocounter{figure}{-1}
	\begin{figure}
    	\begin{center}
			\includegraphics[width=\columnwidth]{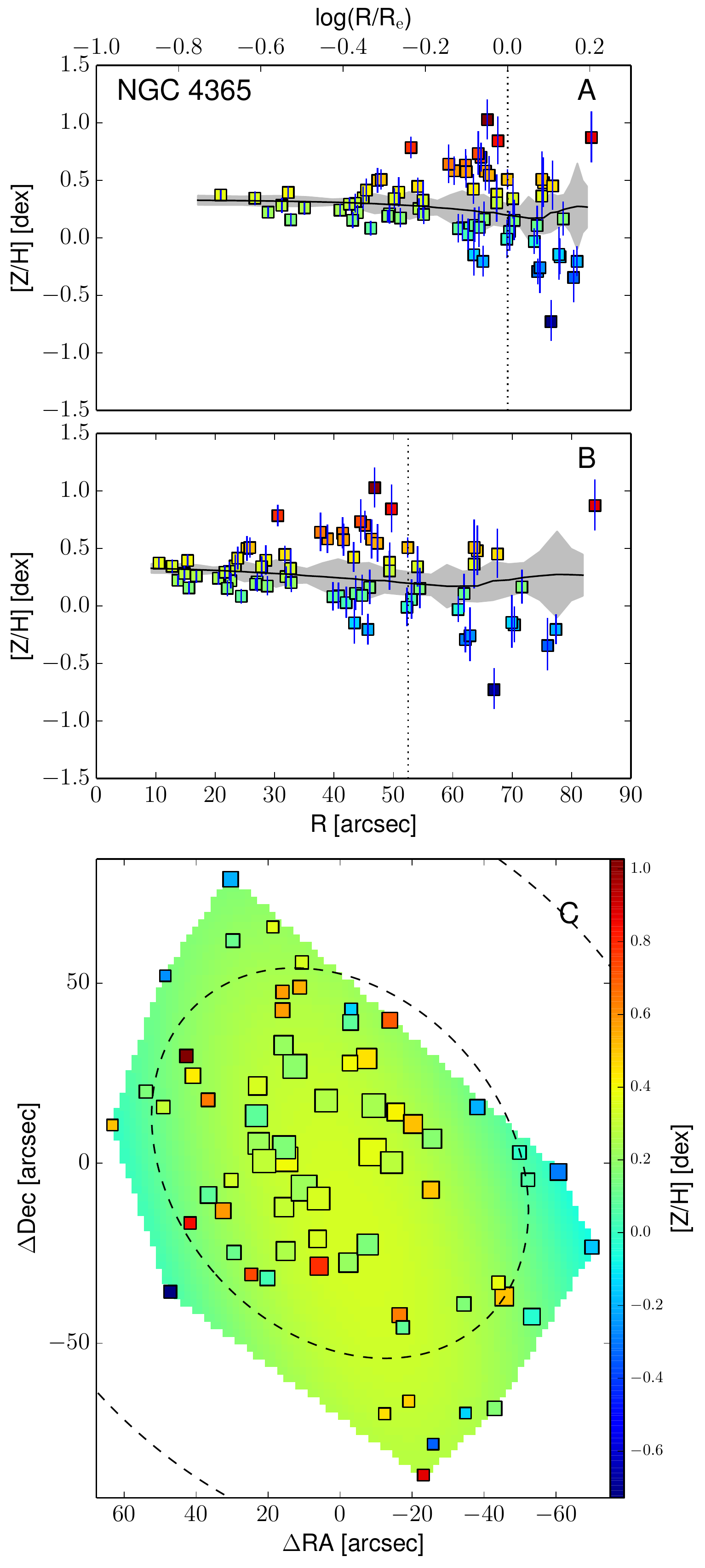}
	\end{center}
	    \caption[]{Continued. 
	    }
  \end{figure}
  
  \addtocounter{figure}{-1}
	\begin{figure}
    	\begin{center}
			\includegraphics[width=\columnwidth]{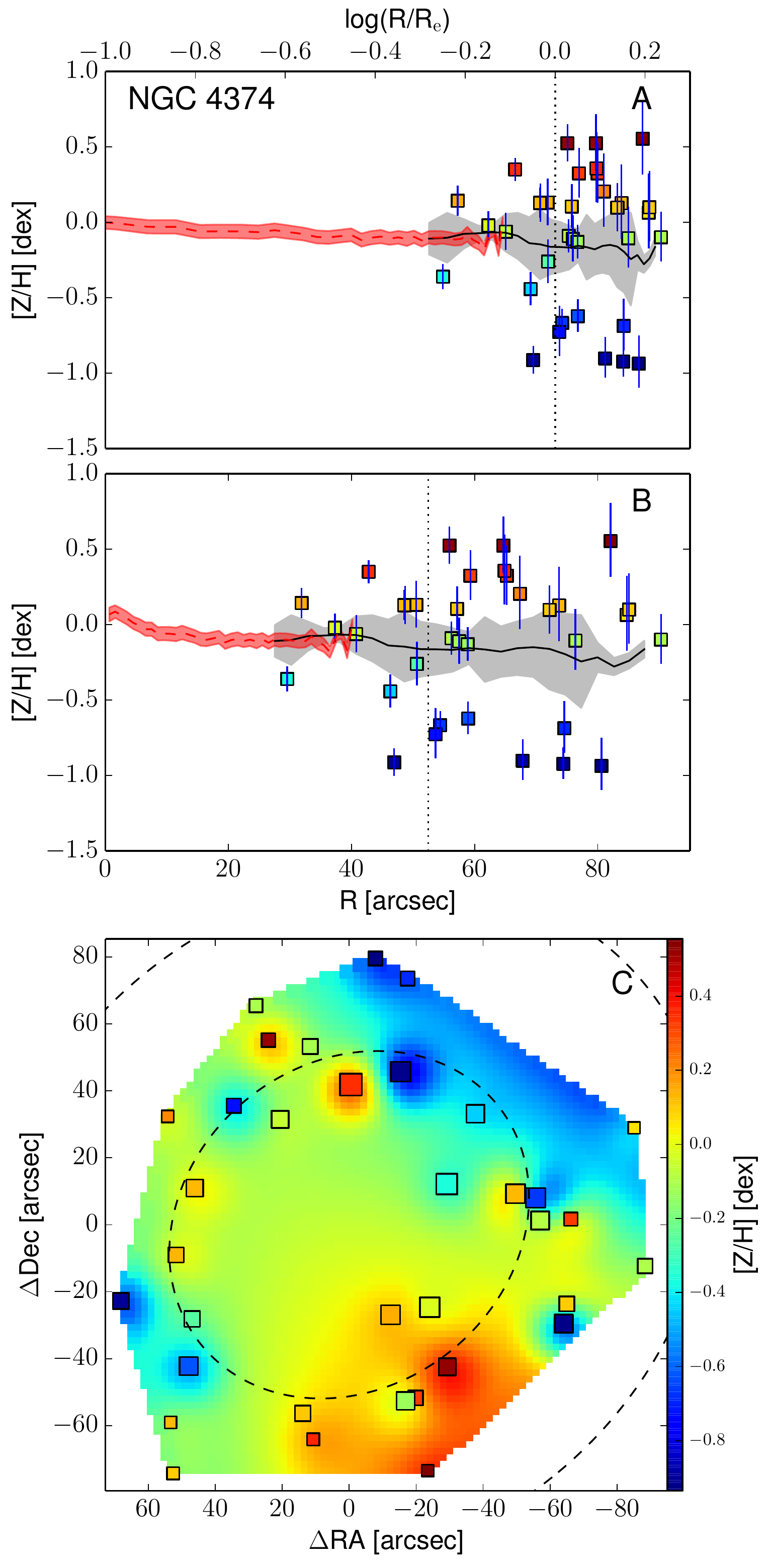}
	\end{center}
	    \caption[]{Continued. 
		The red dashed line shows the metallicity radial profile extracted from the 2D \sauron\ 
		metallicity map. 
	    }
  \end{figure}

 	\addtocounter{figure}{-1}
 		\begin{figure}
    	\begin{center}
			\includegraphics[width=\columnwidth]{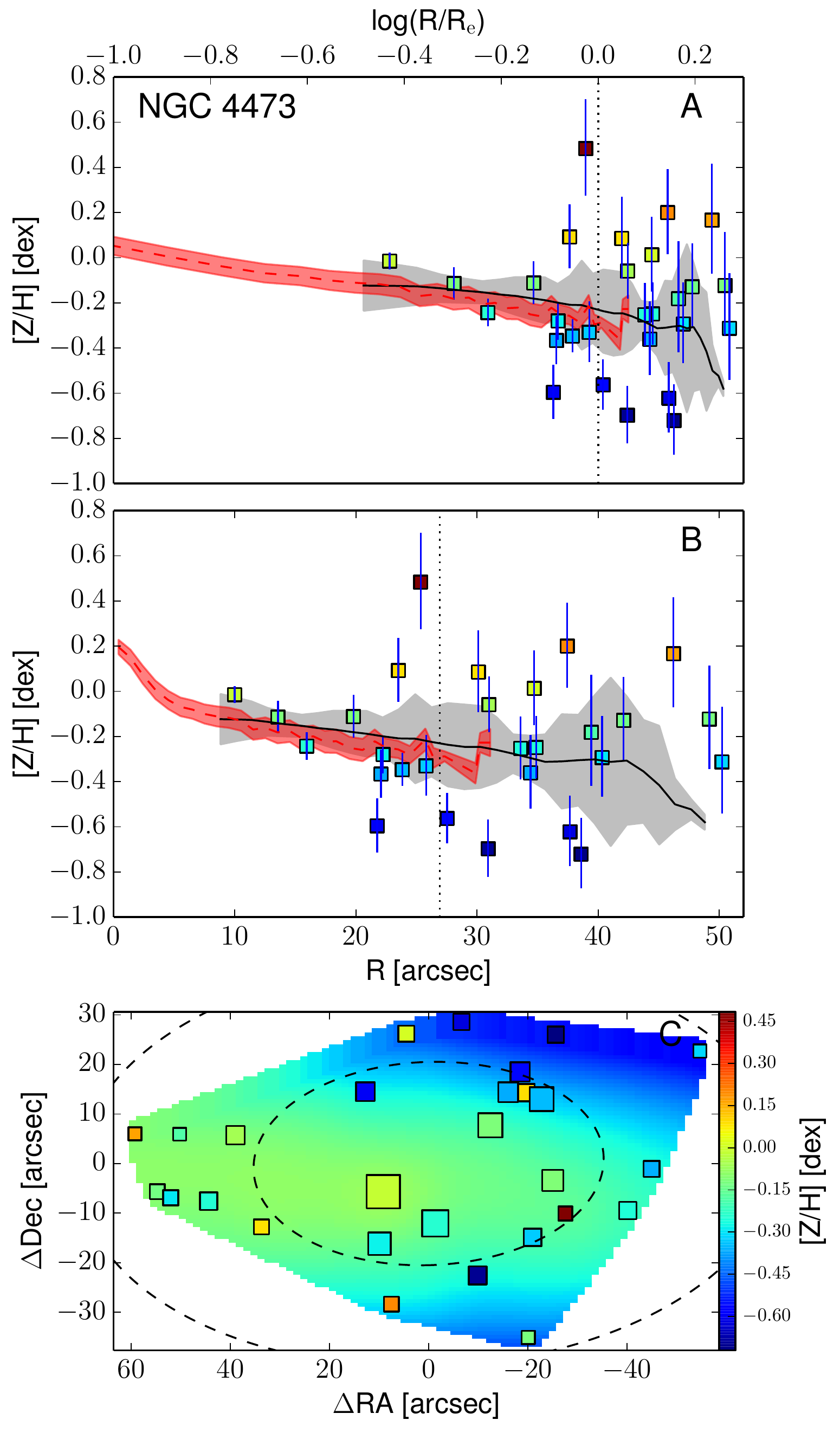}
	\end{center}
	    \caption[]{Continued. 
		The red dashed line shows the metallicity radial profile extracted from the 2D \sauron\ 
		metallicity map. 
	    }
  \end{figure}

 	\addtocounter{figure}{-1}
	\begin{figure}
    	\begin{center}
			\includegraphics[width=\columnwidth]{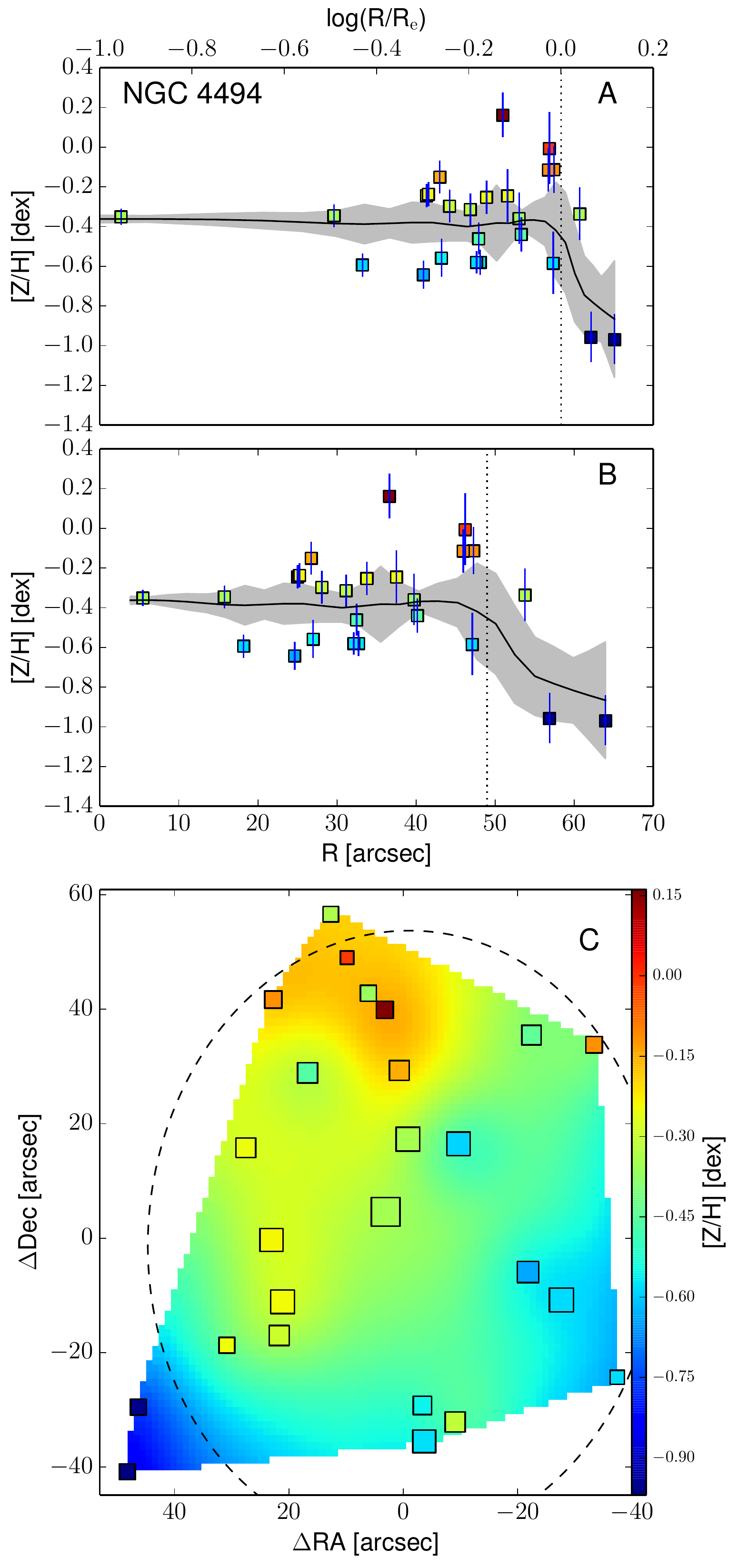}
	\end{center}
	    \caption[]{Continued. 
	    }
  \end{figure}

  \clearpage
  \addtocounter{figure}{-1}
	\begin{figure*}
    	\begin{center}
			\includegraphics[width=2\columnwidth]{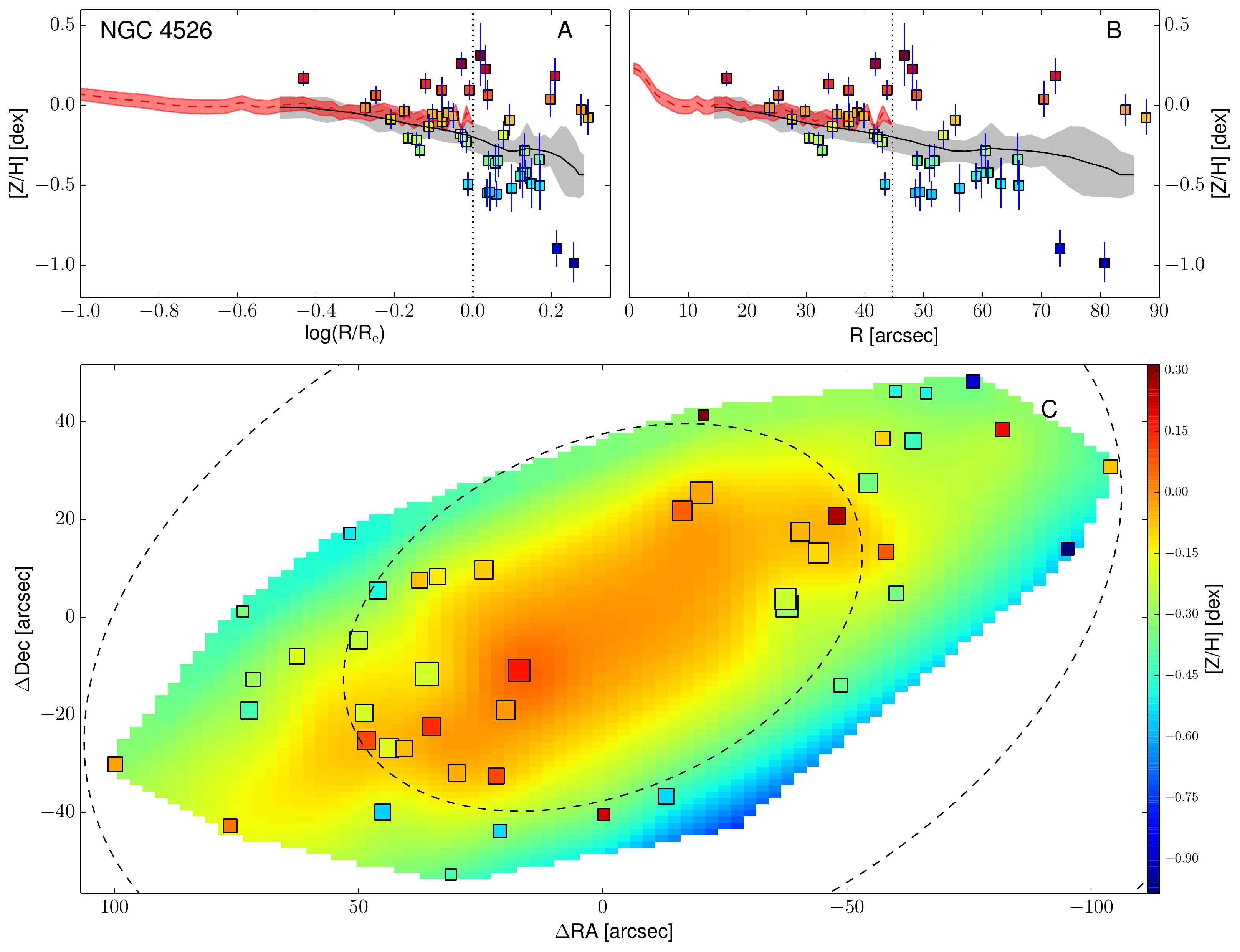}
	\end{center}
	    \caption[]{Continued. 
		The red dashed line shows the metallicity radial profile extracted from the 2D \sauron\ 
		metallicity map. 
	    }
  \end{figure*}
 \clearpage
 
  \addtocounter{figure}{-1}
	\begin{figure}
    	\begin{center}
			\includegraphics[width=\columnwidth]{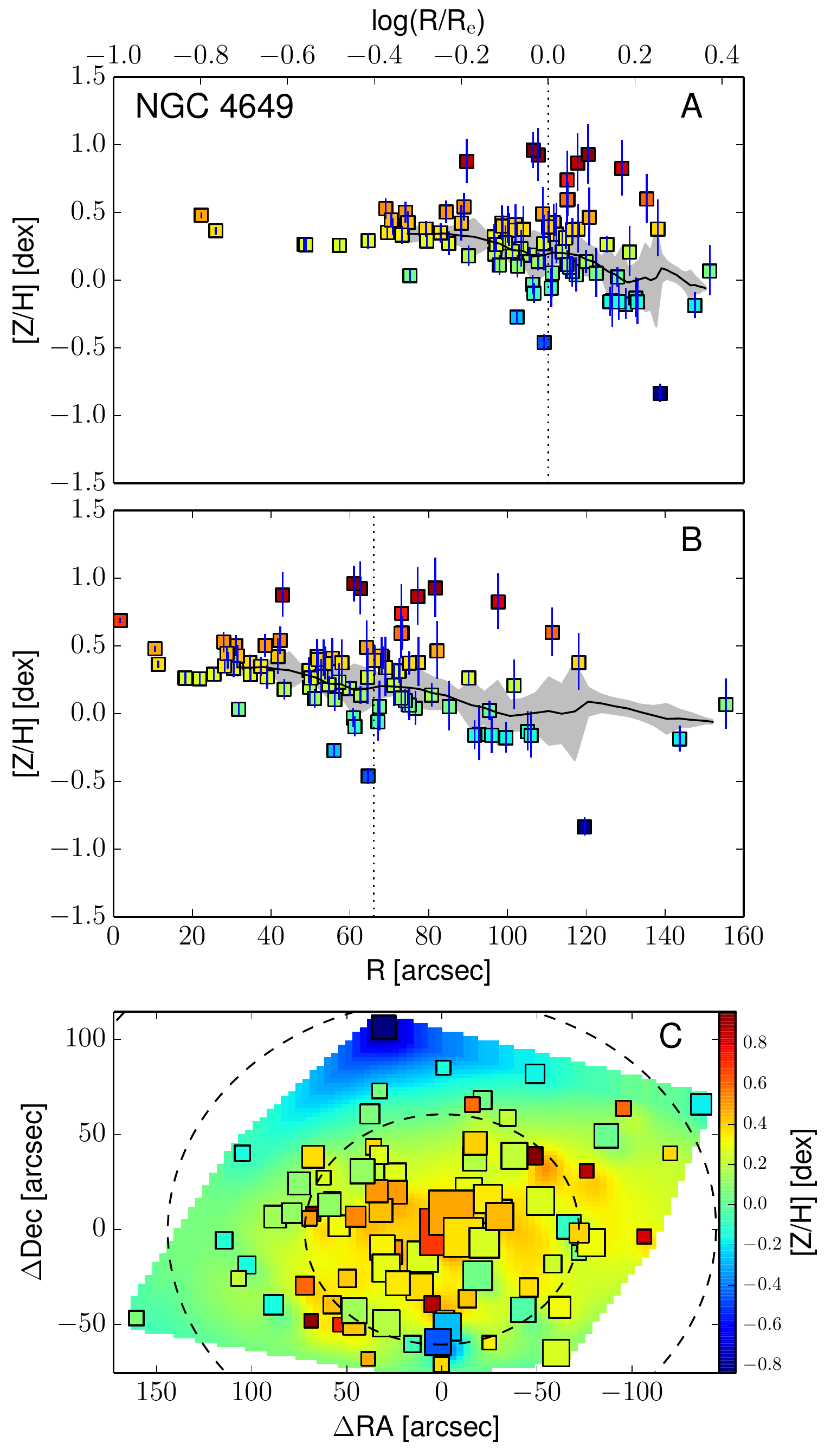}
	\end{center}
	    \caption[]{Continued. 
	    }
  \end{figure}
  
  \addtocounter{figure}{-1}
	\begin{figure*}
    	\begin{center}
			\includegraphics[width=\textwidth]{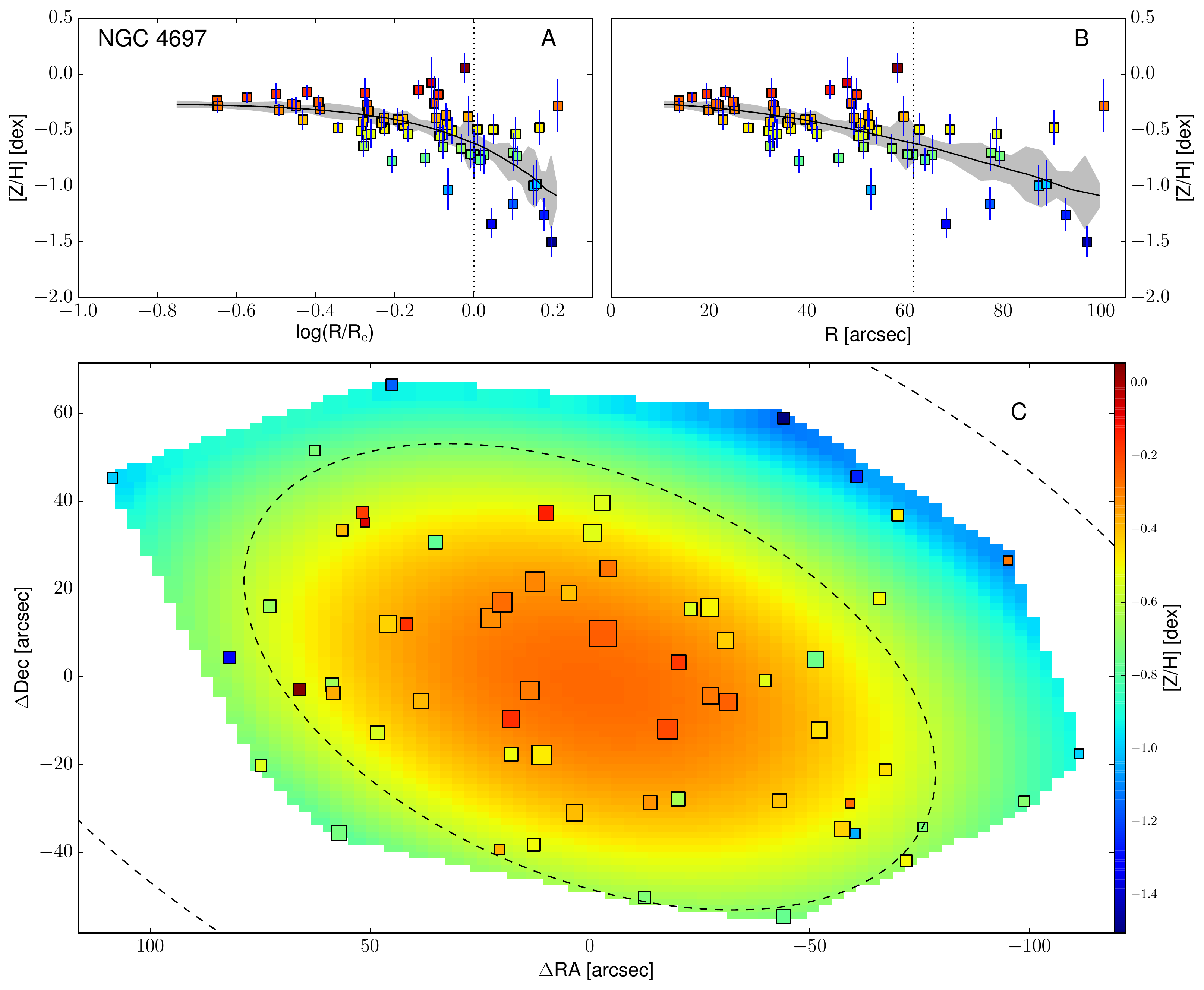}
	\end{center}
	    \caption[]{Continued. 
	    }
  \end{figure*}

 \clearpage

  \addtocounter{figure}{-1}
	\begin{figure}
    	\begin{center}
			\includegraphics[width=\columnwidth]{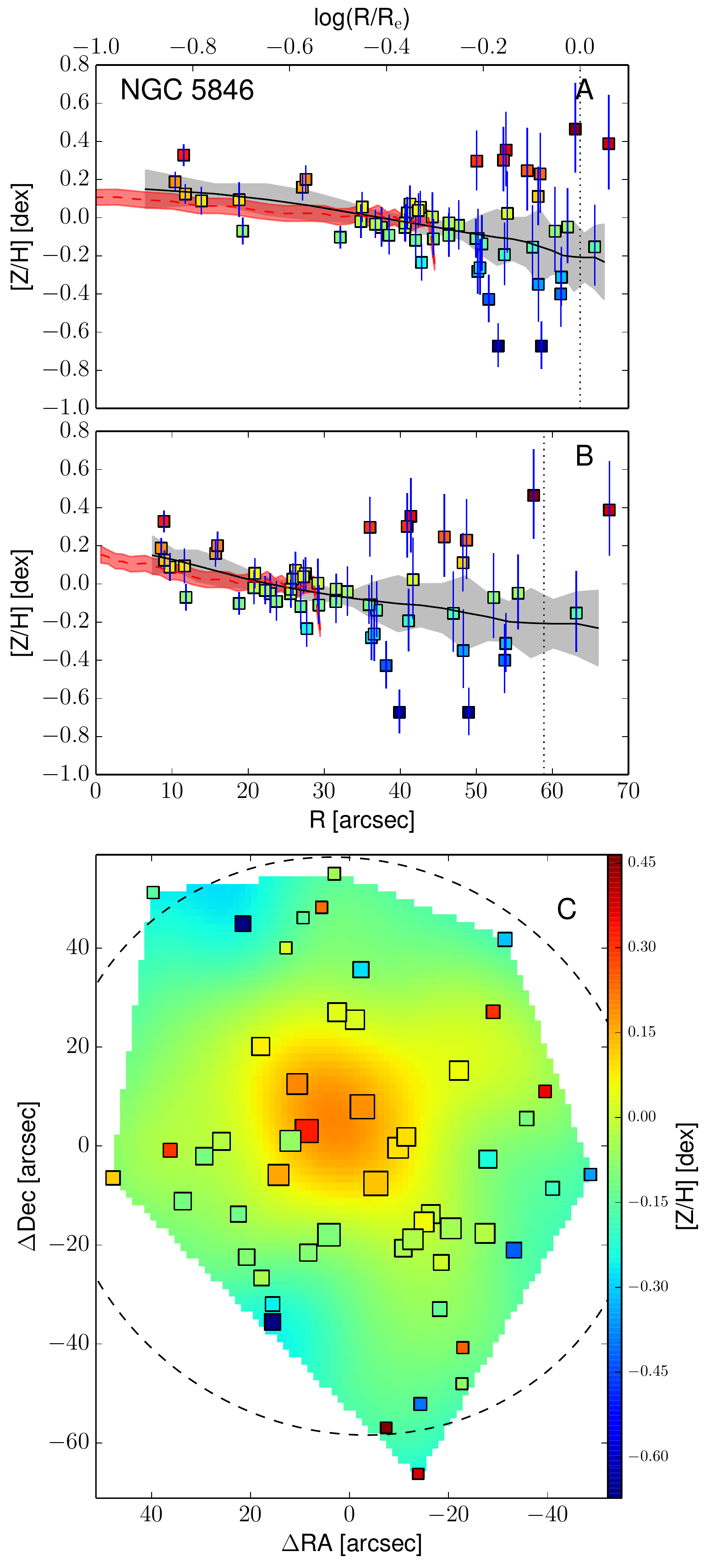}
	\end{center}
	    \caption[]{Continued. 
		The red dashed line shows the metallicity radial profile extracted from the 2D \sauron\ 
		metallicity map. 
	    }
  \end{figure}

  \addtocounter{figure}{-1}
	\begin{figure}
    	\begin{center}
			\includegraphics[width=\columnwidth]{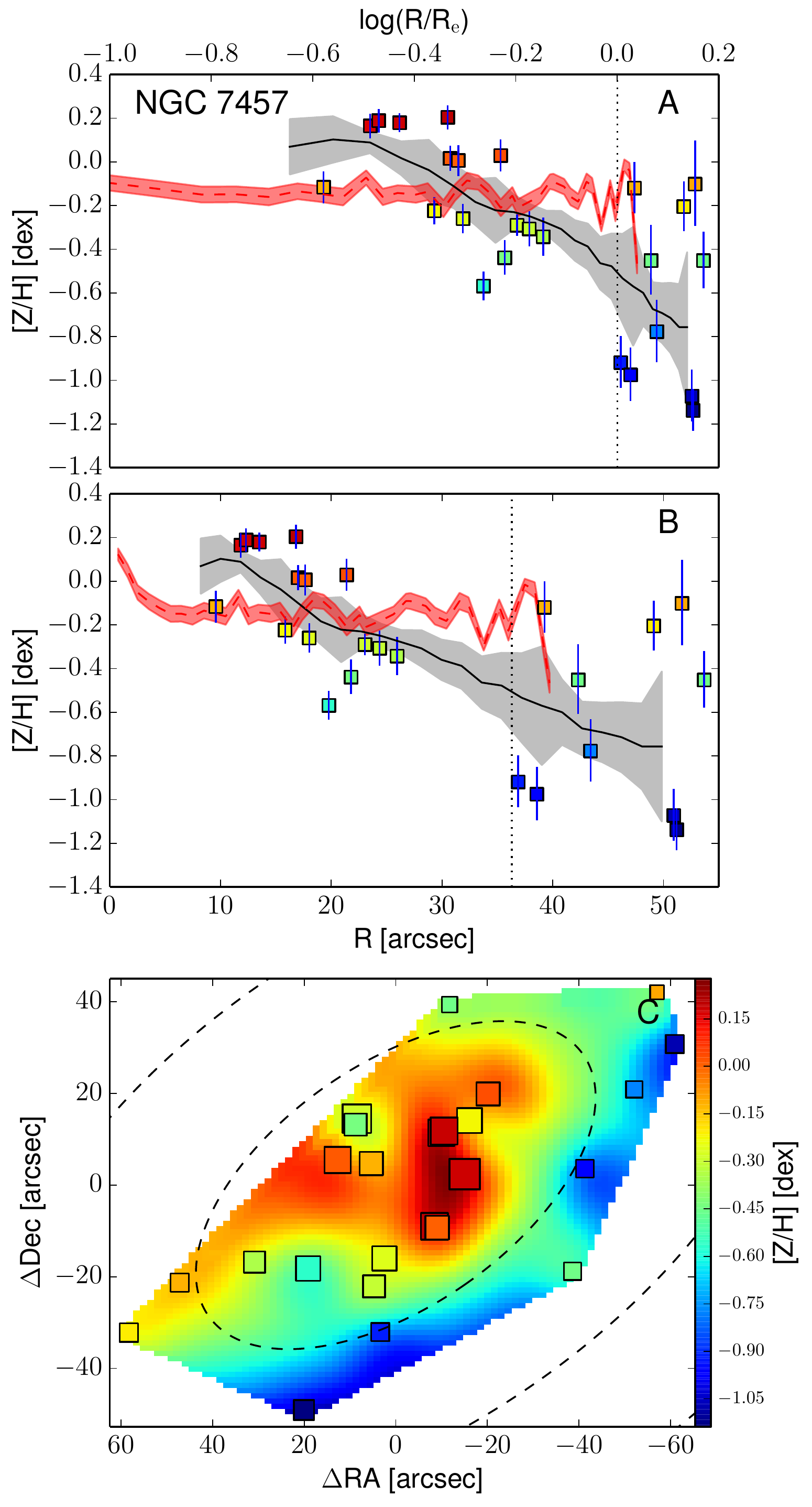}
	\end{center}
	    \caption[]{Continued. 
		The red dashed line shows the metallicity radial profile extracted from the 2D \sauron\ 
		metallicity map. 
	    }
  \end{figure}

 }

\newcommand{\placefigRanalysis}{
	\begin{figure*}
    	\begin{center}
			\includegraphics[width=\textwidth]{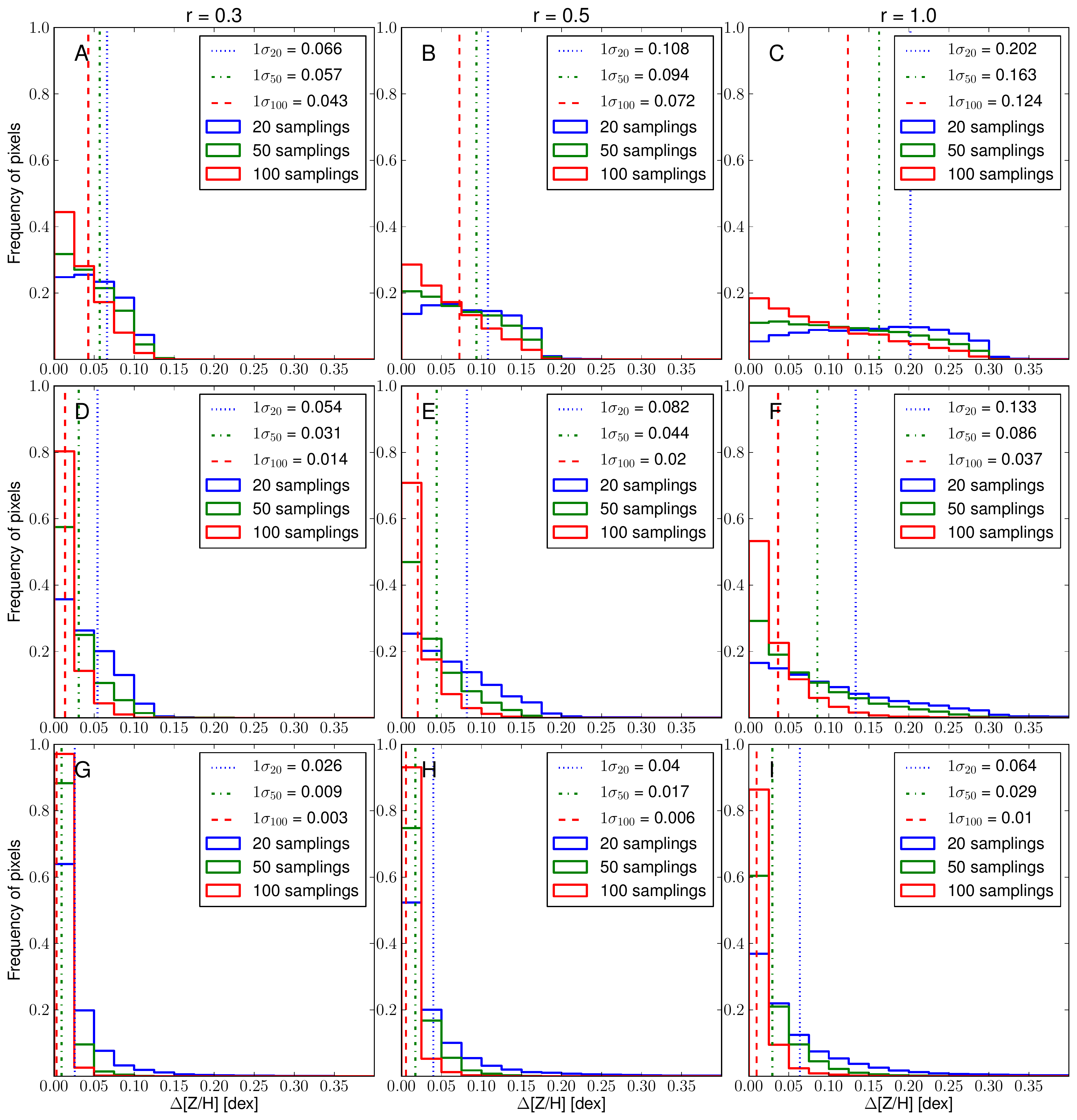}
     	\end{center}
	    \caption[]{Kriging substructures test results. 
	    The histograms present the distribution of the pixels along with the absolute difference between the 
	    original substructure value and the retrieved kriging value, in bins of $0.025~\rm{dex}$. 
	    Each panel shows the coaddition of 100 different statistical realizations of the sampling set for the 3 different 
	    cases of $20$, $50$ and $100$ sampling pixels, with, respectively, a blue, a green and a red histogram. 
	    The different panels are relative to the different combinations of the substructure parameters $r$ and $\sigma$, 
	    linked with the central value and the size of the substructure (see text).
	    The red dashed, the green dash-dotted and the blue dotted lines show the ranges where the 68\% of the pixels are 
	    enclosed in the $20$, $50$ and $100$ samplings cases respectively. 
	    These values are also presented in the legend of each panel.
}
    \label{fig:R}
  \end{figure*}
}

\newcommand{\placefigSanalysis}{
	\begin{figure*}
    	\begin{center}
			\includegraphics[width=\textwidth]{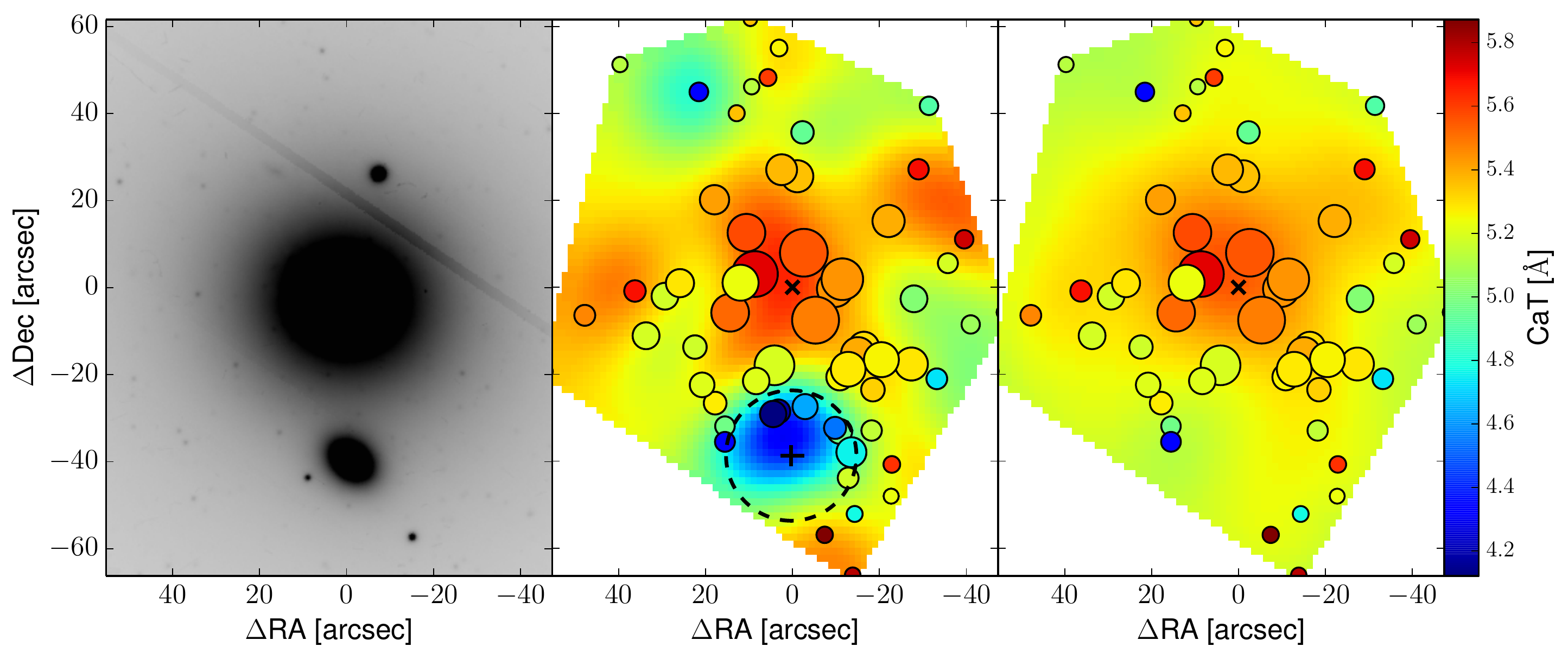}
	\end{center}
	    \caption[]{NGC~5846. 
	    The \textit{left} panel shows a DSS image of the NGC~5846 system, oriented North-up and East-left. 
	    The main object in the centre is NGC~5846, while the nearby companion galaxy NGC~5846A is 
	    a compact elliptical $\approx40"$ south of the NGC~5846 centre. 
	    On the central and right panels the CaT index kriging maps are presented, together with the 
	    measured points (circles).  
	    The \textit{central} panel shows the CaT index kriging map obtained from the CaT index data points (circles). 
	    The size of each point is inversely proportional to its uncertainty. 
	    Both the map and the points are colour coded accordingly to their CaT index values, following the colour scale 
	    on the right-hand side of the figure. 
	    The diagonal and the vertical crosses show, respectively, the positions of the NGC~5846 and NGC~5846A centres. 
		There is an area with low CaT index values (i.e. blue) around the centre of NGC~5846A. 
	    The black dashed line encloses the points within 15 arcsec of NGC~5846A. 
	    These points have been removed, being the most affected by NGC~5846A light. 
	    The \textit{right} panel shows the CaT index kriging map obtained after the removal of NGC~5846A points, 
	    together with the remaining data points (circles). 
	    Similarly to the \textit{central} panel, both map and points are colour coded according to their CaT index values, 
	    scaled as shown in the colour bar on the right. 
	    Again, the point sizes are inversely proportional to their associated CaT index uncertainties. 
	    After excluding data associated with NGC~5846A, the low-metallicity area is no longer found in 
	    the kriging map and we can obtain a metallicity map for just NGC~5846. 
	    }
    \label{fig:S}
  \end{figure*}
}

\newcommand{\placefigUdiscussion}{
	\begin{figure}
    	\begin{center}
			\includegraphics[width=\columnwidth]{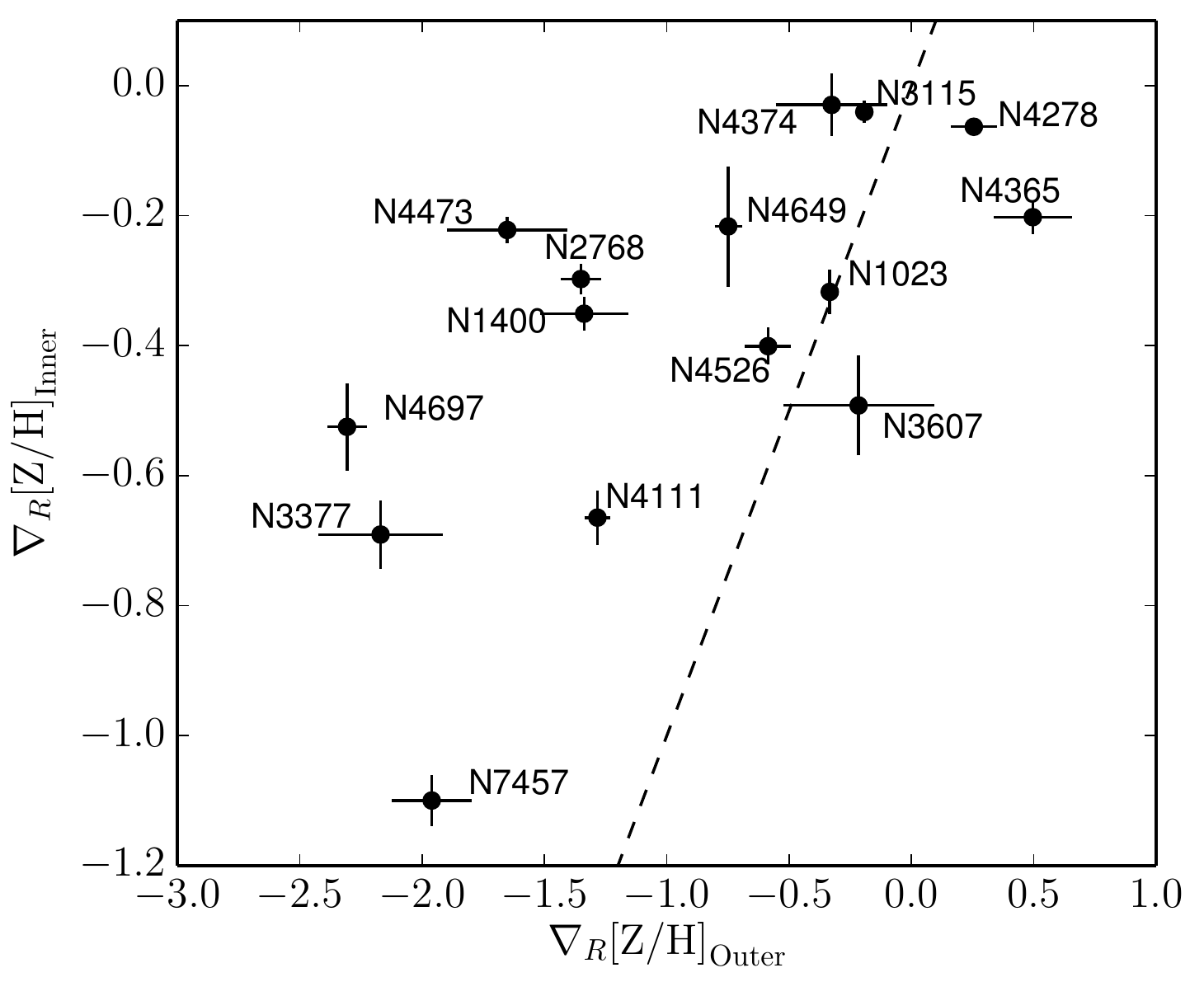}
	\end{center}
	    \caption[]{Comparison between the inner and the outer metallicity gradients. 
	    The inner metallicity gradients ($0.32 < R \leq 1~\rm{R_{\rm{e}}}$) 
	    are presented against those measured in the outer regions 
	    ($1 < R \leq 2.5~\rm{R_{e}}$). 
	    The dashed line is the 1:1 relation. 
	    Galaxies scatter about the 1:1 line, but most show a steeper metallicity gradient in the 
	    outskirts. 
	    }
    \label{fig:U}
  \end{figure}
}

\newcommand{\placefigVdiscussion}{
	\begin{figure*}
    	\begin{center}
     		\includegraphics[width=\textwidth]{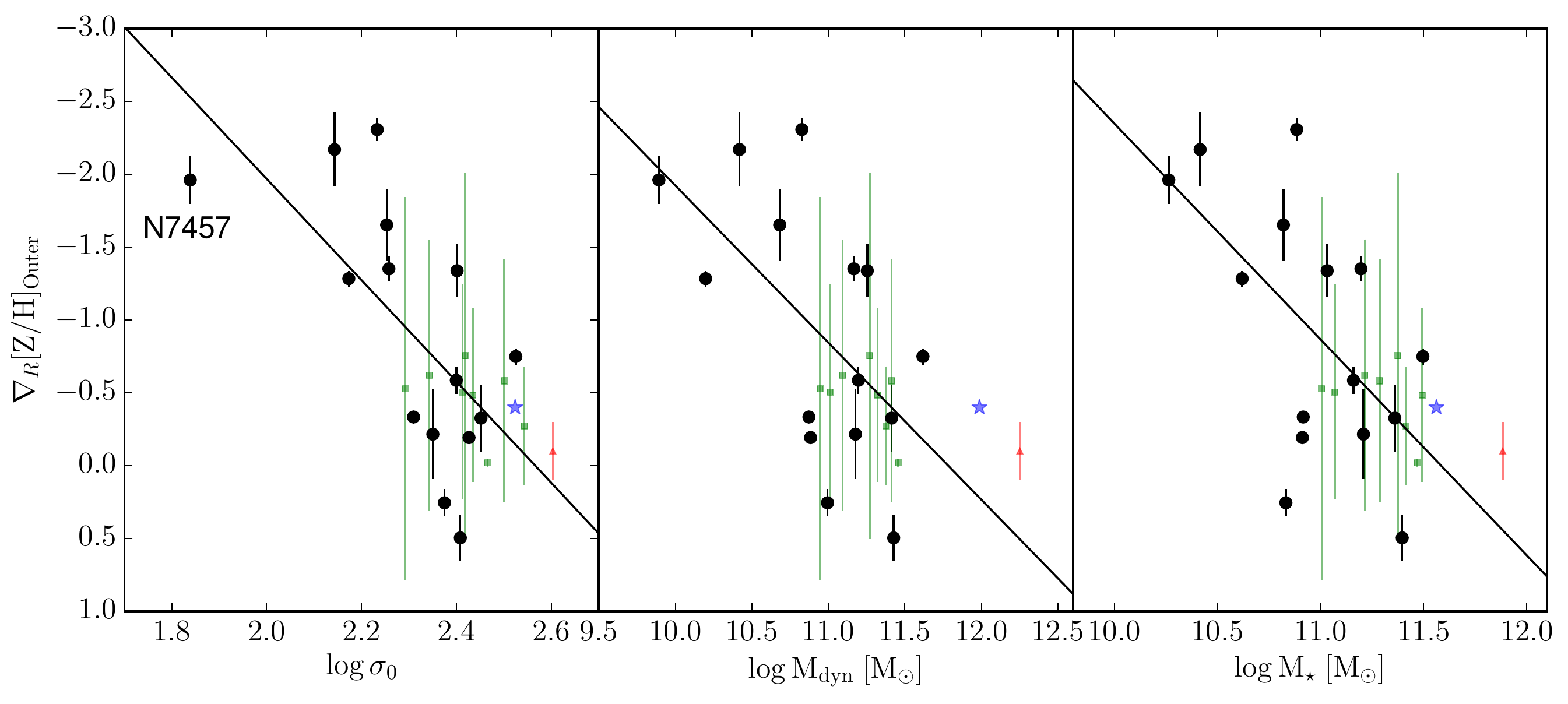}
	\end{center}
	  \caption{
	  	 	Outer ($1-2.5~\rm{R_{e}}$) metallicity gradient trends with galaxy mass proxies.
		   The outer metallicity gradient $\nabla_R\rm{[Z/H]_{Outer}}$ is shown 
		   against the central stellar velocity dispersion $\sigma_{\rm{0}}$, the 
		   dynamical mass $M_{\rm{dyn}}$ and the stellar mass $M_{\rm{\star}}$ 
		   from the \textit{K}-band luminosity, in the \textit{left}, 
		   \textit{central} and \textit{right} panel respectively. 
		   Steeper metallicity gradients are upwards on the plot. 
		   The black circles show the galaxies in our sample, while the green squares are the 
		   values from \citet{Greene12}, the blue star is NGC~4472 from \citet{Mihos13} 
	           and the red triangle is NGC~4889 from \citet{Coccato10}. 
		   The black solid lines are the linear fits to our sample (including NGC~7457) in all the cases.  
		   Thanks to the range in mass covered by this study, we are able for the first time to investigate the correlation 
		   between outer metallicity gradients and galaxy mass. 
		  In particular, each panel shows that steeper outer metallicity profiles are found in low-mass ETGs and shallower in giant ETGs. 
		   }
	    \label{fig:V}
  \end{figure*}
}

\newcommand{\placefigWanalysis}{
	\begin{figure}
    	\begin{center}
			\includegraphics[width=\columnwidth]{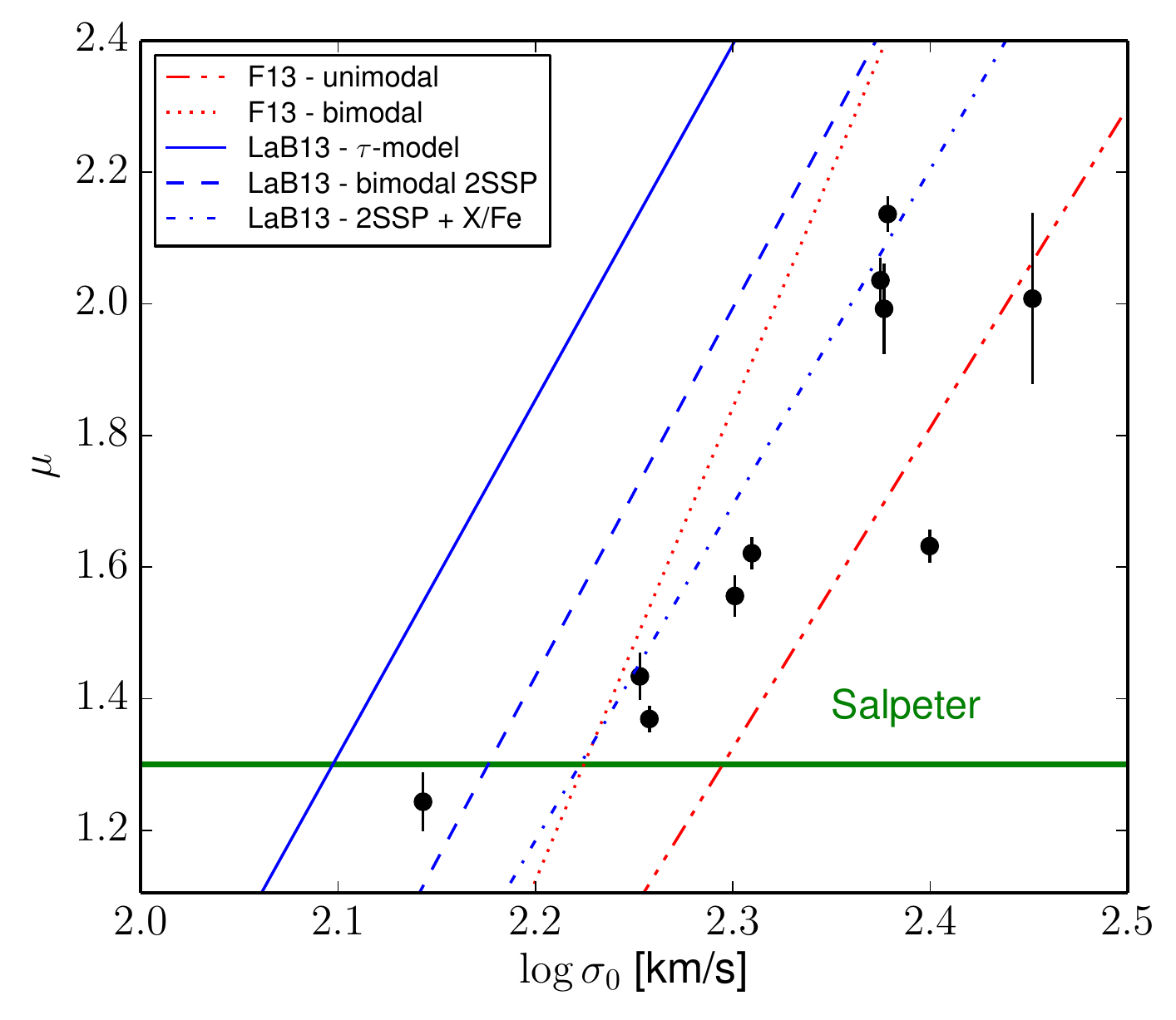}
		\end{center}
	    \caption[]{
	    The IMF slope trend with central velocity dispersion. 
	   	The IMF slope $\mu$ needed to compensate for the [Z/H] offset at $1~\rm{R_{e}}$ between the \sauron\ 
	   	profiles and ours is shown on the vertical axis for our sample galaxies (black circles).  
	   	On the horizontal axis the central velocity dispersion of the galaxies (proxy of the galaxy mass) 
	   	is shown. 
	   	A trend is noticeable, with high-mass galaxies having steeper IMFs.
	   	The green thick solid horizontal line shows the slope of the \citet{Salpeter55} IMF (i.e. $\mu = 1.3$). 
	   	The only point consistent with a slope shallower than a \citet{Salpeter55} IMF is NGC~3377. 	 
	   	The red dotted and dash-double dotted lines present the relations found by \citet{Ferreras13} with, 
	   	respectively, a bimodal and a unimodal IMF. 
	   	The blue solid, dashed and dot-dashed lines show, respectively, the relations found by 
	   	\citet{LaBarbera13} with their $\tau$, 2SSP and 2SSP+X/Fe models. 
	   	Despite the uncertainties and systematics involved in our measurements, our points show 
	   	a general agreement with the literature models. 
	   	}
    \label{fig:W}
  \end{figure}
}

\newcommand{\placefigXanalysis}{
	\begin{figure}
    	\begin{center}
			\includegraphics[width=\columnwidth]{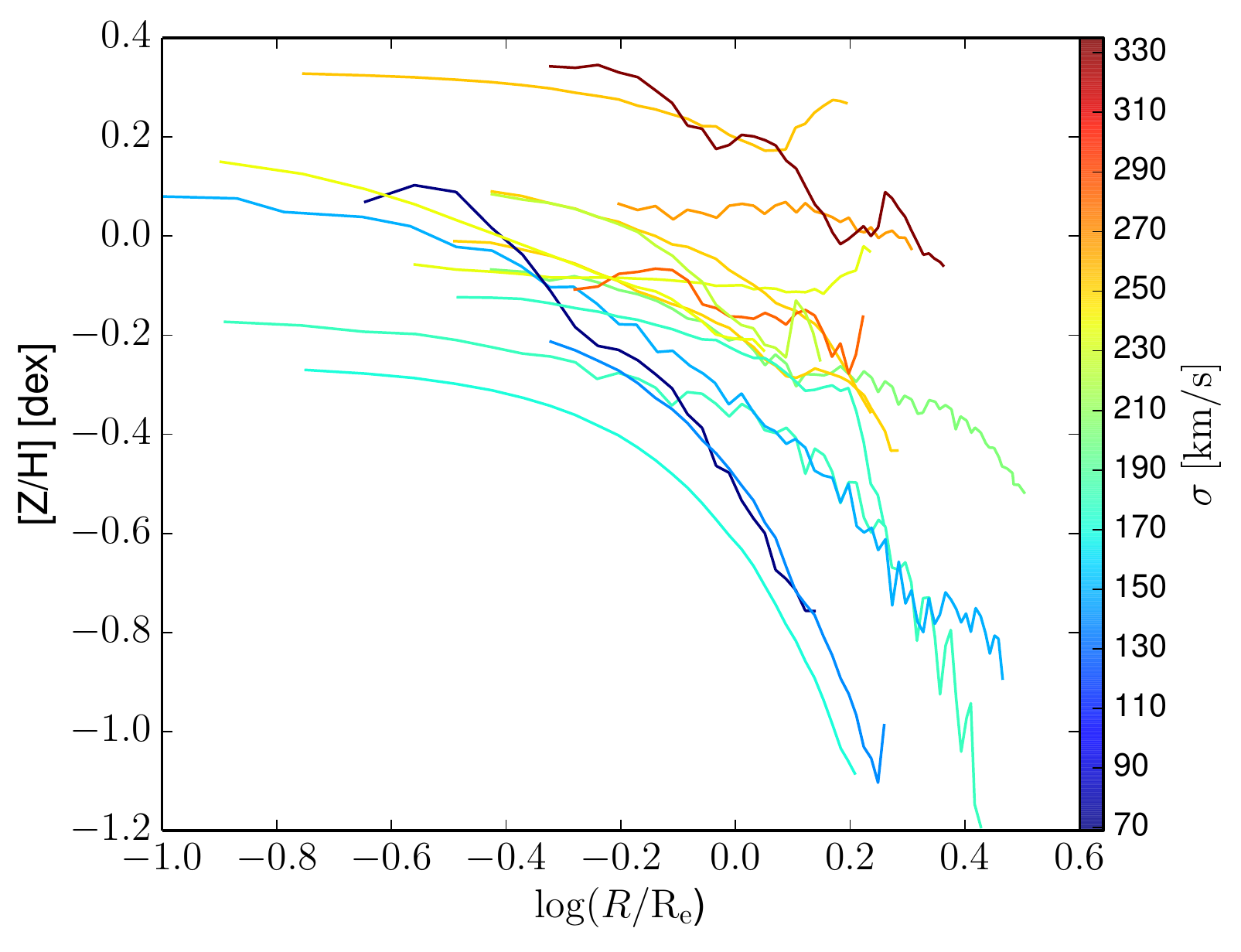}
		\end{center}
	    \caption[]{Metallicity radial profiles for the galaxies in our sample. 
	    We excluded NGC~1407 and NGC~4494 
	    and the outer regions of NGC~3115 and NGC~4111 due to large scatter. 
	    The plot shows metallicity versus galactocentric radius scaled by $\rm{R_{e}}$. 
	   	Each curve presents the radial metallicity profile of a galaxy, colour coded according to the galaxy 
	   	central velocity dispersion defined in Table \ref{tab:A}.
	    }
    \label{fig:X}
  \end{figure}
}

\newcommand{\placefigYanalysis}{
	\begin{figure}
    	\begin{center}
			\includegraphics[width=\columnwidth]{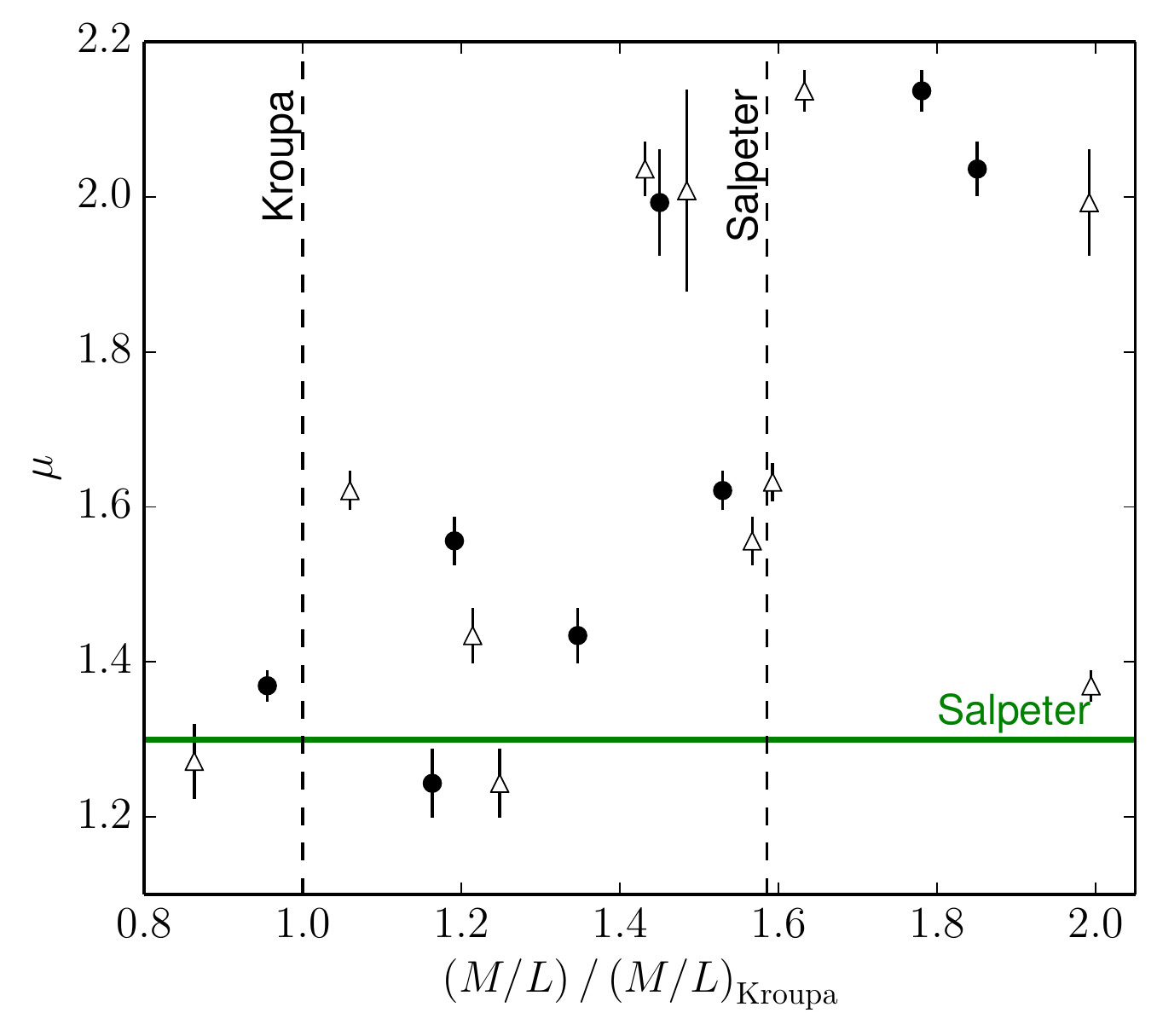}
		\end{center}
	    \caption[]{
	    The IMF slope trend with the $M/L$ normalized to the $M/L$ measured assuming a \citet{Kroupa01} IMF. 
	   	The IMF slope $\mu$ needed to compensate for the [Z/H] offset at $1~\rm{R_{e}}$ between the \sauron\ 
	   	profiles and ours is shown on the vertical axis for our sample galaxies.  
	   	The filled circles are the values of $(M/L)$ and $(M/L)_{\rm{Kroupa}}$ measured by \citet{Conroy12b} in the $K$-band. 
		The open triangles are the values for $(M/L)$ and $(M/L)_{\rm{Kroupa}}$ measured by \citet{Cappellari13} in the SDSS $r$-band. 
	   	The green thick solid horizontal line shows the slope of the \citet{Salpeter55} IMF (i.e. $\mu = 1.3$). 
	   	The vertical dashed lines show the $M/L$ obtained for a fixed \citet{Kroupa01} (\textit{Kroupa}) and a fixed \citet{Salpeter55} 
	   	(\textit{Salpeter}) IMF. 
	   	A trend is noticeable for our sample galaxies, with steeper IMFs corresponding to higher dynamically (i.e. \citealt{Cappellari13} 
	   	and spectroscopically (i.e. \citealt{Conroy12b}) measured mass-to-light ratios. 

	  	}
    \label{fig:Y}
  \end{figure}
}

\title[Large radii metallicity maps]{The SLUGGS survey: Exploring the metallicity gradients of nearby early-type galaxies to large radii}

\author[Pastorello et al.]
  {Nicola~Pastorello$^{1}$\thanks{email: npastorello@swin.edu.au},  
  {Duncan~A.~Forbes$^{1}$}, 
  {Caroline~Foster$^{2}$},
  {Jean~P.~Brodie$^{3}$},
\newauthor
  {Christopher~Usher$^{1}$},
  {Aaron~J.~Romanowsky$^{3,4}$},
  {Jay~Strader$^{5}$},
  {Jacob~A.~Arnold$^{3}$}
  \\
   $^1$Centre for Astrophysics \& Supercomputing, Swinburne University, Hawthorn VIC 3122, Australia
  \\ $^2$Australian Astronomical Observatory, PO Box 915, North Ryde, NSW 1670, Australia
  \\ $^3$University of California Observatories, 1156 High Street, Santa Cruz, CA 95064, USA
  \\ $^4$Department of Physics and Astronomy, San Jos\'e State University, One Washington Square, San Jose, CA 95192, USA
  \\ $^5$Department of Physics and Astronomy, Michigan State University, East Lansing, MI 48824, USA}
\begin{document}

\date{13$^{\rm{th}}$ February 2014}

\maketitle

\begin{abstract}
	Stellar metallicity gradients in the outer regions of galaxies are a critical tool for disentangling  
	the contributions of in-situ and ex-situ formed stars. 
	In the two-phase galaxy formation scenario, the initial gas collapse creates 
	steep metallicity gradients, while the accretion of stars formed in satellites 
	tends to flatten these gradients in the outskirts, particularly for massive galaxies. 
	This work presents the first compilation of extended metallicity profiles over a wide range of galaxy mass. 
	We use the \deimos\ spectrograph on the Keck telescope in multi-slit mode to obtain radial stellar 
	metallicity profiles for 22 nearby early-type galaxies. 
	From the calcium triplet lines in the near-infrared we measure the metallicity of the starlight up to 
	$3$ effective radii. 
	We find a relation between the outer metallicity gradient and galaxy mass, in the sense 
	that lower mass systems show steeper metallicity gradients than more massive galaxies. 
	This result is consistent with a picture in which the ratio of ex-situ to in-situ formed stars is lower in 
	less massive galaxies as a consequence of the smaller contribution by accretion. 
	In addition, we infer a correlation between the strength of the calcium triplet feature in 
	the near-infrared and the stellar initial mass function slope that is consistent with recent models in the literature. 

\end{abstract}

\begin{keywords}
 galaxies: abundances - 
 galaxies: elliptical and lenticular, cD - 
 galaxies: formation - 
 galaxies: evolution - 
 galaxies: stellar content. 
 
\end{keywords}


\section{Introduction}
\label{sec:introduction}

	Until recently, galaxy formation scenarios fell either into the category of monolithic collapse 
	or hierarchical merging. 
	The first scenario assumes that early-type galaxies (ETGs) formed in a single violent burst 
	of star formation at high redshift, followed by a largely quiescent evolution with few, if any, 
	further star formation episodes \citep{Larson74, Carlberg84, Arimoto87}. 
	By contrast, under the second scenario, larger galaxies are thought to have been built up via successive 
	mergers of smaller systems \citep{Toomre72}. 
	Such merger events are supposed to happen continuously during a galaxy's entire history. 

	\subsection{The two-phase formation scenario}

	Recently a new paradigm has emerged for the formation of massive ETGs, the two-phase 
	formation scenario, which could be considered as a hybrid option between the monolithic collapse and 
	the hierarchical merging models. 
	This scenario is increasingly supported both by theory and by observations, as detailed below.  
	In this two-phase picture, the first phase occurs early ($z>2$) and forms the bulk of the stars through the 
	dissipative collapse of gas. 
	These stars are born in-situ and their formation is driven by the infall of cold gas flows 
	\citep{deLucia07, Dekel09b, Zolotov09, Khochfar09b} or by the cooling of hot gas \citep{Font11}. 

	The second phase involves the accretion of stars formed in smaller satellite galaxies (i.e. ex-situ). 
	This star formation may continue to the current time in these satellite galaxies, regardless of their mass, 
	although they are 	usually gas-poor at the time of merging \citep{Oser10}. 
	In general, the merging satellite systems have a mass that is much lower than the main galaxy, with a typical 
	mass ratio of 1:5 \citep{Oser12}. 
	The accreted systems are added at radii larger than the effective radius $\rm{R_{e}}$ \citep{Naab09b, 
	Font11, Navarro-Gonzalez13}, thus increasing the size of the main galaxy \citep{Naab09b, Oser12, Hilz12, Hilz13}. 
	This gas-poor (dry) accretion phase dominates the galaxy evolution at $z<2$. 

	Observationally, hints of two different formation modes in the Milky Way were first presented 
	by \citet{Searle78}. 
	In their work, they inferred that the inner halo globular clusters (GCs) formed during 
	the collapse of the galaxy central regions, while in the outer halo GCs were accreted. 
	Further evidence supporting two-phase formation has been found 
	from GC systems in nearby galaxies (e.g. \citealt{Arnold11, Forbes11}). 
	\subsection{Stellar metallicity gradients}	
	
	\subsubsection{Predictions from the two-phase formation scenario}
	The formation of galaxies in two phases affects their metallicity ([Z/H]) gradients. 
	Simulations have shown that in-situ formation leads to steep gradients \citep{Kobayashi04, Pipino10}. 
	However, strong AGN feedback can interrupt star formation, causing a flattening of the inner 
	metallicity gradient \citep{Hirschmann12}.

	On the other hand, mixing due to mergers of already formed stellar 
	populations (i.e. ex-situ formation) will modify the metallicity profile 
	\citep{White80, Kobayashi04, diMatteo09a, Font11} in the regions where these processes dominate (i.e. the outer regions). 
	The final metallicity gradients in these outer regions are predicted to show a trend with 
	galaxy mass, with lower mass galaxies having steeper outer metallicity profiles than 
	higher mass systems. 
   	Low-mass galaxies accrete stars mostly from metal-poor low-mass satellites 
    \citep{Naab09b, Lackner12, Hirschmann13}, while the satellites that merge with more massive 
    galaxies are composed, on average, by more metal-rich stellar populations, 
    which will flatten the metallicity profiles in the outer regions. 
	If the mergers are gas-rich, this will mostly affect the metallicity only in the central regions because the gas will 
	sink toward the galaxy centre where it will trigger new star formation, increasing the inner metallicity gradient. 
				
	Thus, with two phases (i.e. dissipative collapse and external accretion) in the formation history 
	of a galaxy, a transition region is expected between the in-situ central and the ex-situ outer regions 
	\citep[see~e.g.~figure~8,][]{Font11}.

	\subsubsection{Observations of inner metallicity gradients}		
	
	The metallicities in the central regions of ETGs have been studied over the years by many different groups, 
	adopting both photometric 	(i.e. colours) and spectroscopic (e.g. absorption line indices) approaches. 
	Hints of a relation between the inner (i.e. $R<1~\rm{R_{e}}$) metallicity gradient and galaxy mass have been found in a number of studies
	\citep{Carollo93, Forbes05, Ogando05}, with increasingly steeper gradients in more massive galaxies. 
	\citet{Sanchez-Blazquez07, Spolaor09, Spolaor10b} and \citet{Kuntschner10} found that such a relation 
	may be valid only for galaxy masses lower than 	$\approx3.5\times 10^{10}~\rm{M_{\odot}}$, while the trend is the 
	opposite for greater masses (i.e. shallower profiles in higher mass galaxies). 
	The pioneering work of the \sauron\ collaboration, using integral field unit (IFU) technology, 
	produced detailed stellar population maps for 48 nearby ETGs up to $1~\rm{R_{e}}$ \citep{Kuntschner10}. 
	In addition to the very valuable 2D spatial metallicity information, their maps confirmed the 
	relation between metallicity gradients and galaxy mass. 
	By contrast, \citet{Koleva11} did not find the same trend connecting metallicity gradient with galaxy mass. 

	The limitation of studies confined to the central parts of ETGs is that they give information predominantly about the 
	stars formed in-situ (and are a mix of those formed during an initial collapse or by later wet mergers). 
	In contrast, outer gradients provide a clearer test of the galaxy formation history as they are sensitive to ex-situ 
	formed stars. 	

	\subsubsection{Observations of outer metallicity gradients}
	Photometric metallicities can be used to extend coverage to larger galactocentric radii, and colours are relatively easy 
	to obtain, but there are serious limitations, as discussed below. 
	The work by \citet{Chamberlain11} measured the average metallicity gradient in a set of lenticular galaxies via the 
	photometric approach out to more than $5~\rm{R_{e}}$. 
	They did not find any correlation of metallicity gradient with galaxy mass, although their metallicity 
	gradient confidence limits were large enough to include both shallow and very steep 
	metallicity trends (i.e.  $\nabla_{R}\rm{[Z/H]} = -0.6~\pm 0.5~\rm{dex/dex}$). 

	A more radially extended study of ETGs in the Virgo Cluster by \citet{Roediger11b} found 
	flat metallicity profiles for the massive galaxies and much steeper mean metallicity 
	gradients in dwarf elliptical galaxies. 
	They found no obvious trend with galaxy mass for either galaxy type. 
	Also using colours, \citet{LaBarbera12} found that 
	metallicity profiles are steep in the outer regions (i.e. $1 < R <8~\rm{R_{e}}$) of both 
	high-mass and low-mass ETGs. 
	However, while for the low-mass ETGs this trend is significant, the gradients in the high-mass ETGs 
	could be affected by a decrease of $[\alpha/Fe]$ at large radii, which is not constrained separately from the 
	metallicity in their work. 
	\citet{LaBarbera12} explained such results as a consequence of the accretion of mostly 
	low-metallicity stars in the outskirts of both giant and low-mass galaxies. 

	The main problem with optical photometric studies is the strong age-metallicity
	degeneracy which, unless there is a homogeneously old stellar population within a galaxy, can lead 
	to wrongly inferred metallicities \citep{Worthey94, Denicolo05a}.  
	In addition, the colour may be affected by dust reddening and the `red halo' effect. 
	This latter issue is a consequence of the dependence on wavelength of the shape of the outer wing of the PSF
	and affects measurements of the external regions of extended objects (see \citealt{Michard02} and references therein). 
	Thus, although colour-based analyses allow estimates of the galaxy metallicity to large 
	radii, a clean separation between age and metallicity is best obtained by spectroscopic analyses 
	\citep{Worthey94}.

	A different approach to exploring the chemical composition of the outer regions of galaxies involves the study of 
	GCs. 
	In general, red GCs follow the kinematic and chemical properties of galaxy field stars 
	(\citealt{Forbes12}, and references therein). 
	Since these objects are compact and bright, their metallicity can be retrieved from photometric colours 
	and from spectroscopy out to more than $10~\rm{R_{e}}$ (see, for example, \citealt{Forbes11} and \citealt{Usher12}) .   

	To date just a handful of works have been able to measure the stellar metallicity 
	in the outskirts of ETGs from spectroscopy. 
	This is because a high signal-to-noise ratio (S/N) is required to obtain reliable estimates 
	of the stellar population parameters, and the outskirts of galaxies are faint.	
	\citet{Weijmans09} carried out one of the first spectroscopic studies exploring the line strength up to $4~\rm{R_{e}}$ 
	in the two galaxies NGC~821 and NGC~3379. 
	In these two ETGs they found hints that the inner line strength gradients remain constant out to such large radii. 
	Similarly, \citet{Coccato10} measured the metallicity of the giant elliptical galaxy NGC~4889 to almost $4~\rm{R_{e}}$, 
	finding in this case that the inner steep metallicity profile becomes shallower outside $1.2~\rm{R_{e}}$.
	With a sample of 33 massive ETGs, \citet{Greene13} found mild 
	metallicity gradients in the outskirts (i.e. up to $2.5~\rm{R_{e}}$), in contrast with the steep inner 
	ones.
	These first results from spectroscopic measurements outside the central regions match quite closely  
	with the prediction of a dissipative collapse model for the innermost stars and an accreted 
	origin for those in the outskirts. 

	\subsection{This paper}
	In this work, we expand the sample of ETGs for which outer metallicity gradients 
	have been spectroscopically measured. 
	In particular, for the first time such extended metallicity profiles are measured over a wide range of galaxy masses. 
	Specifically, we take advantage of the calcium triplet (CaT) lines in the near-infrared (i.e. at $8498$, $8542$ and $8662~\rm{\AA}$)
	to measure the metallicity of the integrated stellar population out to $2.5~\rm{R_{e}}$.
	We use spectra obtained with the DEep Imaging Multi-Object Spectrograph 
	(\deimos) on Keck \citep{Faber03} as part of the SLUGGS survey\footnote{http://sluggs.swin.edu.au} \citep{Brodie14}. 
	\deimos\ is a very efficient instrument in the spectral region of the
	CaT lines. 
	This spectral feature has been long known as an indicator of metallicity \citep{Armandroff88, Diaz89, Cenarro01} 
	that is only minimally affected by the stellar age \citep{Schiavon00, Vazdekis03} 
	and thus is useful in breaking the age-metallicity degeneracy. 
	The method used to extract the galaxy component from the background of \deimos\ spectra was developed 
	by \citet{Proctor09} and used by \citet{Foster09} and \citet{Foster11} to obtain stellar metallicity radial profiles in 3 ETGs. 
	From the metallicity measured at different spatial locations we create 2D metallicity maps for each galaxy in 
	our sample. 
	These maps are then used to extract metallicity gradients both inside and outside $1~\rm{R_{e}}$. 
	
	The structure of this paper is as follows. 
	In Section \ref{sec:data} we present the data reduction and the method used to measure 
	the metallicity from the CaT index. 
	Section \ref{sec:analysis} focuses on the production of 2D metallicity maps and 
	the measurement of radial metallicity profiles for the galaxies in our sample, as well as the estimation 
	of the metallicity gradients inside and outside $1~\rm{R_{e}}$.	
	Section \ref{sec:results} discusses the comparison between these inner and the outer 
	gradients, and their trends with the galaxy mass. 
	In Section \ref{sec:discussion} 
    we discuss our findings in relation to predictions in the literature, and in 
    Section \ref{sec:conclusions} we provide a summary of the results. 
	In addition, Appendix \ref{sec:Appendix_A} explains in detail our 2D mapping technique and 
	discusses its general applicability to astronomical data. 
	In Appendix \ref{sec:Galaxies}, individual sample galaxies are discussed.


\section{Data} \label{sec:data}
	\subsection{Observations}
		%
		In this paper we present 1D radial metallicity profiles for 22 galaxies, most of them observed 
		as part of the ongoing SAGES Legacy Unifying Globulars and GalaxieS (SLUGGS) survey.		
		For 18 of these galaxies we have been able to extract 2D metallicity maps. 
		
		As presented in Table \ref{tab:A}, this survey includes nearby ($D < 30~\rm{Mpc}$) 
		ETGs over a range of luminosities, morphologies and environments.
		In addition, the last two rows of the table present two extra galaxies  
		(i.e. NGC~3607 and NGC~5866), also observed and analysed in the same manner 
		as those in the SLUGGS survey. 		
		\placetabA
		One of the aims of the SLUGGS survey is to study GC systems around these 
		galaxies \citep{Brodie14} using specifically designed multi-slit masks on the 
		\deimos\ spectrograph mounted on the Keck II telescope.
		The \deimos\ field-of-view has a rectangular shape of $16.7 \times 5~\rm{arcmin^2}$ 
		in which we include up to 150 slits, targeting GC and/or galaxy stellar light. 
		The data analysed in this paper have been obtained over the course of 8 years and 23 observing runs. 

	\subsection{Reduction}\label{sec:reduction}
		The \deimos\ data are reduced using a modified version of the IDL \texttt{spec2D} pipeline 
		\citep{Cooper12, Newman13}, as described in \citet{Arnold14}. 
		From each \deimos\ slit it is possible to retrieve both the target object (i.e. the globular 
		cluster) light and the background light. 
		The background light consists of the galaxy stellar light plus the sky.
		In order to extract only the galaxy component from the \deimos\ spectra, the ``Stellar Kinematics 
		with Multiple Slits" (SKiMS) technique described in \citet{Norris08}, \citet{Proctor09} and 
		\citet{Foster09} has been used. 
		In addition, the modification to the pipeline provides the inverse variance for each pixel of the 
		spectra. 
		This is used in the following analysis to obtain an estimate of the continuum level.

		The first step of the procedure to retrieve the galaxy light from the spectra consists 
		of identifying the sky contribution to the total background.
		Thanks to the large \deimos\ field-of-view, the slits at larger angular distances from the 				
		galaxy centre contain a negligible contribution from the galaxy light, and therefore can 
		be considered as pure sky. 
		To measure the sky contribution on each spectrum, we follow the procedure in \citet{Proctor09}. 
		In particular, we define a sky index as the ratio of the flux in a central sky-dominated band ($8605.0$ to 
		$8695.5~\rm{\AA}$) to the flux in two side bands ($8526.0$ to $8536.0~\rm{\AA}$ 
		and $8813.0$ to $8822.0~\rm{\AA}$), representing the continuum. 
		Higher sky indices correspond to spectra with a higher sky contribution. 
		The spectra with the highest sky indices are then used as templates to fit the sky 
		component in each slit. 
		In fact, with the weighted combination of these spectra we model in each slit a unique sky spectrum, 
		using the penalized maximum likelihood \texttt{pPXF} software \citep{Cappellari04}. 				

		After the subtraction of this sky spectrum, the same software is used to fit the resulting spectrum 
		(containing the galaxy stellar light) with a set of weighted template stars (obtained with 
		the same instrument setup). 
		This code returns the line-of-sight velocity distribution (LOSVD) Gauss-Hermite moments 
		(mean velocity $V$, velocity dispersion $\sigma_{\rm{V}}$, skewness $h_3$ and kurtosis $h_4$), plus 
		the relative contribution of the templates to the final fitted spectrum.
		In this paper we use the velocity dispersion values obtained from \texttt{pPXF}, while the 
		detailed analysis and discussion of the stellar kinematics for our sample of galaxies can be    
		found in \citet{Arnold14}.

		\footnotetext[2]{http://ned.ipac.caltech.edu}
		\footnotetext[3]{\citet{Paturel03}, http://leda.univ-lyon1.fr}

		\subsubsection{CaT index}
			In order to obtain the stellar metallicity, we first measure the CaT indices from each stellar 				%
			spectrum in our sample and, then, we apply a velocity dispersion correction to these values, 
			similar to that done in \citet{Foster09}. 
			The adopted CaT index definition is from \citet{Diaz89}:
			\begin{eqnarray}
				\rm{CaT} = 0.4\times Ca1 + Ca2 + Ca3
			\end{eqnarray}
			where $Ca1$, $Ca2$ and $Ca3$ are the equivalent widths of the three Ca{{\sc $\,$ii}} lines at 
			$8498$, $8542$ and $8662$~\AA, respectively. 

			To measure the equivalent widths of these lines we follow the method described in 
			\citet[Appendix A2]{Cenarro01}. 
			We first need an estimate of the continuum level, which is obtained by interpolating 
			across selected passbands \citep{Foster09}.
			These spectral ranges are defined in order to avoid regions heavily affected by 
			residual sky lines in the galaxies in our sample (i.e. with recession velocities between 
			$558$ and $2260~\rm{km/s}$), and are fitted with a straight line adopting the 
			values of the variance as weights for each pixel. 
			Similarly, three spectral intervals are defined for the CaT lines (Table \ref{tab:B}). 

			\placetabB 

			The CaT index can be measured on either the real spectra or the fitted spectra 	
			obtained from the \texttt{pPXF} fitting code (also used for the kinematic analysis). 
			This latter is the approach taken by the GC CaT studies of \citet{Foster10} 
			and \citet{Usher12}.  
			While the former could be affected by poor sky-subtraction and 
			noise, the latter needs further assumptions for the template selection. 
			For this reason we test both possibilities for one galaxy (NGC~5846), finding that for spectra with S/N $> 30$ 
			the difference between the two methods is smaller than the associated error 
			(Figure \ref{fig:B}). 
			In fact, considering only the data points above this S/N cut, the standard deviation from a perfect 
			match is $0.22~\rm{\AA}$, while the mean error for this subset of values is 
			$\Delta \rm{CaT} = 0.30~\rm{\AA}$.
			Similar values are obtained in the other galaxies of our sample. 
			We choose therefore to continue the analysis measuring the CaT indices on the real spectra, 
			to be consistent with \citet{Foster09}. 
			 \placefigBdata	

			Another issue in the measurement of line indices is the presence of weak absorption lines within the 
			passbands of the index definition, which can potentially alter the final result. 
			In the case of the CaT index, the most prominent of them is the Fe line at 
			$8688~\rm{\AA}$, lying completely in the reddest continuum passband. 
			The consequent underestimation of the continuum level potentially lowers the measured value of 
			the CaT index. 
			We can test how this issue could affect the final metallicity measurements.
			We experiment with measuring the CaT index with and without masking the iron line, finding that the final 
			extracted metallicities are not affected by this spectral feature. 
			We thus maintain the original passbands definition for the rest of the 
			analysis, remaining consistent with the method adopted in \citet{Foster09}. 

			At this stage, the obtained CaT index values need to be corrected for the velocity 
			dispersion line broadening in order to be comparable with the values obtained from the models. 
			Spectra with a larger $\sigma_{\rm{V}}$ have broader absorption lines that can exceed the defined 
			passbands, causing an underestimation of the CaT index. 
			To correct for this effect, we convolve the \citet{Vazdekis03} old age (i.e. $12.6~\rm{Gyr}$) and 
			\citet{Salpeter55} initial mass function (IMF) single stellar population (SSP) model spectra 
			by a set of Gaussians with a range of $\sigma_{\rm{V}}$ in 
			the interval $[0, 400]$~\kms. 
			These SSP models span a range of $\rm{[Z/H]}$ values from $-2.32~\rm{dex}$ to $-0.05~\rm{dex}$.
			On these spectra we measure the CaT index, finding its relation with the velocity dispersion 
			of the spectra (Figure \ref{fig:D}). 
			With such a relation, we can correct the indices measured on real spectra, 
			according to their velocity dispersion. 
			\placefigDdata
			
			To obtain reliable uncertainties for the CaT index measurements, we carry out a Monte Carlo simulation 
			in a fashion similar to the kinematic uncertainties estimation in \citet{Arnold14}. 
			For each spectrum we obtain 100 statistical realizations, adding noise to the best-fit curve from 
			\texttt{pPXF}. 
			The noise is added using the inverse variance array, which is an output of the modified \texttt{spec2D} 
			reduction pipeline. 
			We then measure the CaT index in each realization, before computing the standard deviation.

			One mask for NGC~2768 was observed on two different nights, allowing for 
			a test of the measurement repeatability. 
			As shown in Figure \ref{fig:E}, the CaT index values for these two masks are consistent only at high 	
			S/N and, for this reason, 	we choose to apply a selection in all our datasets, keeping only the 	
			spectra with S/N $> 30$.
			The standard deviation of these selected values from the perfect match is 
			of $0.22~\rm{\AA}$.
			\placefigEdata				
		
		\subsubsection{Metallicity} \label{sec:metallicity}
		    The relation between the CaT index and metallicity has been derived in a number of papers 
		    for both individual stars and globular clusters in many galaxies (see \citealt{Usher12} and 
		    references therein).
			Although the conversion is straightforward, obtaining the metallicity [Z/H] from the CaT index 			%
			needs several assumptions. 
			In order to obtain the converted values, we have to derive the relation between CaT indices and 
			metallicities from template spectra, for which we already know the nominal metallicities. 
			We choose SSP models of \citet{Vazdekis03} mostly because the spectral resolution is comparable 
			with our data. 
			In addition, the \citet{Vazdekis03} models cover a wide range of metallicities and IMFs. 
			The primary issue is the choice of the stellar age for these SSP spectra. 
			In general we could safely adopt an old stellar population (age = 12.6~Gyr) as representative 
			of that found in the haloes of ETGs.
			In fact, as discussed in \citet{Foster09}, the use of an old SSP stellar library 
			instead of a younger one leads to insignificant differences in the inferred metallicities, as the CaT 
			lines are quite insensitive to the age of the stellar population, if older than a few Gyr. 
			A second assumption regards the IMF adopted for the templates. 
			As shown in \citet{Vazdekis03} and \citet{Conroy12b}, the CaT strength depends on the 
			giant star contribution in the stellar population. 
			For a constant metallicity, a ``bottom-heavy" IMF (e.g. rich in dwarf stars) will lead to 
			lower measured CaT indices than a ``bottom-light" one.
			A way to break this degeneracy between IMF and metallicity is to analyse spectral features more 
			dependent on the dwarf star component of the stellar population (e.g. Na{{\sc $\,$i}} doublet and 
			FeH Wing-Ford band).
			Unfortunately, this analysis is not possible with our spectra, because both the Na{{\sc $\,$i}} doublet 
			and the Wing-Ford band lie outside the wavelength range of our data. 

			In this work, we assume a canonical \citet{Salpeter55} IMF, to be consistent with 
			previous literature studies. 
			We thus measure the CaT index on the 7 stellar population models, with a constant age of $12.6~\rm{Gyr}$ and 
			different metallicities. 
			The available \citet{Vazdekis03} spectra have metallicities from 
			$\rm{[Z/H]} = -2.32~\rm{dex}$ to $\rm{[Z/H]} = -0.05~\rm{dex}$. 
			These metallicities are obtained correcting the values in \citet{Vazdekis03} using the empirical 
			correction found  by \citet{Usher12}. 
			A caveat to keep in mind is that the abundance ratios of [Ca/Fe] may differ 
			between the \citet{Vazdekis03} models and our target ETGs. 
			Another issue related with this procedure is that the available stellar population models  
			lack supersolar metallicities, and thus the relation we measure between 
			CaT and metallicity has to be extrapolated for $\rm{[Z/H]} > 0~\rm{dex}$. 
			Once fitted with a second order polynomial, we then obtain the relation between the measured and
			velocity dispersion-corrected CaT index values and metallicity:
			\begin{eqnarray}
				\rm{[Z/H]} = 0.11 \cdot \rm{CaT}^{2} - 0.37\cdot \rm{CaT} -2.13 
			\end{eqnarray}
			which is shown in Figure \ref{fig:F}.
			
			\placefigFdata			

			We will later find an empirical correction for the metallicity after comparing our measured values with the 
			\sauron\ inner profiles obtained using Lick indices and assuming a relation between the IMF steepness and 
			the galaxy mass \citep{Cenarro03}.
			A more complete discussion of this can be found in Section \ref{sec:IMFdependency}.


\section{Analysis} \label{sec:analysis}
  	In order to measure the metallicity gradients of the galaxies in our sample up to $2.5~\rm{R_{e}}$, we must first explore 
  	and understand the underlying 2D distribution using our sparse metallicity values.
 	 The presence of contaminant neighbouring galaxies, or substructures, could affect the final integrated 
  	metallicity radial profile obtained.
	To identify these, for all the galaxies in our sample we analyse 2D maps of the CaT index distribution, 
	obtained by interpolating the slit points as described in Section \ref{sec:2dmaps}. 
	Evaluating case by case, we exclude the slits which are probably not related to the galaxy under study.
	In particular, the only galaxy for which we identify contaminated slits is NGC~5846. 
	From the remaining slits, we obtain radial metallicity profiles (Section \ref{sec:radialprofile}) which are used to 
	compare our metallicity values with \sauron\ data in the overlapping regions. 
	In Section \ref{sec:Offset} we present the metallicity offsets between \sauron and our own metallicity values, discussing 
	their possible causes. 
	We then obtain an empirical correction for our values in Section \ref{sec:empcorrection}. 
	Finally, from the new corrected metallicity values, we create 2D metallicity maps from which, in Section 
	\ref{sec:virtualslit}, we extract azimuthally averaged 1D metallicity profiles. 
	Such profiles are used in Section \ref{sec:gradients} to measure reliable metallicity gradients within and beyond $1~\rm{R_{e}}$ 
	for most of our galaxies.

  	\subsection{2D CaT index maps}\label{sec:2dmaps}
	    Thanks to the wide field-of-view of \deimos\, we are able to probe several square arcminutes of the galaxy 
	    metallicity spatial distribution. 
  		The drawback is that our slits are not uniformly distributed nor do they cover a contiguous portion of the field. 
		Instead, they are spread around the field, primarily targeting bright objects like GCs. 
		To partially solve this problem and be comparable with the results from IFU spectroscopy, we need to use 
		a 2D interpolation technique in order to retrieve 2D maps of the CaT index and the stellar 
		metallicity.
		In this work we choose to adopt the kriging technique, described in detail in Appendix \ref{sec:Appendix_A}. 
		As demonstrated for the metallicity case in Section \ref{sec:testKriging}, kriging is a powerful method that is 
		able to recover the overall 2D structure for a variable, while it is also able to at least spot small scale structures 
		if there are enough sampling points in the field.
		To date, in astronomy only a handful of cases using kriging have been published 
		(e.g. \citealt{Platen11, Berge12, Gentile13, Foster13}). 
		Kriging is very useful in our case, where we aim to map the outskirts of galaxies looking for metallicity 
		trends. 
		In particular, we adopt the kriging code included in the package \texttt{fields} \citep{Furrer09}, written in the 
		statistical programming language \texttt{R}. 

		Analysing the CaT index kriging map of each galaxy, we find a few slits that we need to exclude. 
		In particular, in NGC~5846 several slits were placed on the companion satellite 
		NGC~5846A and on NGC~5845. 
		While the NGC~5845 slits are easily excluded because of their distance from the centre of 
		NGC~5846, slits near NGC~5846A are identified on the 2D CaT index map based on their significantly 
		lower value of the CaT index (and metallicity).  
		In Figure \ref{fig:S}, the presence of NGC~5846A is noticeable as a lower-metallicity/CaT index substructure 
		with respect to the main galaxy. 
		To be conservative and avoid contamination by this galaxy, we discard all the slits within $15$ arcsec 
		of the centre of NGC~5846A. 
		We do not find other such contaminated galaxies in the rest of our sample. 
		
		\placefigSanalysis

	\subsection{1D metallicity radial profiles}\label{sec:radialprofile}
	From the metallicity data points of each galaxy we obtain radial metallicity profiles. 
	To extract these profiles, we find the ellipse-based circular-equivalent radius from the centre of 
	all the points, calculated as in \citet{Romanowsky12}. 
	Firstly, we project the $RA$ and $Dec$ coordinates along the galaxy's principal axes, applying 
	a simple rotation of the coordinates by an angle equal to the galaxy's position angle $PA$:
	\begin{equation}\label{EqnDistance1}
	\left\{
		\begin{aligned}
			x=RA\cdot \cos(PA) - Dec \cdot \sin(PA) \\
			y=RA\cdot \sin(PA) + Dec \cdot \cos(PA)
		\end{aligned}
	\right.
	\end{equation}
	where $x$ and $y$ are the new coordinates along the major and the minor axes, respectively. 
	The circular-equivalent radius $R$ of each point is then defined as:
	\begin{eqnarray}\label{EqnDistance2}
		R = \sqrt{x^2 \cdot q + y^2 / q}
	\end{eqnarray}
	where $q$ is the photometric axial ratio of the galaxy. 
	We then include also radial metallicity profiles from the literature (see Appendix \ref{sec:Galaxies}). 
	We obtain \sauron\ 1D [Z/H] profiles from the 2D metallicity maps of 
	\citet{Kuntschner10}, using the same techniques.
	The galactocentric radius of each point in these maps is obtained as per Equation \ref{EqnDistance2}. 

	\subsubsection{Radial coverage}
	As a consequence of the S/N limits and the lack of observations in the very centre of our galaxies, 
	the typical radial coverage of our measurements is $0.32<R<2.5~\rm{R_{e}}$. 
	This differs from most previous spectroscopic work, which had single integrated central 
	measurements or complete coverage up to $1~\rm{R_{e}}$ (e.g. the typical \sauron\ 
	radial coverage is about $0.6~\rm{R_{e}}$). 
	Comparing with \sauron\ galaxies, there are 11 galaxies with overlapping coverage 
	in the range $0.32\lesssim R \lesssim1~\rm{R_{e}}$. 

	\subsection{Metallicity offset with \sauron}\label{sec:Offset}
	Most of the galaxies in common between us and the \sauron\ sample show an offset in 
	metallicity, with our CaT index-derived metallicities systematically lower. 
	In panel $A$ of Figure \ref{fig:H} an example of these metallicity offsets (i.e. NGC~5846) is shown. 
	In order to deal with this issue, we first explore the causes of such offsets and then apply a 
	correction to obtain the metallicity values on the \sauron\ scale. 
	 
	\placefigHanalysis	 
	 
	\subsubsection{Possible explanations for the metallicity offset}
	We first discuss all the 
	possible non-astrophysical reasons for the observed difference between \sauron\ and our 
	metallicity profiles. 
	We do not consider the possibility of the offsets due to overestimated \sauron\ metallicities, 
	which we assume as correct. 
	We check if the metallicity offsets could be explained by stellar population effects (i.e. 
	$\rm{\alpha}$-enhancement and age), limits in the CaT-metallicity relation and/or 
	issues related with the observation or the reduction of the data. 
	
	Calcium is an $\rm{\alpha}$-element and therefore its abundance is linked to the 
	$\rm{[\alpha/Fe]}$ ratio of the galaxy \citep{Thomas03a}. 
	However, \citet{Brodie12} found that the relation between the CaT index and metallicity is  
	insensitive to the $\rm{\alpha}$-enhancement of the stellar population. 
	In addition, the calcium abundance seems to trace the iron abundance (i.e. $\rm{[Ca/Fe]} 
	\approx 0$), independently of galaxy mass \citep{Conroy14}. 

	Age has a negligible influence on the CaT strength for stars older 
	than 3~Gyr \citep{Vazdekis03}, 
	which are dominant in ETGs. 

	The CaT-metallicity relation we adopt is extracted from SSP model spectra. 
	Another possible explanation for the metallicity offset is that the CaT is a feature not well 
	reproduced in such models.
	For example, if the CaT lines in the models saturate for metallicities greater than a certain value, 
	we would not be able to measure such metallicities with our method (\citealt{Foster09} 
	estimated the saturation limit for the CaT as $\leq 6.2~\rm{\AA}$).
	However, it was demonstrated in \citet{Usher12} that the models we are adopting are not affected 
	by saturation at least up to the solar metallicity value ($\rm{[Z/H]} = 0$). 
	They argued that the apparent saturation observed by \citet{Foster09} is probably due to the presence of 
	weak metal lines in the high metallicity model spectra. 
	Because of these lines, the pseudocontinuum level is underestimated (lowering the CaT index measure) and the expected 
	increase in the CaT index due to the higher metallicity is compensated for, mimicking a saturation effect. 
	In our case we do not see this saturation effect, measuring in several cases super-solar metallicities 
	($\rm{[Z/H]} > 0$) from the CaT. 

	As an example of this,	in Figure \ref{fig:G} we present two fitted NGC~2768 spectra with super-solar and 
	sub-solar metallicities. 
	The spectra are normalized with respect to the continuum level in order to compare the CaT 
	equivalent widths. 
	The dashed spectrum has a velocity dispersion $\sigma_{\rm{V}} = 151$~\kms, a 
	S/N $= 69$ and a measured CaT index $= 5.25~\rm{\AA}$, while the solid spectrum has a 
	comparable velocity dispersion $\sigma_{\rm{V}} = 147$~\kms, a 
	S/N $= 71$ and a measured CaT index $= 6.78~\rm{\AA}$.
	Assuming a unimodal \citet{Salpeter55} IMF, the CaT index value corresponding 
	to a solar metallicity is $\approx 6.5~\rm{\AA}$ (this value is smaller for 
	steeper IMFs, see Section \ref{sec:IMFdependency}).
	The presence of spectra with higher CaT indexes (and, thus, with a super-solar metallicity)  allows us to  
	exclude the saturation of the CaT lines as a possible cause for the offset in metallicity we observe. 

	\placefigGanalysis

	In addition, we test the analysis pipeline, running the independent analysis code of \citet{Usher12}  
	on a subsample of our dataset, adopting the same CaT index definition and retrieving the same 
	metallicities.
	We also measure the CaT index for a random sample of spectra using the IRAF package \texttt{splot}, 
	confirming again our previous measurements.

	We try also to exclude issues linked with the observations or the reduction of the 
	data (i.e. sky subtraction, flat field division) which could result in an bad estimation of 
	the continuum level and, consequently, in a miscalculation of the CaT equivalent width.
	We find that the observed offset in metallicity could be caused by a total flux overestimation of 
	$\approx 10$\% in each spectrum (i.e. the CaT index measure would be underestimated by 
	$\approx 0.5~\rm{\AA}$).
	This can be ruled out because the data of the galaxies showing an offset come from different masks.
	It is improbable that the same extra flux is present in spectra obtained during different 
	nights, with different observing conditions and different positions of the masks 
	on the sky.
	We also verify our results with the metallicities obtained by applying a similar method to 
	longslit data for NGC~4278, using the same instrumental setup.
	Therefore, some extra flux seems unlikely to cause the observed offset.

	In summary, we verified that the offset is not caused by instrumental or data reduction effects. 
	We have also excluded stellar population parameters (e.g., $\rm{\alpha}$-enhancement or stellar age) as 
	causing the metallicity difference we observe with respect to the \sauron\ values. 

	\subsubsection{Initial Mass Function dependency}\label{sec:IMFdependency}
	After ruling out the above conceivable causes for the metallicity offsets, we 
	assume that they have an astrophysical origin. 
	We propose that it is caused by the IMF (and, in particular, by the ratio between dwarf and 
	giant stars).
	As well as metallicity, the CaT strength depends on the stellar surface gravity (which is stronger in giant stars 
	and weaker in dwarf stars).
The IMF slope $\mu$ definition adopted in this work follows the formalism of \citet{Vazdekis03}. 
	In particular, in the mass interval $[m, m+dm]$ the number of stars $\Phi (m)$ is:

	\begin{eqnarray}
	\Phi(m) \propto m^{-\left( \mu +1\right)}. 
	\end{eqnarray}
	With this definition, the \citet{Salpeter55} IMF corresponds to $\mu=1.3$. 

	If the IMF is dominated by low-mass stars (i.e. a bottom-heavy IMF), its slope $\mu$ in the low mass star regime will be 
	steeper. 
	As a consequence, the relation between CaT index and metallicity will be steeper (i.e. the same 
	measured CaT index would correspond to a higher metallicity).
	Thus, a variable IMF (with different slopes) could be responsible for a Ca under-abundance, 
	explaining the offsets we observe \citep{Cenarro03}. 
	We exclude the possibility that this variable IMF could also affect the \sauron\ metallicity measurements. 
	In fact, \sauron\ metallicities are derived from the Lick indices \Hb, Fe5012, Mg$b$ and Fe5270, 
	which use spectral features unaffected by different IMF slopes \citep{Vazdekis01}. 

		For all the galaxies in common with the \sauron\ sample, some of our data points radially overlap with the \sauron\ metallicity 
		profiles \citep{Kuntschner10}. 
		In each case we measure the metallicity offset as the value minimizing the function: 
		\begin{eqnarray}
		\chi^2 = \sum_{n=1}^{N} \frac{\left( \rm{[Z/H]_{SKiMS,~\it{n}}} - \rm{[Z/H]_{SAURON}} \right)^{2}}{\sqrt{\left(\rm{\Delta[Z/H]_{SKiMS,~\it{n}}}\right)^{2}+ \left(\rm{\Delta[Z/H]_{SAURON}}\right)^{2}}}
		\end{eqnarray}		 
		where $N$ is the total number of overlapping data points, $\rm{[Z/H]_{SKiMS,~\it{n}}}$ and $\rm{\Delta[Z/H]_{SKiMS,~\it{n}}}$, respectively, 
		the metallicity of the $n$-th SKiMS data point and its uncertainty. 
		Similarly, $\rm{[Z/H]_{SAURON}}$ and $\rm{\Delta[Z/H]_{SAURON}}$ are, respectively, the metallicity and the metallicity 
		uncertainty of the corresponding point of the \sauron\ metallicity profile. 
		The uncertainty on the offset is calculated via bootstrapping of the data points overlapping with the \sauron\ 
		profile. 

		Assuming that the metallicity offsets are due entirely to the variation of IMF slope, 
		we calculate the IMF slope that would be necessary to reproduce the \sauron\ metallicities in the different galaxies. 
		We use the \citet{Vazdekis03} old age (i.e. $12.6~\rm{Gyr}$) SSP models to recover a relation between the CaT index, 
		the metallicity and the IMF slope. 
		This relation is obtained by fitting the models with a second order polynomial 
		(as done in Section \ref{sec:metallicity}). 
		In these stellar models the CaT is not well constrained for IMF slopes steeper than $\mu = 2.8$. 
		An additional issue is linked to the limited number of discrete IMF slopes for which the models are available. 
		To help overcome this limitation, we interpolate the available models in order to predict the CaT index to metallicity conversion 
		for all possible IMF slopes between $\mu=0.3$ and $\mu=2.8$ (Figure \ref{fig:J}).
		For each galaxy we find the IMF slope that would convert the measured CaT index at $R=1~\rm{R_{e}}$ into the 
		\sauron\ metallicity extrapolated at $R=1~\rm{R_{e}}$. 
		To estimate the uncertainties of the IMF slope, we propagate those on the metallicity extrapolated value at 
		$R=1~\rm{R_{e}}$. 
		
		\placefigJanalysis		
		
		For the galaxies in common with \sauron, we find that the IMF slopes are in the range $1.25<\mu<2.15$,  
		which is steeper than the \citet{Salpeter55} IMF slope (i.e. $\mu=1.3$) in all but one case (i.e. NGC~3377). 
		Again, this result is obtained under the assumption that the offset we observe in metallicity is entirely a 
		consequence of a non-universal IMF. 
		In this sense, our values are an upper limit to the real IMF slope. 

		\citet{Cenarro03} found a strong anti-correlation between the CaT index and the galaxy central velocity 
		dispersion ($\sigma_{\rm{0}}$). 
		In recent years a number of other studies have confirmed a relationship between the shape of the IMF (in the low stellar mass  
		regime) and galaxy mass, velocity dispersion and elemental abundance 
		\citep{vanDokkum10, Treu10, Auger10, Graves10, Thomas11, Sonnenfeld12, Cappellari12, Dutton12, Dutton13a, 
		Dutton13b, Conroy12b, Conroy14, Ferreras13, Smith12, Spiniello12, Spiniello13, LaBarbera13, Tortora13, Geha13}. 
		For example, \citet{Ferreras13} found a relation between $\mu$ and $\sigma_{\rm{0}}$ from the 
		analysis of three IMF-sensitive spectral features, i.e. TiO$_1$, TiO$_2$ and Na8190. 
		
		In Figure \ref{fig:W} we show our IMF slopes against the central velocity dispersion for the galaxies 
		in common with the \sauron\ sample. 
		The $\sigma_{\rm{0}}$ values are taken from Table \ref{tab:A}.
		Since NGC~7457 has a low central velocity dispersion, with a poorly estimated metallicity 
		offset (and therefore a poorly estimated $\mu$), 
		we exclude it from the plot. 
		A trend with the central velocity dispersion (a proxy for the galaxy mass) is seen. 
		We find that higher-mass galaxies have steeper IMFs than low-mass 
		galaxies. 
		In the unique case of NGC~3377, the inferred IMF is shallower than a \citet{Salpeter55} IMF. 
		However, the uncertainties shown in Figure \ref{fig:W} are underestimated as they are the 
		confidence limits of our IMF slope extraction in CaT-metallicity space and 
		they do not include the uncertainties on the CaT index measurements, nor the uncertainties in the \citet{Vazdekis03} 
		SSP models. 
		In the typical case of a CaT index measure at $1~\rm{R_{e}}$ of $5.5~\pm0.2~\rm{\AA}$ we estimate a 
		propagated uncertainty on the IMF slope of the order of $\Delta \mu = \pm0.15$. 

		Together with the data points, in Figure \ref{fig:W} we show several $\mu$-$\sigma_{\rm{0}}$ relations 
		found by \citet{Ferreras13} and \citet{LaBarbera13}. 
		We are not able to quantify the confidence limits of such relations from their papers. 
		Even though the IMF slopes we measure are upper limits, our points lie in a region 
		of the $\mu$-$\sigma_{\rm{0}}$ space compatible with these literature relations. 

		\placefigWanalysis
					
		In addition, we also compare our IMF slope upper limits with the mass-to-light ($M/L$) ratios presented in 
		\citet{Conroy12b} and \citet{Cappellari13} for the galaxies in common. 
		In Figure \ref{fig:Y} we plot the IMF slope values against the \citet{Conroy12b} 
		spectroscopic best-fit $K$-band $M/L$ in terms of the best-fit $K$-band $M/L$ measured 
		assuming a fixed \citet{Kroupa01} IMF $\left( M/L \right)_{\rm{Kroupa}}$ for the 8 galaxies in common with our 
		sample. 
		We also plot the IMF slope values against the \citet{Cappellari13} 
		dynamical best-fit SDSS $r$-band $M/L$ in terms of $M/L_{\rm{Kroupa}}$ for the 
		11 galaxies in common with our sample. 
		The \citet{Cappellari13} $M/L$ values are obtained from galaxy kinematics and are presented together 
		with the $M/L$ obtained from spectral fitting assuming a \citet{Salpeter55} IMF  $\left( M/L \right)_{\rm{Salpeter}}$. 
		In order to have the ratios expressed in terms of the $(M/L)/(M/L_{\rm{Kroupa}})$, we multiply the given 
		$(M/L)/(M/L_{\rm{Salpeter}})$ values by $1.65$ \citep{Conroy10}. 
		The ratios $(M/L)/(M/L_{\rm{Kroupa}})$ are sensitive only to the IMF. 
		From the plot it is possible to see a correlation between our CaT-derived IMF slopes and both the independent IMF measures 
		of \citet{Conroy12b} and \citet{Cappellari13}. 
		However, it is worth noting that a significant scatter is visible between \citet{Conroy12b} and \citet{Cappellari13} on 
		a galaxy-by-galaxy basis, in agreement with the comparisons of \citet{Smith14}. 
				
		Both the relations found for the IMF slope with the galaxy $\sigma_{\rm{0}}$ 
		(Figure \ref{fig:W}) and $M/L$ (Figure \ref{fig:Y}) 
		provide some confidence to our claim that the metallicity offsets compared to the \sauron\ metallicities are mostly due 
		to real IMF variations among the galaxies.  
		
		\placefigYanalysis
	
		\subsection{Empirical correction}\label{sec:empcorrection}
		Considering the limitations of the analytic approach to the metallicity 
		offset issue (i.e. adopting different IMF slopes for 
		our galaxies), we use an empirical correction for our metallicity profiles. 							
		In order to simplify the task, we assume that the offsets we observe between our 
		CaT-derived metallicity profiles and the \sauron\ ones are constant with radius and 
		depend exclusively on the galaxy mass.  
		Under the first assumption, we sum the inferred offset to all the metallicity 
		points we measure from the CaT index in the 
		galaxies in common with the \sauron\ sample.
		In panel $B$ of Figure \ref{fig:H} the corrected metallicity values are shown together with the profile extracted 
		from the 2D \sauron\ metallicity map. 
		The second assumption allows us to calibrate an empirical relation between the central velocity dispersion $\sigma_{\rm{0}}$ 
		presented in Table \ref{tab:A} 	(proxy for the galaxy mass) and the metallicity offset. 
		We exclude from this relation NGC~7457 because it deviates from a linear relation. 
		In Figure \ref{fig:K} we show the metallicity offsets against the galaxy central velocity dispersion. 

		\placefigKanalysis				

		The empirical linear correction is: 
		\begin{eqnarray}\label{eqn:correction}
			\rm{\Delta [Z/H]} = (3.78\pm 0.92)\cdot \log \sigma_{\rm{0}} - (8.08\pm 2.15)
		\end{eqnarray}	
		with an rms of $0.04$.

	The empirically corrected 1D radial metallicity values are plotted in Figure \ref{fig:O}, together with the 
	available literature values and profiles.	
	In the cases of NGC~821 \citep{Proctor05}, NGC~1400, NGC~1407 \citep{Spolaor08b} and NGC~3115 
	\citep{Norris06} the longslit metallicity profiles are extracted directly from literature plots. 
	Unfortunately, this method prevents us from quantitatively estimating the metallicity 
	offset with our uncorrected metallicity values.  
	It is, however, remarkable that our metallicity profiles for these four galaxies qualitatively match with the 
	longslit data after the empirical correction of Equation \ref{eqn:correction} is applied. 
	The works of \citet{Proctor05}, \citet{Spolaor08b}, \citet{Norris06} and \citet{Kuntschner10} all derive metallicities 
	from the Lick indices using the $\chi^2$ technique \citep{Proctor02, Proctor04}, but adopt different stellar population models.  
	While \citet{Proctor05}, \citet{Spolaor08b} and \citet{Norris06} fit their indices to the \citet{Thomas03a} SSP models (with the 
	former correcting the metallicities to include the oxygen abundance variations in the stars used to define such models), 
	\citet{Kuntschner10} use the SSP models by \citet{Schiavon07}.

	Literature metallicity values at a given radius 
	are often not directly comparable with ours, because the 
	metallicity in many cases is reported as an averaged value within an area 
	around the galaxy centre. 
	Usually this area corresponds to the spectroscopic aperture or to the fibre size. 
	In these cases we plot the literature points at scaled galactocentric radii. 
	Assuming a de Vaucouleurs profile, we calculate the total luminosity within the averaged area and we assign 
	the radius within which half of this luminosity is included as the $x$ coordinate of the value.
	The scatter between the literature values can be easily explained by the adoption of different techniques and SSP 
	models in the metallicity measurements.
		
	After applying the metallicity corrections (i.e. Equation \ref{eqn:correction}), we can obtain metallicity 
	maps for most of the galaxies in our sample. 
	No fitting technique is able to reliably retrieve values in an under-sampled area, and 
	kriging is no exception. 
	For this reason we exclude the galaxies with a combination of few measured data points and/or insufficient azimuthal 
	sampling coverage of the field (i.e. NGC~720, NGC~821, NGC~2974 and NGC~5866). 
	As a first pass, we reject the galaxies with fewer than 16 metallicity data points. 
	As a second pass, we measure the angular separation between the data points in the field of each galaxy. 
	If the sum of the two widest angular separations is greater than $160~\rm{degrees}$, we consider the 
	azimuthal sampling coverage of the field as insufficient. 

	All the reliable kriging metallicity maps are presented in the bottom panels of Figure \ref{fig:Q} together with 
	the individual data points. 
	Map pixels and data points are colour coded according to their metallicity, consistently with the colour bar 
	on the right hand side of the map. 
	The surface brightness isophotes at $1$, $2$ and $3~\rm{R_{e}}$ are also shown as dashed black ellipses. 

	\placefigOanalysis	
	\placefigQanalysis
	\clearpage
	
	\subsection{Radial profile extraction from 2D metalliciy maps}\label{sec:virtualslit}
	The 2D metallicity maps obtained from the kriging interpolation are very useful for spotting substructures and 
	for visualizing the 2D metal distribution of a galaxy. 
	These maps could be used to compare the observations with 2D metallicity distribution predictions from 
	future simulations. 
	However, the current available simulations of ETGs predict only radial metallicity profiles. 	
	In order to compare our results with such simulation profiles, we extract azimuthally averaged 
	1D metallicity profiles from our kriging metallicity maps. 
	Knowing the coordinates of each pixel, we find the circular-equivalent galactocentric radius following Equations \ref{EqnDistance1} and \ref{EqnDistance2}.
	We then azimuthally average the metallicity values within circular bins in the new circular-equivalent 
	space, adopting a bin size $\delta R = 0.05~\rm{R_{e}}$.

	An important simplification we make is that the stellar metallicity 2D profile follows the brightness 
	profile in a galaxy. 
 	Specifically, we measure the circular-equivalent radius of each map pixel 
 	assuming PA and axial ratio values obtained from the shape of the isophotes.  
	In principle, the stellar metallicity may not follow the light distribution, and this 
	could lead to systematic errors in the metallicity profile extrapolation from the maps. 
	For example, a different ellipticity of the metallicity 2D distribution could explain 
	the metallicity bump in NGC~4111 at $4~\rm{R_{e}}$. 
	For consistency, we keep the standard approach (i.e. adopting the photometric PA and 
	axial ratio for the metallicity distribution), acknowledging that this could cause systematic errors in the profile 
	extraction of some of our galaxies.

    The confidence limits of the metallicity profiles are obtained via bootstrapping. 
    For each galaxy we sample with replacement the original dataset 1000 times, maintaining 
    the same number of data points.
 	In order to avoid degeneracies in the kriging interpolation, 
	if in the same dataset the same original point is chosen more than once, we shift its spatial position, adding a 
	random value in the range $-0.5<\Delta \overline{x} < 0.5$ arcsec to both the $RA$ and $Dec$ coordinates. 
	This addition is physically negligible, considering that the typical \deimos\ slit is much longer (i.e. $\geq4~\rm{arcsec}$). 
	From these new datasets, we obtain 1000 different kriging maps from which we extract the radial profiles.
    For each radial bin we extract the histogram of the profile values and, assuming a Gaussian 
    distribution, we find the boundary values which include 68\% of the distribution.    
    These values are then used as confidence limits for the real radial profiles.
	The top panels of Figure \ref{fig:Q} present these metallicity radial profiles in both $\rm{log} (R/\rm{R_{e}})$ and 
	linear spaces, together with the values and the positions of the measured data points. 
	Here, measured points are shown as squares, colour-coded according to their metallicity value  
	(see the colour bar). 
	The black lines show the metallicity profiles extracted from the kriging map. 
	For the galaxies in the \sauron\ sample we also present the metallicity radial profile extracted from 
	the \sauron\ 2D metallicity maps. 
	For NGC~1400 and NGC~1407 the radial metallicity profiles extracted along the major axis by 
	\citet{Spolaor08b} are shown as point-dashed blue lines. 
	Similarly, for NGC~3115 we present the metallicity profiles obtained by \citet{Norris06} along the major axis, 
	as a dot-dashed line. 

	In Figure \ref{fig:X} we present the metallicity profiles we obtained, colour coded 
	according to the galaxy central velocity dispersion $\sigma$ (i.e. proxy for the galaxy mass). 
	We exclude from the figure NGC~1407 and NGC~4494 and the outermost parts of NGC~3115 and NGC~4111 
	due to their large scatter. 
 	\placefigXanalysis
	
	\subsection{Metallicity gradients}\label{sec:gradients}
	The radial extent of our datasets allows us to probe the stellar metallicity beyond the effective radius in 
	most cases. 
	The metallicity profiles in the outer regions, in fact, can be compared with predictions from simulations in order 
	to infer the scenario in which the galaxies formed.
	Furthermore, a comparison between the inner and the outer metallicity gradients can reveal the 
	importance of feedback processes in galaxy formation. 

	Using radial profiles extracted from the kriging maps, we are able to measure the metallicity gradients 
	up to several effective radii. 
	However, in order to have a clean set of profiles, we exclude from the sample NGC~1407, NGC~4494 and NGC~5846. 
	In the first case, the galaxy shows strong substructures in metallicity, which makes the measure of a metallicity 
	gradient unreliable. 
	Moreover, in NGC~1407 the metallicity profile extracted from the kriging map is dominated by a single point for 
	$R>\rm{R_{e}}$ (see Figure \ref{fig:Q}).
	Similarly, NGC~4494 presents a very steep metallicity profile for $R>1~\rm{R_{e}}$, driven by two single 
	points in the South-East region of the field (see Figure \ref{fig:Q}).
	Lastly, NGC~5846 has only two data points at $R>1~\rm{R_{e}}$ and, thus, we are able to reliably 
	measure only its inner metallicity gradient. 
	
	The final sample for which we obtain the outer (inner) metallicity gradients contains 15 (16) galaxies. 
	For these, we measure the gradients by performing a weighted linear least-squares fit for the data points 
	in the logarithmic space $\rm{[Z/H]}$-$\log\left( R/\rm{R_{e}}\right)$ in two different radial ranges. 
	Because the sample includes galaxies of different galaxy sizes, in order to compare these  
	objects (and be consistent with the literature studies) we define such radial ranges with 
	respect to $\rm{R_{e}}$. 
	In particular we consider the inner gradients measured at $ -0.5 < \log \left( R/\rm{R_{e}}\right) 
	\leq 0$ (i.e. corresponding to $0.32<R\leq 1~\rm{R_{e}}$) and the outer gradients measured at 
	$0 < \log\left( R/\rm{R_{e}}\right) \leq 0.4$ (i.e. corresponding to $1<R \leq 2.5~\rm{R_{e}}$).
	The gradients are presented in Table \ref{tab:C}. 
	The uncertainties on the metallicity gradients presented in the table have been estimated via bootstrapping and represent 
	the $1\sigma$ confidence limit. 
		
	\placetabC


\section{Results} \label{sec:results}
	In Figure \ref{fig:U} we plot the inner and the outer metallicity gradients measured in our galaxies.
	The black dashed line in the plot shows the locus of points where the inner and the outer gradients are equal. 
	From the plot it is noticeable that very few galaxies maintain the same inner gradient outside the effective 
	radius.
	In particular, most of the galaxies in our sample have an outer gradient which is much steeper than the inner one, 
	while a somewhat reversed trend is noticeable in just three galaxies: NGC~3607, NGC~4278 and NGC~4365.
	NGC~3607 is consistent with having a flat outer gradient within the uncertainty 
	and NGC~4278 has a slight positive outer metallicity profile. 
	For NGC~4365 the apparent positive outer 
	gradient could be driven by just the outermost data point. 
	One can see that in Figure \ref{fig:Q} an overall flat metallicity gradient (in both the inner and 
	the outer regions) could exist within the uncertainties associated with the metallicity profile. 

	\placefigUdiscussion

	In our sample there are several galaxies (e.g. NGC~3115 and NGC~4374) for 
	which the kriging maps show the presence of metallicity substructures (see Figure \ref{fig:Q}). 
	Such metallicity substructures could be the cause of the mild metallicity gradients 
	we measure in the outer regions of these galaxies. 
	Since most of these substructures in the kriging maps are obtained from a 
	significant number (i.e. $N>5$) of nearby data points, we believe they 
	may be real. 
	Higher spatial density coverage is required to confirm these substructures. 
	
	Our data expand the mass range in the literature for which the outer metallicity gradients have 
	been measured. 
	In Figure \ref{fig:V} we plot the metallicity gradients measured in the outer regions of our 
	galaxies along with the stellar central velocity dispersion $\sigma_{\rm{0}}$, the total dynamical mass 
	$M_{\rm{dyn}}$ and the total stellar mass $M_{\rm{\star}}$. 

	In a random motion dominated stellar system such as an ETG, the stellar central velocity 
	dispersion is a proxy for the gravitational potential (and thus for the total galaxy mass).  
	We find that a correlation exists between this value and the outer metallicity gradient, with the higher central 
	velocity dispersions corresponding to shallower outer metallicity profiles. 
	Fitting all our measurements with a linear relation, we find:
	\begin{eqnarray}
	  \nabla_{R}\rm{[Z/H]_{Outer}} = (3.48 \pm 1.19)\cdot \log\sigma_{\rm{0}} - (8.92\pm 2.75) 
	\end{eqnarray}
	where $\sigma_{\rm{0}}$ is the central stellar velocity dispersion in $\rm{km/s}$ and $\nabla_{R}\rm{[Z/H]_{Outer}}$ 
	the outer metallicity profile.
	The rms of this relation is $0.18$ and the fit is presented as a black dashed line. 
	The statistical significance of the fit is almost $3\sigma$.
	We note that the NGC~7457 central velocity dispersion has a large range of values in the literature, from 
	the lowest $\sigma_{\rm{0}}=  23.0~\pm  16.0~\rm{km/s}$ \citep{DalleOre91} to the highest 
	$\sigma_{\rm{0}}=136.0~\pm  11.0~\rm{km/s}$ \citep{Dressler83}. 
	Thus, we fit all the measurements excluding NGC~7457, obtaining the relation:
	\begin{eqnarray}
	  \nabla_{R}\rm{[Z/H]_{Outer}} = (5.23 \pm 1.81)\cdot \log\sigma_{\rm{0}} - (13.06\pm 4.24).
	\end{eqnarray}	
	The rms is $0.12$. 
	Another measure of the strong correlation is the Spearman rank correlation coefficient, which is 
	$r_{\rm{S}} = 0.69$ with a significance of $p=99.6\%$ if we fit all the points.
	Excluding NGC~7457, $r_{\rm{S}} = 0.64$ with $p=98.6\%$.

	In the left panel of Figure \ref{fig:V}, we also overplot the value for NGC~4889 obtained by \citet{Coccato10}, the value for NGC~4472 
	obtained from \citet{Mihos13} and the gradients 
	of another 8 massive galaxies from \citet{Greene12}. 
	To obtain the outer gradient from the \citet{Mihos13} plots, we assume $[\alpha/\rm{Fe}]$ as radially constant and adopt their 
	colour/metallicity relation for a 10~Gyr old stellar population, obtaining $\nabla_{R}\rm{[Z/H]} \approx  -0.4~\rm{dex/dex}$ in 
	the radial range $1<R<2.5~\rm{R_{e}}$.
	In the case of the \citet{Greene12} values, we convert the iron abundances [Fe/H] into metallicities after adopting the \citet{Thomas03a} 
	relation and assuming that the [Mg/Fe] gradient traces the $\rm{[\alpha/Fe]}$ one. 
	The central velocity dispersion $\sigma_{\rm{0}}$ for the \citet{Greene12} galaxies is calculated 
	from their measurements within 	$1~\rm{R_{e}}$ (i.e. $\sigma_{\rm{e}}$) after adopting the equation in 
	\citet{Cappellari06}: 
	\begin{eqnarray}\label{eqn:sigma}
		\sigma_{\rm{R}} = \sigma_{\rm{e}} (R/\rm{R_{e}})^{-0.066}
	\end{eqnarray}		 
	where $\sigma_{\rm{R}}$ is the velocity dispersion averaged within a distance $R$ from the centre. 
	In our case, to obtain the central value of the velocity dispersion while avoiding degeneracies, we assume 
	$\sigma_{\rm{0}} \approx \sigma_{\rm{1"}}$ (i.e. the velocity dispersion measured at 1 arcsec from 
	the galaxy centre). 
	These central values are, on average, 15\% higher than those measured within $1~\rm{R_{e}}$. 
	We do not include these literature values in our fit because the radial ranges in which they are 
	defined are different from ours (i.e. $R > 1.2~\rm{R_{e}}$). 

	On the central panel of Figure \ref{fig:V} we present the outer metallicity gradients plotted against the total dynamical 
	mass of our galaxies. 
	To calculate the dynamical mass $\rm{M_{dyn}}$ we adopt the equation in \citet{Cappellari06}:
	\begin{eqnarray}
	M_{\rm{dyn}} = \frac{\beta \rm{R_{e}}\sigma_{\rm{e}}^{2}}{G}
	\end{eqnarray}
	where the scaling factor $\beta=5.0$, $\rm{R_{e}}$ is the effective radius, $\sigma_{\rm{e}}$ is the velocity dispersion within 
	$1~\rm{R_{e}}$ and $G$ is the gravitational constant. 
	As in \citet{Cappellari06} we assume a constant $\beta$ for all our galaxies for simplicity. 
	However, we are aware that this may be an oversimplification \citep{Courteau13}. 

	In order to calculate $\sigma_{\rm{e}}$ we correct the central $\sigma_{\rm{0}}$ 
	by inverting Equation \ref{eqn:sigma}. 
	A similar procedure allows us to overplot the values from \citet{Coccato10} as a red triangle, \citet{Mihos13} as a blue star
	and \citet{Greene12} as green squares. 
	The linear fit of our points returns the relation:
	\begin{eqnarray}
		\nabla_{R}\rm{[Z/H]_{Outer}} = (1.08 \pm 0.45)\cdot \log M_{\rm{dyn}} - (12.70\pm 4.92) 
	\end{eqnarray}
	where $\nabla_{R}\rm{[Z/H]_{Outer}}$ is the outer metallicity gradient and $M_{\rm{dyn}}$ 
	the total dynamical stellar galaxy mass expressed in solar masses. 
	The rms of this relation is $=0.47$ and the statistical significance of the line slope is $2.4\sigma$.
	The Spearman index value we find for all our points is $r_S = 0.56$ with a confidence $p=96.9\%$. 

	The right panel of Figure \ref{fig:V} shows the outer metallicity gradients against the total 
	stellar mass of our galaxies measured from the total \textit{K}-band absolute magnitude. 
	These are given in Table \ref{tab:A}. 
	For the \citet{Coccato10}, \citet{Greene12} and \citet{Mihos13} galaxies the \textit{K}-band absolute magnitudes 
	are from 2MASS \citep{Jarrett00}. 
	In the case of \citet{Greene12}, the distances are from HyperLeda archive, while for NGC~4889 and NGC~4472 we 
	adopt the distances given in \citet{Coccato10} and \citet{Mihos13}, respectively. 
	These magnitudes are then converted to stellar mass assuming $(M/L)_{K} = 1$, which is consistent 
	with that used by \citet{Forbes08} for an old (i.e. 12.6 Gyr) stellar population with nearly solar metallicity.
	This value is obtained for a \citet{Chabrier03} IMF, which is essentially identical to the \citet{Kroupa02} IMF.  
	In order to obtain the mass values for a \citet{Salpeter55} IMF, 
	we have to add $0.2~\rm{dex}$ to the mass in logarithmic space (see \citealt{Conroy12b}). 
	A different $(M/L)_{K}$ would not affect the relation, as long as this ratio does not vary 
	between the galaxies. 
	In fact, this value is expected to be universal for ETGs within a $10\%$ confidence level \citep{Fall13}.
	In addition, $(M/L)_{K}$ is very insensitive to metallicity at old ages (e.g. a change of $\pm 0.5~\rm{dex}$ 
	in $\rm{[Z/H]}$ corresponds to a variation $\Delta M/L_K$ of only $\sim0.01$). 
	However, if there is a systematic IMF change with mass, $(M/L)_{K}$ would be affected by it. 
	In particular, higher mass galaxies (with steeper IMFs) would have higher $(M/L)_{K}$. 

	Comparing the metallicity gradients with the stellar mass, we find that the relation  
	is slightly less significant than with $\sigma_{\rm{0}}$ and $M_{\rm{dyn}}$.  
	However, we still find a good correlation between $\nabla_{R}\rm{[Z/H]}_{\rm{Outer}}$ and 
	$M_{\rm{\star}}$. 
	We fit the distribution in logarithmic space with a linear relation:
	\begin{eqnarray}
	   \nabla_{R}\rm{[Z/H]_{Outer}} = (1.48 \pm 0.61)\cdot \log M_{\rm{\star}} - (17.17\pm 6.64) 
	\end{eqnarray}
	where $\nabla_{R}\rm{[Z/H]_{Outer}}$ is the outer metallicity gradient and $M_{\rm{\star}}$ the total 
	stellar galaxy mass expressed in solar masses. 
	The rms of this relation is $0.38$ and the statistical significance of the line slope is $2.43\sigma$.
	The Spearman index value we find for all our points is $r_S = 0.53$ with a confidence $p=95.5\%$. 
	Such a relation would be shallower under the assumption of higher $(M/L)_{K}$ in higher-mass galaxies.

		In summary, the  outer metallicity gradients of the galaxies in our sample correlate with their stellar mass. 
		Most of the galaxies in our sample present a steeper metallicity profile for $R>1~\rm{R_{e}}$ 
		with respect to the inner regions. 
		Moreover, the gradients show a correlation with three different mass proxies, 
		with the smaller galaxies showing increasingly steeper metallicity gradients 
		in the outskirts. 
		Comparison with theoretical predictions is discussed in Section \ref{sec:discussion}. 
		
		\subsection{Metallicity gradients outside $2.5~\rm{R_{e}}$}\label{sec:extreme}
		We can also probe the metallicity profiles outside $2.5~\rm{R_{e}}$ in 
		the four galaxies NGC~1023, NGC~2768, NGC~3115 and NGC~4111 up to, respectively, $3.2$, $2.7$, 
		$3.6$ and $4.7~\rm{R_{e}}$. 
		Unfortunately, only NGC~1023 is not affected by 
		poor azimuthal coverage of the 2D field outside $2.5~\rm{R_{e}}$. 
		For this galaxy we measure a gradient $\nabla_R = -1.47~\pm0.06~\rm{dex/dex}$ in the region 
		$2.5<R<3.2~\rm{R_{e}}$, which is much steeper than 
		the value measured within $1<R<2.5~\rm{R_{e}}$. 
		With just one profile we can not make any statistically 
		meaningful statement about the metallicity trends in this radial range. 
	
	\placefigVdiscussion 
	
	%


\section{Discussion} \label{sec:discussion}
	In this section we discuss theoretical predictions for the metallicity gradients 
	of ETGs in the inner ($R<1~\rm{R_{e}}$) and outer ($R>1~\rm{R_{e}}$) 
	regions from different galaxy formation models. 
	We also discuss observational results from the literature, comparing them with our findings. 
	Historically, metallicity gradients have been mostly measured on a logarithmic radial scale, 
	in order to have scale-free values which follow the galaxy light distribution. 
	For literature metallicity gradients measured on a linear scale, we convert them to a 
	logarithmic scale for consistency purposes. 
		
\subsection{Inner and Outer metallicity gradients}
	\subsubsection{Theoretical predictions}
	Different formation scenarios predict different metallicity gradients in the inner and outer 
	regions of galaxies. 
	If a galaxy forms its stars mostly via dissipative collapse, its metallicity gradient shows a rapid decline 
	with radius \citep{Carlberg84}. 
	As gas falls toward the gravitational centre due to dissipation, it is 
	chemically enriched by contributions from evolved stars so that new stars in 
	the centre are more metal-rich than at larger radii \citep{Chiosi02}.
	This enrichment process is more efficient in the stronger gravitational potential of large galaxies, so both the steepness 
	of the metallicity gradient and the central metallicity are expected to increase 
	with the mass of the galaxy.	
	Hydrodynamical simulations by \citet{Pipino10} showed that, in the context of quasi-monolithic 
	formation, inner metallicity gradients can be as steep as $\nabla_{R}\rm{[Z/H]} =
	 -0.5~\rm{dex/dex}$, with a typical value of  $\nabla_{R}\rm{[Z/H]} = -0.3~\rm{dex/dex}$. 
	Similarly, the simulations of \citet{Kawata03} predict in this scenario 
	an average metallicity gradient $\nabla_{R}\rm{[Z/H]} = -0.3$ 
	for the typical galaxies, which slightly steepens in case of high-mass galaxies. 
	The simulations of \citet{Kobayashi04} 
	returned steeper metallicity gradients (i.e. $\nabla_{R}\rm{[Z/H]} = -0.46~\rm{dex/dex}$) 
	in the range $0.03 < R < 2~\rm{R_{e}}$ for galaxies formed monolithically. 
	
	On the other hand, dry mergers between galaxies of comparable size (i.e. major mergers) 
	flatten the pre-existing gradient 
	up to $3~\rm{R_{e}}$ \citep{diMatteo09a}. 
	In this case, \citet{Kobayashi04} measured a typical metallicity gradient of $\nabla_{R}\rm{[Z/H]} = -0.19~\rm{dex/dex}$ in the range 
	$0.03 < R < 2~\rm{R_{e}}$. 
	In the same work, it was shown that lower mass ratio mergers progressively flattened the metallicity gradients. 

	If the mergers are rich in gas, the gas sinks to the centre of the gravitational 
	potential, where it triggers new star formation. 
	Such a process will enhance the metallicity in the innermost regions, increasing the metallicity gradient within 
	$1~\rm{R_{e}}$. 
	\citet{Hopkins09} found that these new inner metallicity gradients can be as steep as $\nabla_{R}\rm{[Z/H]} = 
	-0.8~\rm{dex/dex}$. 
	However, the outer regions are little affected by gas accretion and the outer metallicity gradients are flattened by the 
	violent relaxation process. 	

	The two-phase model predicts that the early stages of galaxy formation resemble the dissipative collapse model. 
	Eventually, most galaxies are supposed to increase their size by accreting low metallicity stars in 
	their outer regions via minor mergers, i.e. ex-situ stars \citep{Lackner12, Navarro-Gonzalez13}. 
	If the accreted galaxies are rich in gas, the predictions for the inner metallicity gradients resemble those from 
	the gas-rich major merger picture (e.g. \citealt{Navarro-Gonzalez13}). 
	From the plots of \citet{Navarro-Gonzalez13}, in the range $0.2<R<1~\rm{R_{e}}$, one expects a tiny increase 
	of the inner metallicity profile's steepness of 	$-0.03~\rm{dex/dex}$ due to the new star formation 
	in the central regions. 
	This value may be even steeper if the innermost regions ($R<0.1~\rm{R_{e}}$, where the metallicity 
	profile peaks) are included. 

	The two-phase scenario predictions for the metallicity in the outer regions vary between different 
	studies. 
	\citet{Cooper10} found from simulations that a high number of mergers is necessary in order to 
	completely wash out the pre-existing metallicity gradient in these regions. 	
	According to \citet{Navarro-Gonzalez13}, ex-situ formed stars contribute no more than 10-50\% to the final galaxy stellar mass. 
	For this reason, the metallicity gradients outside $1~\rm{R_{e}}$ should not be noticeably different from those of galaxies 
	experiencing a quiescent evolution (i.e. without mergers). 
	From their plots we estimate the outer metallicity gradients in the range $1<R<2.5~\rm{R_{e}}$ for both the 
	merger-dominated and the quiet evolution cases, as $\nabla_{R}\rm{[Z/H]} \approx  -0.3~\rm{dex/dex}$. 

	The simulations in a cosmological context by \citet{Lackner12} roughly agree on the total stellar mass 
	contribution by accreted stars (i.e. 15-40\%) and their external location in the final galaxy. 
	However, this work found that the low-metallicity of accreted stars may create strong outer metallicity gradients. 
	Taking into account metal-cooling and galactic winds in their (minor merger) simulations, 
	\citet{Hirschmann13} showed that the 
	metallicity gradients in the outskirts depend on the mass of the accreted satellites and, consequently, 
	on the mass of the main galaxy. 
	They thus predict steeper outer metallicity gradients in low mass galaxies (this is discussed further in Section \ref{sec:metTrendsMass}). 

	In summary, stars formed via dissipative collapse should result in a steep metallicity profile 
	in both the inner and outer regions of ETGs.  
	In general, major mergers flatten the metallicity profile in both inner and outer regions. 
	However, if these mergers are gas-rich, they can steepen the slope of the metallicity gradients within $1~\rm{R_{e}}$. 
	On the other hand, dry minor mergers may either flatten or steepen metallicity gradients in the outer regions. 
	
	\subsubsection{This work}
	In Section \ref{sec:gradients} we measured the metallicity gradients in both the inner 
	($0<R<1~\rm{R_{e}}$) and outer ($1<R<2.5~\rm{R_{e}}$) regions of our galaxies. 
	In Figure \ref{fig:U} we compared the inner to the outer metallicity gradients. 

	Most galaxies in our sample have steeper outer metallicity gradients than the inner ones, 
	resembling the results of 
	\citet{Lackner12} and \citet{Hirschmann13} hydrodynamical simulations. 
	Both these works take into account metal cooling and predict the  
	outer regions to be populated by accreted low-metallicity stars.  
	In addition, the \citet{Hirschmann13} simulations also include galactic wind 
	feedback, which affects the star formation efficiency of the 
	accreted objects that have different initial masses. 
	Eventually, this extra feedback process leads to a wide range of outer metallicity gradients in the 
	final galaxy. 
	This range could be even wider if there is a radially variable IMF (see Section \ref{sec:discussionIMF}). 

	Predictions from gas-poor major merger simulations agree on a flattening of the 
	metallicity gradients in both the inner and the outer regions, due to violent relaxation. 
	The majority of our galaxies do not present both flat inner and outer gradients. 
	Our data thus suggest that dry major mergers are infrequent in the evolutionary history of ETGs, 
	consistent with the predictions of \citet{Lackner12} (see also \citealt{Scott13}). 
	Flat inner and outer profiles are noticeable in only two ETGs in our sample: NGC~4278 and NGC~4365. 
	Both galaxies show signs of recent merger events. 
	NGC~4278 hosts a massive distorted \Hi\ disc \citep{Knapp78} which is misaligned with respect to both stellar 
	kinematics and the photometric axis \citep{Morganti06}. 
	In the case of NGC~4365, we have strong indications that a minor
	merger event is ongoing \citep{Blom12b, Blom14}, while the presence 
	of a third GC subpopulation may indicate a past major merger event. 
	
	In the case of gas-rich major mergers, the inner metallicity gradient could be steepened by new star formation, but
	in the outer regions a flat metallicity gradient is still expected. 
	One galaxy in our sample (i.e. NGC~3607) with such features may have undergone a 
	gas-rich merger event. 
	The central region of NGC~3607 hosts a stellar-gaseous polar disc with hints of ongoing star 
	formation \citep{Afanasiev07}. 
	In general, because we do not probe the very central regions of our galaxies (i.e. where the metallicity peaks), 
	our inner measurements should be considered as lower limits to the real metallicity profile's steepness. 
	However, it is possible to notice in Figure \ref{fig:Q} that the available literature metallicity inner 
	profiles do not show radial gradients significantly different from those we have extracted at $R<1~\rm{R_{e}}$ in 
	all the galaxies in common, except NGC~7457. 
 	Such galaxies do not show a steep metallicity gradient in the very central regions, which argues against 
 	gas-rich mergers in their recent formation histories.

\subsection{Metallicity gradient trends with galaxy mass}\label{sec:metTrendsMass}
	\subsubsection{Theoretical predictions}
	Depending on the adopted galaxy formation mode, different relations between the metallicity 
	gradients and the final galaxy mass are predicted. 
	The simulations of \citet{Pipino10} showed that a 
	mild metallicity gradient trend with mass 
	may exist in quasi-monolithic formation, with more massive galaxies having steeper gradients. 
	On the contrary, if the main galaxy formation mode involves gas-rich major mergers, no clear trends of the 
	metallicity gradient with mass are expected \citep{Hopkins09}. 
	In this case, the lack of a mass/metallicity gradient relation is supported by the predictions of 
	\citet{Kobayashi04} in the range $0.03< R < 2~\rm{R_{e}}$. 

	In the two-phase formation scenario, a relation between the outer metallicity gradient and 
	the galaxy mass could be linked either with the number or the mass 
	of the accreted satellites. 
	In the first case, \citet{Cooper10} showed how a low number of mergers preserves a pre-existing 
	metallicity gradient, while many accretion episodes can completely erase it. 
	In the second case, \citet{Hirschmann13} predicted that, on average, massive galaxies accrete higher mass satellites. 
	Because these satellites can retain more of their own gas against stellar winds, their stars have generally higher 
	metallicities than smaller satellites and, once accreted by the main galaxy, contribute more to the flattening of 
	the pre-existing metallicity profile (M. Hirschmann, private communication). 	
	In both these cases, one should expect shallower gradients in the outer regions of high-mass galaxies, and 
	steep metallicity profiles in 	low-mass galaxies. 	
	
	\subsubsection{Previous observations}
	In their sample of ETGs, \citet{Spolaor10b} found that the metallicity gradients measured 
	within $R < 1~\rm{R_{e}}$ correlate with $\sigma_{\rm{0}}$ in the 
	mass range $1.6 < \log\sigma_{\rm{0}} < 2.15$ with lower mass galaxies showing shallower profiles than 
	higher mass galaxies. 
	In addition, they found that for $\log\sigma_{\rm{0}} > 2.15$ an anticorrelation exists and the more massive 
	galaxies have shallower profiles than the less massive ones. 
	The authors claimed that the two different mass regimes (i.e. low-mass and high-mass) 
	are linked with two different galaxy formation modes. 
	In the low-mass regime the formation may be dominated by the initial gas collapse and the star formation 
	efficiency is linked with the gravitational potential (i.e. the higher the mass, the deeper the potential and hence the steeper 
	the metallicity gradient).  
	On the other hand, during their evolution, high-mass galaxies experience a high number of mergers.
	The frequency of these events increases with galaxy mass and, if such mergers 
	flatten the metallicity gradient also in the inner regions, it is not surprising that 
	the higher is the galaxy mass, the shallower is the metallicity distribution. 
	While the relation \citet{Spolaor10b} found in the low-mass regime is tight, in the high-mass regime there is a 
	clear scatter. 
	This scatter has been justified as the consequence of different interplaying variables during merger processes. 
	Conversely, \citet{Koleva11} found a variety of metallicity gradients for dwarf galaxies and no 
	evidence of connection with the galaxy mass. 
	In recent years a few studies have been able to spectroscopically probe the metallicity gradients at 
	$R>1~\rm{R_{e}}$. 
	The 8 massive ETGs in \citet{Greene12} revealed shallow metallicity gradients up to $2.5~\rm{R_{e}}$ 
	with a typical value $\nabla_{R}\rm{[Z/H]} \approx -0.2~\rm{dex/dex}$ in their mass range 
	$2.2 < \log\sigma_{\rm{0}} < 2.5$. 

	With a larger sample of galaxies, 
	\citet{Greene13} did not find noticeable differences in the outer ($R>2~\rm{R_{e}}$) 
	metallicity gradients of galaxies in their mass range (i.e. $1.6 < \log\sigma_{\rm{0}} < 2.6$). 
	\citet{Coccato10} measured the metallicity gradient of NGC~4889 within $R<1.2~\rm{R_{e}}$ and 
	in the range $1.2<R<4~\rm{R_{e}}$. 
	For this massive galaxy, they found a steep metallicity gradient in the central region (i.e. $\nabla_{R}\rm{[Z/H]} 
	= -0.5~\rm{dex/dex}$) and a very shallow gradient outside it (i.e. $\nabla_{R}\rm{[Z/H]} 
	= -0.1~\rm{dex/dex}$). 

	To explore the metallicity at even larger radii the only viable approach is through deep photometric imaging. 
	With this approach, \citet{LaBarbera12} found steeply declining metallicity profiles in both giant and low-mass 
	ETGs, suggesting that the accreted stars are more metal-poor than those formed in-situ in 
	the external regions, regardless of the galaxy mass. 
	In agreement with \citet{LaBarbera12}, \citet{Chamberlain11} did not find a correlation between
	metallicity gradients and galaxy mass in their sample of lenticular galaxies probed out to $5~\rm{R_{e}}$. 

	For the massive galaxy NGC~4472, \citet{Mihos13} found a very steep colour gradient up to $7~\rm{R_{e}}$. 
	Such a colour gradient can be explained only as a very strong negative metallicity gradient, which, 
	in the case of an old stellar population, can be as steep as $\nabla_{R}\rm{[Z/H]} \approx  -3~\rm{dex/dex}$ 
	at $1000~\rm{arcsec}$ from the centre. 
	Even though the main colour of the accreted satellite is bluer than that of the main galaxy, 
	the in-situ formed stars in the outer regions should be more metal-poor than the accreted ones. 
	Accretion may thus increase the metallicity in the outer regions, decreasing the metallicity 
	gradient. 
	For this reason, \citet{Mihos13} claimed that the steep metallicity gradient in the halo of NGC~4472 
	can only be formed during the initial collapse phase, with negligible contributions by later accretion.

	\subsubsection{This work}
		
	In this section we discuss our results in relation to the predictions from simulations and compare 
	them to previous observations. 
	The metallicity gradients measured in the range $1<R<2.5~\rm{R_{e}}$ 
	strongly correlate with galaxy mass, with steeper gradients 
	associated with low-mass galaxies (Figure \ref{fig:V}). 
	This reinforces the scenario in which low-mass galaxies are mostly composed of stars formed in-situ, 
	and thus the chemical enrichment (i.e. the metallicity) strongly decreases with galactocentric radius. 
	Such a trend would be conserved by minor mergers, which contribute low-metallicity stars to the outermost 
	regions. 
	For high-mass galaxies the probability of merging events during their evolution 
	is higher. 
	In Figure \ref{fig:V}, higher mass galaxies have flatter outer metallicity gradients 
	(and, thus, a higher number of accreted stars). 
	This is consistent with the predictions of the two-phase formation scenario and, in particular, 
	with the simulations of \citet{Hirschmann13}. 

	NGC~1023 is the only galaxy for which we have a reliable measurement of the metallicity gradient 
	beyond $2.5~\rm{R_{e}}$ (see Section \ref{sec:extreme}). 
	This gradient is very steep (i.e. $\nabla_{R}\rm{[Z/H]} 	= -1.47~\rm{dex/dex}$) and might suggest 
	a change in metallicity gradients for radii larger than 2-3 $\rm{R_{e}}$. 

	\subsubsection{Three different radial ranges in ETGs?}
	
	We may speculate at this point that the metallicity gradient has different behaviours in the 
	different radial regions of ETGs. 
	In the inner regions (i.e. $R<1~\rm{R_{e}}$), the gradient may correlate with galaxy mass, 
	following the relation found by \citet{Spolaor10b}. 
	An exception are the few galaxies that experienced major mergers in their history. 
	These galaxies show flat metallicity gradients, regardless of their mass. 
	At larger radii (i.e. $1< R < 2.5~\rm{R_{e}}$) the metallicity gradient strongly depends on the accretion of 
	ex-situ formed stars and, thus, is steeper in low-mass galaxies and shallower in high-mass 
	galaxies. 
	At even larger radii, the limited longslit and photometric observations suggest 
	very steeply declining metallicity profiles. 
	These regions are populated by a very metal-poor stellar population
	which could be the original component of the main system and may even be similar 
	for all galaxies. 
	This component did not chemically enrich itself because of its low stellar density. 
	However, direct measurements of the stellar population at these galactocentric radii can be 
	pursued only through deep photometric observations and may be affected by a number of 
	issues (e.g. age/metallicity degeneracy, 
	red halo effect, sky-subtraction, etc.).  

\subsection{Radial IMF variations}\label{sec:discussionIMF}
	In Section \ref{sec:analysis} we found evidence for a relation between the CaT index 
	and the slope of the stellar IMF, which may be linked via the mass of the galaxy (Figure \ref{fig:K}). 
	While Lick indices are largely independent of the IMF, the CaT is strongly 
	affected by the presence of giant stars in the stellar population. 
	Thus, a steeper IMF slope 
	is associated with more dwarf stars and 
	lower CaT indices for a given metallicity.
	Because the IMF steepness correlates with the total galaxy mass, it is not surprising 
	that the difference between CaT-derived and Lick-derived metallicities 
	is higher in more massive galaxies (as seen in Figure \ref{fig:K}). 

	An important caveat here is that we assume a radially constant IMF for each galaxy. 
	We can speculate about how our metallicity gradients will change with a radially changing IMF slope. 
	Recent studies have shown how low-mass galaxies tend to have a more bottom-light IMF than 
	massive galaxies \citep{Cenarro03, Thomas03b, Conroy12b, Ferreras13}. 
	Thus, if the IMF varies with radius, it should be more bottom-light at 
	larger radii (e.g. where ex-situ formed stars dominate). 
	This radial dependence of the IMF has also been proposed by \citet{Carollo14} to 
	explain the different carbon-enhanced metal-poor stellar populations they observed in the
	inner and outer halo of the Milky Way. 
	
	The corrections we apply to our CaT-derived metallicities are derived from offset measures 
	with Lick indices-derived metallicities (i.e. from \sauron) within $1~\rm{R_{e}}$. 
	In the case of a more bottom-light IMF at large radii, such a correction is an overestimation 
	of the real offset between the two metallicity measurements. 
	Thus, the true metallicities in the massive galaxies' outskirts would be lower and, in general, 
	the outer metallicity gradients slightly steeper in all galaxies. 
	This effect would be most pronounced for the most massive ETGs in our sample.  
	
	We qualitatively address this issue by estimating the metallicity gradient that a low- and a 
	high- mass galaxy in our sample (i.e. NGC~3377 and NGC~4649), would have under the assumption 
	of a pure \citet{Salpeter55} IMF at $R=2.5~\rm{R_{e}}$. 
	We also assume that the metallicity offset between ours and \sauron\ galaxies is valid at $R=1~\rm{R_{e}}$ 
	and is entirely due to a different IMF slope. 
 	From the value of the metallicity at $R=2.5~\rm{R_{e}}$ (assuming a fixed IMF slope of $\mu\equiv1.3$) 
 	and the corrected metallicity at $R=1~\rm{R_{e}}$ (with a steeper IMF slope of $\mu>1.3$) we can 
 	thus re-estimate the outer metallicity gradients for an IMF slope 
 	which varies with radius.  
 	 	 	
 	In particular, for NGC~3377 we have a corrected metallicity of $\rm{[Z/H]} = -0.49~\rm{dex}$ at $1~\rm{R_{e}}$
 	and an $\mu\equiv1.3$ metallicity of $\rm{[Z/H]} = -1.38~\rm{dex}$ at $2.5~\rm{R_{e}}$. 
 	The metallicity gradient is therefore $\nabla_{R}\rm{[Z/H]} = -2.24~\rm{dex/dex}$, i.e. consistent with the one in 
 	the case of a radially constant IMF (i.e. $\nabla_{R}\rm{[Z/H]} = -2.17~\pm0.25~\rm{dex/dex}$). 
 	This is not surprising, considering that the IMF we extracted for this galaxy in Section \ref{sec:IMFdependency} 
 	was already close to a \citet{Salpeter55} IMF (see Figure \ref{fig:W}). 
 	
 	For NGC~4649, a massive galaxy, the applied empirical correction is 
 	larger than for NGC~3377. 
 	The corrected metallicity is $\rm{[Z/H]} = +0.19~\rm{dex}$ at $1~\rm{R_{e}}$ 
 	and the $\mu\equiv1.3$ metallicity is $\rm{[Z/H]} = -1.58~\rm{dex}$ at $2.5~\rm{R_{e}}$ . 
	The outer metallicity gradient of NGC~4649 would be 
	$\nabla_{R}\rm{[Z/H]} 	= -4.45~\rm{dex/dex}$, i.e. much steeper than 
	in case of a radially constant IMF (i.e. $\nabla_{R}\rm{[Z/H]} 	= -0.75~\pm0.05~\rm{dex/dex}$).
 	
	In conclusion, if the metallicity offset between our profiles and those of \sauron\ 
	are completely caused by the different (galaxy mass dependent) IMF, 
	we find that our metallicity gradients depend on the radial behaviour of the IMF. 
	In massive galaxies, a radially changing IMF slope would lead to steeper inferred outer metallicity profiles  
	than in low-mass galaxies, potentially cancelling out the relation between the outer metallicity gradient and the galaxy 
	mass we observe (Figure \ref{fig:V}).  	
	However, we note that the outer metallicity gradients for the most massive galaxies in our sample (under the 
	assumption of a radially constant IMF) are similar to the gradients measured by \citet{Coccato10}, 
	\citet{Greene12} and \citet{Mihos13} for their massive galaxies using optical lines that are largely IMF invariant. 
	Thus, a strong radial decrease of the IMF slope is unlikely, and the relation between outer metallicity gradient 
	and galaxy mass will be preserved.


\section{Conclusions} \label{sec:conclusions}
	In this work we have explored the stellar metallicity in 22 nearby ETGs out to 
	several effective radii. 
	Adopting the SKiMS technique for multi-slit wide-field spectroscopic data, we have been able to 
	obtain stellar spectra in the outer regions of these galaxies. 
	From the spectra we extracted the CaT index, which we converted into total 
	metallicity $\rm{[Z/H]}$ assuming a \citet{Salpeter55} IMF. 
	The CaT feature is largely insensitive to age in old stellar populations, which is generally
	the case for ETGs. 
	We compared these raw CaT-derived metallicities with those obtained from 
	Lick indices by the \sauron\ survey for the galaxies in common between the two samples, 
	finding a noticeable offset in most of the cases. 
	This comparison led to a correlation between the galaxy central velocity dispersion (a proxy for galaxy mass) 
	and the difference between these two different estimates of the metallicity. 
	We used this relation to empirically correct the CaT index-derived metallicity values in each of our galaxies. 
	This relation supports the idea that the CaT index is influenced by the IMF 
	of a galaxy. 
	In addition, we obtained the IMF slopes necessary to produce the observed metallicity offsets. 
	Such slopes show a trend with galaxy mass that is consistent with recent models in the literature, 
	with higher mass galaxies having the steeper IMFs. 

	From the empirically corrected metallicity values we obtained 2D metallicity maps, using 
	the kriging interpolation technique. 
	From these maps, we extracted azimuthally averaged metallicity radial profiles.
	For most of our galaxies these profiles extend to $2.5~\rm{R_{e}}$, allowing us to 
	compare the metallicity gradients between the inner and the outer regions with the 
	predictions from simulations. 

	We found evidence that the low-mass galaxies in our sample have steep inner and outer 
	gradients. 
	This is qualitatively consistent with the model in which most of their stars formed in-situ, while 
	the few accreted satellites have low metallicity because of the relatively strong stellar wind feedback. 
	Massive galaxies show metallicity gradients consistent with an increased 
	contribution of accreted stars. 
	While the inner profiles have a steepness comparable with those of most low-mass galaxies, the outskirts become 
	shallower with increasing galaxy mass. 
	This behaviour of massive ETGs resembles the results from simulations, in which stars in their outskirts are mostly accreted from 
	satellites massive enough to have self-enriched. 
	In this way, the accreted stars, once mixed with those formed in the galaxy outskirts, partially erase 
	any pre-existing metallicity gradient.  
	Our findings may provide important constraints to distinguish between different stellar and AGN 
	feedback models in low-mass galaxies and in the satellites later accreted by the present-day massive ETGs. 
	In the future, it will be possible to extend this study integrating photometric, kinematic and dynamical constraints 
	(e.g. \citealt{Arnold11}). 
	The addition of large-radius optical spectroscopy will also help to deal with age and IMF uncertainties, while refined simulations 
	are needed for comparison.

\section*{Acknowledgements}
	We want to thank the (anonymous) referee for the useful comments 
	and suggestions which helped to improve this paper. 
	We are also grateful to S. Courteau, M. Hirschmann, S. Kartha, J. Roediger, P. S\' anchez-Bl\' azquez 
	and L. Spitler for comments and suggestions. 
	Moreover, we wish to thank N. Scott, E. Emsellem, H. Kuntschner and the \atlas\ team 
	for sharing unpublished data from their survey. 
	We acknowledge the Department of Physics and Astronomy at University of 
	Padova for the helpful environment provided during part of this work. 
	Some of the data presented herein were obtained at the W. M.
	Keck Observatory, operated as a scientific partnership among the
	California Institute of Technology, the University of California and
	the National Aeronautics and Space Administration, and made possible 
	by the generous financial support of the W. M. Keck Foundation. 
	The authors wish to recognize and acknowledge the very significant 
	cultural role and reverence that the summit of Mauna Kea
	has always had within the indigenous Hawaiian community. 
	The analysis pipeline used to reduce the \deimos\ data was developed
	at UC Berkeley with support from NSF grant AST-0071048. 
	This research has made use of the NASA/IPAC Extragalactic
	Database (NED) which is operated by the Jet Propulsion Laboratory,
	California Institute of Technology, under contract with the National
	Aeronautics and Space Administration. We acknowledge the usage
	of the HyperLeda data base (http://leda.univ-lyon1.fr).
	DF thanks the ARC for support via DP130100388. 
	JB acknowledges support from NSF grant AST-1109878. 
	This work was also supported by National Science Foundation grant 
	AST-0909237. 
	
\bibliographystyle{mn2e}
\bibliography{bibliography}{}

\appendix
\section{Kriging} \label{sec:Appendix_A}
  	In this work we chose to adopt the kriging spatial interpolation technique, initially used in geology \citep{Krige51}.
	Originally kriging was used to identify underground mineral deposits from the analysis of 
	samplings obtained through randomly placed drillings on the surface.
	The power of this fitting technique, which justifies its use in astronomy, is that it assumes a physical relation between 
	the values of a variable (e.g. metallicity) and the different positions in the explored field. 
	Kriging is a method for optimal interpolation based on the linear regression against the observed points, 
	weighted according to the spatial covariance values \citep{Matheron63} and defined as the 
	best linear unbiased estimator \citep{Cressie88}. 
	All interpolation algorithms estimate the real value of a function $\widehat{f}$ at a given location 
	$\widehat{\bf{r}}$ as a weighted linear combination of nearby data points $f(\textbf{r}_i)$ at the 
	$N$ locations $\textbf{r}_i$:
	\begin{eqnarray}\label{eqn:kriging}
		\widehat{f} \left(\widehat{\textbf{r}} \right) \approx f\left(\widehat{\textbf{r}} \right) =\sum^N_{i=1}\lambda_i f(\textbf{r}_i)
	\end{eqnarray}
	and usually the weights $\lambda_i$ are assigned according to functions that give a decreasing 
	weight with increasing separation distance. 
	In the kriging case, instead, weights depend on a data-driven weighting covariance function 
	$\Pi$, rather than an arbitrary one. 
	We will use the fit of a `semivariogram' to find this covariance function, as shown in Section \ref{sec:SV}. 
	The weights calculated in this way minimize the error with respect to the 	real data according to the mean 
	square variation. 
	The kriging equations from which it is possible to obtain the weights $\lambda_j$ are:
	\begin{eqnarray}
  		\sum^N_{j=1} \textbf{C}( \textbf{r}_i,\textbf{r}_j)\lambda_j = \textbf{c}(\textbf{r}_i,\widehat{\bf{r}})
	\end{eqnarray}
	where $\bf{C}$ is the matrix:
	\begin{eqnarray}
		\textbf{C}( \textbf{r}_i,\textbf{r}_j)=\Pi( f(\textbf{r}_i), f(\textbf{r}_j))
	\end{eqnarray}
	and $\bf{c}$ the vector:
	\begin{eqnarray}
  		\textbf{c}(\textbf{r}_i,\widehat{\textbf{r}})=\Pi(f(\textbf{r}_i), f(\widehat{\textbf{r}} ) ).
	\end{eqnarray}
	From this linear system of $N$ equations the determination of the $N$ unknown weights $\lambda_i$ is 
	straightforward. 
	The covariance matrix $\textbf{C}$ has to be inverted only once, but the weights have to be specifically computed 
	for each interpolation site $\widehat{r}$. 
	In many cases, kriging results are comparable with the ones obtained from other interpolation techniques \citep{Isaaks89}. 
	Kriging helps to compensate for the effect of data clustering because individual points within a cluster tend to 
	be less weighted 	than isolated ones. 
	Furthermore, this method provides an estimate of the interpolation uncertainties, producing a kriging map of the 
	residuals between the obtained map values and the initial ones in the sampled locations. 
	In order to test the reliability of kriging in retrieving the original values we have run several simulations, which 
	are presented in Section \ref{sec:testKriging}.
	
	\subsection{The semivariogram}\label{sec:SV}
		Usually the covariance function $\Pi(f(\textbf{r}_i), f(\textbf{r}_j ) )$ depends only on the distance between 
		the points $d=|\textbf{r}_i-\textbf{r}_j|$. 
		It is possible to describe this spatial dependence of a data set by characterizing it using a semivariogram 
		$\gamma(\textbf{r}_i,\textbf{r}_j)$. 
		This function is the mean square variation of the values as a function of distance:
		\begin{eqnarray}
			\gamma(\textbf{r}_i,\textbf{r}_j) \equiv \frac{\Pi(|f(\textbf{r}_i)- f(\textbf{r}_j )|^2 )}{2}
		\end{eqnarray}
		which, for a stationary random field, reduces to:
		\begin{eqnarray}
  			\gamma(\textbf{h}) \equiv \frac{\Pi(|f(\textbf{r})- f(\textbf{r} + \textbf{h} )|^2 )}{2}
		\end{eqnarray}
		where $\textbf{h}$ is the distance between two points. 
		The semivariogram represents the spatial autocorrelation of the measured data points. 
		Once each pair of locations is plotted, a model is fit through them. 
		There are three parameters that are commonly used to describe these models. 
		The \textit{range} parameter defines the distance over which the semivariogram model can be assumed 
		as constant. 
		Pairs of points separated by distances smaller than this value are considered partially autocorrelated, 
		whereas spatial positions farther apart are not. 
		The value of the semivariogram at the distance defined by the \textit{range} value is named the \textit{sill}. 
		The third parameter is the \textit{nugget} which defines the autocorrelation value for infinitesimally small 
		separation distances ($h\approx0$). 
		An example of semivariogram fitting is presented in Figure \ref{fig:P}. 
		Once the semivariogram is fitted, given the distance between the map points and each single point in the 
		data set it is easy to extrapolate the value $\widehat{\gamma}$ and the covariance function. 
		With these, an estimate of the values in all the map points $\widehat{f} (\textbf{r}_j  )$ can be obtained 
		from the Equation \ref{eqn:kriging}. 
			
		\placefigPanalysis

		\subsection{Testing kriging}\label{sec:testKriging}
			A key assumption of kriging is that the correlation between the values of two sampled points 
			depends only on their relative distance, and not on their locations in the field. 
			In most astronomical applications of the kriging technique the spatial distribution of the 
			measured values is not homogeneous (e.g. metallicity has usually a peaked 2D distribution) and it depends on 
			variables other than distance (e.g. stellar population,	galaxy inclination, etc.).		
			We thus chose to test the quality of a 2D field kriging reconstruction without this assumption from simulated 
			observations, in order to quantify the reliability of the maps we will obtain from real metallicity 
			data.

			We created a total of 9 two-dimensional (100$\times$100 pixels) metallicity distributions assuming 
			different radial profiles, axial ratios and centre positions. 
			The adopted metallicity radial profiles are either analytic (i.e. linear or exponential) or derived from galaxy evolution 
			simulations (i.e. \citealt{Hopkins09}) and scaled arbitrarily.  		
			We also span a range of axial ratios and different positions in the field in order to simulate different galaxy 
			inclinations and spot misleading artefacts. 
			The three cases we tested are: a single galaxy, two galaxies with different central metallicity 
			profiles and a single galaxy with a ring-shaped bump in the 2D metallicity profile. 

			To simulate the metallicity error for each position in the grid we assumed a distribution inversely proportional to the 
			light flux, which follows a  \citet{deVaucouleurs48} radial profile. 
			We scaled this error distribution in order to have the minimum equal to $\delta\rm{[Z/H]} = 0.05~\rm{dex}$ 
			where the light flux peaks and 
			the maximum $\delta\rm{[Z/H]} = 0.50~\rm{dex}$ in the minimum flux pixel (the values are taken from an approximation of the actual 
			metallicity errors we retrieve for NGC~5846). 
			With these assumptions we have a reasonable, if simplistic, error distribution that we can use to weigh the data points 
			in kriging.
						
			After the construction of the mock 2D metallicity and metallicity error distributions, we extracted from them $n$ randomly 
			positioned samplings. 
			Applying kriging on these points we obtained the interpolated map, which we then subtracted from the original distribution 
			to evaluate the residuals. 
			These residuals are then compared with the associated uncertainties, calculated as the sum in quadrature of the assigned 
			metallicity uncertainties and the kriging estimated uncertainties (Figure \ref{fig:L}). 

			\placefigLanalysis																																			
			
			At this point we want to quantify the reliability of kriging in reconstructing the original distribution of the variable we are 
			probing. 
			In particular, we will analyse how well kriging is able to return the overall 2D distribution and with which limits it can 
			reveal substructures in the field. 

			To address the first part of the question we assume that the quality of the map as a whole depends on just the number of 
			samplings, if uniformly distributed on the field.
			Considering a simple 2D distribution (i.e. single galaxy case), we created 100 sampling realizations for the three cases with, 
			respectively, a high ($n=100$), an average ($n=50$) and a low ($n=20$) number of points. 
			For all the pixels in each realization we calculated the absolute residual as the absolute value of the original map 
			value subtraction from the kriging extrapolated one.
			We then divided this residual by the pixel associated total uncertainty, which is resulted from the sum in quadrature of the 
			assigned intrinsic measurement error with the kriging interpolation estimated uncertainty. 
			We found that in the high and average sampling cases, 68\% of the kriging pixels have an absolute difference respect to 
			the original map ones, respectively, $\leq52.7$\% and $\leq55.7$\% of the associated metallicity error. 
			Also in the low number case, the 68\% of the pixels are enclosed within a fraction of the associated error, even if with a 
			higher percentage ($\leq0.75$\%) respect to the other two cases (Figure \ref{fig:N}).	
			\placefigNanalysis

			We also verified that the kriging maps we obtained are good enough to extrapolate trends in the field. 
			With this purpose we extract the values within a virtual long slit centred on the mock galaxy (or on both the galaxies 
			in the double galaxy case) in both the original and the kriging retrieved maps. 
			We found that, even in the poorly sampled case, the overall trend is well reproduced and the mismatching areas are the ones 
			near the peak of the 2D distribution (Figure \ref{fig:M}).  
			
			\placefigManalysis

			The second issue to investigate is how kriging is able to spot, if not reconstruct, substructures in the field. 
			As seen in Figure \ref{fig:M}, where the distribution profile is steeper, kriging does not seem able to return the correct shape 
			without an adequate and dense sampling, which in our tests is particularly true in the centre of the galaxies. 
			To explore the quality of the substructure mapping we created a mock field of $100\times100$ pixels in which a 
			single galaxy is present.
			We then added a substructure shaped as a 2D Gaussian profile in which the axial ratio is stretched to $b/a=0.5$. 
			The position of this substructure in the field is random, as well as the position angle of its axes, while its 
			scale (described by a Gaussian $\sigma$) and its central value are the parameters we explored. 
			In particular, the peak value of the Gaussian substructure is expressed as a ratio of the corresponding metallicity value in the 
			same pixel of the original map without the addition of the substructure, obtaining:
			\begin{eqnarray}
				\rm{[Z/H]}(x,y)_{0,\rm{S}} = \rm{[Z/H]}(x,y)_{0,\rm{NS}}\cdot(1+r)
			\end{eqnarray}
			with $\rm{[Z/H]}(x,y)_{0,\rm{S}}$ as the substructure central pixel value in the final map, 
			$\rm{[Z/H]}(x,y)_{0,\rm{NS}}$ the value of the same 
			pixel in the map without the substructure and $r$ the parameter we change. 
			From the total field, $n$ pixels are randomly chosen as samplings and a kriging map is built. 
			Considering only the pixels enclosed in a square centred on the peak of the Gaussian and with size $2\sigma\times2\sigma$,
			we computed the reconstructed substructure metallicity difference in each of them as:
			\begin{eqnarray}
				\Delta \rm{[Z/H]}_{S} = (O_{\rm{S}} - O_{\rm{NS}}) - (K_{\rm{S}} - K_{\rm{NS}})
			\end{eqnarray}	
			where $O$ refers to the original map, $K$ to the kriging map and the subscripts $\rm{S}$ and $\rm{NS}$ refer to the 
			cases with and without the substructure, respectively.
			We obtained $100$ statistical realizations for each combination of the parameters $\sigma$, $r$ and $n$, changing the 
			sampling pixels and the position and position angle of the substructure. 
			From these realizations we obtained the histograms presented in Figure \ref{fig:R}. 
			In general we observed that a higher number of samplings allows us to obtain a better extrapolation of the substructure 
			2D shape, but the final result depends also on the radial profile of the feature. 
			In fact, if the radial profile is steep (i.e. the substructure has a higher central value and a small scale dimension), 
			the majority of the points in the kriging map has a discrepancy greater than $\Delta\rm{[Z/H]}_{\rm{S}} = 0.1~\rm{dex}$ even 
			in the high number sampling case (Figure \ref{fig:R}, panel C).  
			On the other hand, keeping the central value constant and increasing the scale $\sigma$, the radial distribution of the 
			substructure is smoother and kriging is able to reproduce it with residuals $\Delta\rm{[Z/H]}_{\rm{S}} \leq 0.1~\rm{dex}$ 
			for most pixels.  

			\placefigRanalysis
			
			From these tests we can conclude that the kriging method is able to retrieve the overall 2D distribution of a 
			variable with a high degree of accuracy even with a low number of samplings.
			However, in the case of substructures, one needs more samplings for increasingly 
			steeper profiles.

\section{Individual Galaxies}\label{sec:Galaxies}
	\subsection{General remarks on literature values}
	In this section we will discuss the metallicity kriging maps we obtained and 
	the available literature metallicity values and profiles for all the galaxies in our sample. 
	In the literature several different ways of reporting the metallicity of a stellar population exist. 
	In order to have some comparison with the literature, we selected only the 
	works which report either the total metallicity [Z/H] or both the measured iron abundance [Fe/H] 
	and the $\rm{\alpha}$-elements abundance $\rm{[\alpha/Fe]}$.
	It is possible to retrieve the total metal abundance adopting the relation 
	[Z/H]=[Fe/H]+0.94$\rm{[\alpha/Fe]}$ \citep{Thomas03a}.
	In the case of \citet{Conroy12b} values, we considered $\rm{[Mg/Fe]} \approx \rm{[\alpha/Fe]}$. 
	Most of the literature give central values within $\rm{R_{e}}/2$, $\rm{R_{e}}/8$ or $\rm{R_{e}}/16$, but 
	often the reference value for the effective radius is different. 
	For this reason we scaled the radii of these values with 
	respect to our assumed effective radii. 
 	All literature values are plotted in Figure \ref{fig:O}, together with our metallicity values. 
 	In the plots, the literature values referring to an average metallicity in the centre are placed at the 
 	distance from the centre which encloses half of the light in the averaged region, assuming 
 	a de Vaucouleurs profile. 
	On the other hand, we found only a handful of available radial profiles for the galaxies in our sample. 
	The \sauron\ radial profiles have been extracted from the \sauron\ metallicity maps \citep{Kuntschner10}.
	In the case of the other metallicity radial profiles, we extracted them from the plots in the literature 
	papers. 

	\subsection{NGC~720}
		The radial metallicity profile (Figure \ref{fig:O}) 
		is flat in the range ($-0.5<\log\left(R/\rm{R_{e}} \right) < -0.2$), 
		with some scatter at larger distances from the centre. 
		We excluded this galaxy from following analyses of the metal distribution because of both the low number of 
		measured metallicities (i.e. 20) and the inhomogeneous azimuthal coverage of the 2D field, which leads to an 
		unreliable kriging map. 
	\subsection{NGC~821}
		This object is part of the \sauron\ sample. 
		\citet{Proctor05} measured the metallicity radial profile using Lick indices. 
		The metallicity in the inner region of NGC~821 has been measured by several studies 
		\citep{Trager00a, Denicolo05b, Howell05, Conroy12b}, 
		all consistent with a central value $0.1<\rm{[Z/H]}_{\rm{inner}}<0.5 \rm{dex}$.
		Our data points match quite well with both the \sauron\ and \citet{Proctor05} radial 
		profiles (Figure \ref{fig:O}). 
		As for NGC~720, the low number of points (i.e. 17) and inhomogeneous azimuthal coverage of the field 
		inhibit the extraction of a reliable kriging metallicity map. 
		In any case, with just one data point outside $1~\rm{R_{e}}$ we would not be able to extract the outer metallicity 
		gradient necessary for the further steps of our analysis. 

		\subsection{NGC~1023}
		In this case the measured metallicity points are consistent with the \sauron\ profile (Figure \ref{fig:O}). 
		The available literature values for the inner metallicity are quite consistent with the 
		results of \sauron\ \citep{Silchenko06, Conroy12b}.
		The slight increase of the \sauron\ profile for $\log\left( R/\rm{R_{e}} \right) > -0.4$ also 
		matches the behaviour of the values we measure in the inner South-West regions of the galaxy. 
		However, the final kriging-extracted profile is substantially offset with respect to the 
		\sauron\ outermost values. 
		The extracted kriging map extends outside $3~\rm{R_{e}}$ because of a single data point at $\approx155$ arcsec 
		from the galaxy centre. 
		Despite this, we still consider the map reliable because of the high number of measured data points and their 
		homogeneous azimuthal coverage of the field. 

		\subsection{NGC~1400}
		For this galaxy, a number of central metallicity measurements are available in literature \citep{Howell05, 
		Barr07, Idiart07, Spolaor08b}, all of which are consistent with a supersolar central metallicity. 
		In addition, \citet{Spolaor08b} also present the radial metallicity profile out to $1~\rm{R_{e}}$. 
		Comparing the distribution of our points with this profile we notice an offset of about $0.25~\rm{dex}$, with 
		our values more metal rich. 
		In this case, it is possible that we overcorrected the CaT-derived metallicities by overestimating the central 
		velocity dispersion for this galaxy (see Section \ref{sec:empcorrection}).  
		The metallicity profiles extracted from the kriging map 
		present higher values than those from \citet{Spolaor08b}. 
		The kriging map for this galaxy shows a metallicity distribution peaked near the galaxy centre and has 
		sufficient data points for azimuthal coverage. 
		The offset between the peak of the 2D metallicity distribution and the photometric centre of NGC~1400 is 
		probably caused by the asymmetric azimuthal distribution of the data points, which is denser in the South-East 
		region of the field.   
		
		\subsection{NGC~1407}
		NGC~1407 is a very well studied galaxy at the centre of the Eridanus A group. 
		Several studies agree with a supersolar central metallicity \citep{Howell05, Cenarro07, Annibali07, 
		Spolaor08b}. 
		\citet{Spolaor08b} also measured the metallicity profile out to about $0.6~\rm{R_{e}}$, which is consistent with 
		our radially overlapping CaT-derived metallicities. 
		A peculiar feature of this galaxy is a bump in the radial metallicity distribution we observe for the measured 
		metallicities between $0.6$-$1~\rm{R_{e}}$. 
		From the kriging map is not clear if this bump corresponds to circular shell around the galaxy centre or is 
		just a feature of the South and East regions of the galaxy. 
		Curiously, in the same region a spike in velocity dispersion and $h_4$ has been noticed in the past 
		\citep{Proctor09, Arnold14}. 
		This higher metallicity region has been chemically probed before only by \citet{Foster09} with a sub-sample 
		of the data presented in this work. 
		However, in such a study the low number of data points did not allow to clearly notice the bump in metallicity. 
		NGC~1407 will be studied in greater detail in a following paper. 

		\subsection{NGC~2768}
		A number of studies agree on a supersolar central metallicity \citep{Howell05, Silchenko06, Idiart07, 
		Serra08, Conroy12b} for this galaxy. 
		\citet{Denicolo05b} measured the average metallicities along both the major and the minor axes of the 
		galaxy, with the latter presenting an higher value. 
		Our measurements are consistent with the radial profile from the \sauron\ survey in the inner regions. 

		\subsection{NGC~2974}
		The central metallicities in literature all agree on a central value slightly supersolar \citep{Denicolo05b, Annibali07, 
		Conroy12b} and are consistent with the \sauron\ profile for this galaxy. 
		The average of our measured metallicities match with \sauron\ in the overlapping radial region. 
		Unfortunately, both the low number of sampling points (i.e. 13) and the inhomogeneous azimuthal coverage of 
		the field indicate an unreliable kriging map. 
		However, the presence of just 2 data points outside $1~\rm{R_{e}}$ would prevent a measure of the
		outer metallicity gradient anyway.  
		
		\subsection{NGC~3115}
		In literature there are three different measures for the central metallicity of NGC~3115, 
		all consistent with a supersolar central 
		value \citep{Howell05, Sanchez-Blazquez06, Idiart07}. 
		\citet{Norris06} measured the metallicity profiles along the major axis of this galaxy.		
		The CaT-derived metallicities of our work are comparable, in the common explored region, with their major axis 
		profile. 
		However, from the kriging extracted metallicity map, it is possible to notice an increase of the stellar 
		metallicity in the North-East region. 
		In addition, we do not see any photometric substructure in correspondence to the higher metallicity region. 
				
		\subsection{NGC~3377}
		The inner metallicity of NGC~3377 has been measured by a number of studies, several of which in agreement 
		with a solar \citep{Trager00a, Sanchez-Blazquez06, Idiart07} or supersolar central metallicity 
		\citep{Howell05, Denicolo05b, Conroy12b}. 
		The photometric study by \citet{Harris07} measured the metallicity of the red-giant stars of NGC~3377 
		up to more than $7~\rm{R_{e}}$.
		They found a metallicity distribution almost identical in their three fields, which are positioned at $1.65$, $2.75$ 
		and $4.6~\rm{arcmin}$ from the galaxy centre. 
		All these metallicity distributions roughly extend from $\rm{[Z/H]} = 1.5~\rm{dex}$ to $\rm{[Z/H]} = -0.3~\rm{dex}$ 
		and peak at $\rm{[Z/H]} =-0.6 ~\rm{dex}$. 		
		The \sauron\ radial profile extends to $1~\rm{R_{e}}$ and is consistent with a slightly supersolar central metallicity 
		and a gentle decline in the probed range. 
		While our data match with those of \sauron\ in the common radial range, outside $1~\rm{R_{e}}$ we 
		observe a steep gradient. 
		This steepness has been also confirmed by the kriging-extracted map. 

		\subsection{NGC~3607}
		For this galaxy, several different measures of the central metallicity have been done, all consistent with 
		a slightly supersolar metallicity \citep{Proctor02, Howell05, Sanchez-Blazquez06, Silchenko06, 
		Idiart07, Annibali07}, except the study by \citet{Denicolo05b} in which the 
		metallicity is considerably higher. 
		Our measured points are enclosed within about $0.4$ and $1.6~\rm{R_{e}}$, and thus we are not able to 
		probe the metallicity in the innermost regions of the galaxy. 
		However, we find from the kriging map a decreasing metallicity to our outermost point. 
		The offset between the metallicity peak in the kriging map and the galaxy photometric centre is probably caused 
		by the inhomogeneous azimuthal data point distribution within $1~\rm{R_{e}}$, with a higher density of points 
		on the South-West region of the probed field.

		\subsection{NGC~4111}
		This object has the smallest angular size in our sample ($\rm{R_{e}}=12.0~\rm{arcsec}$). 
		For NGC~4111 only the recent study of \citet{Silchenko06} has probed the chemical enrichment in the 
		inner regions, finding a value comparable with a solar metallicity. 
		Despite our measurements present quite an important radial scatter, the kriging extrapolation
		shows a centrally peaked metallicity map. 
		The presence of the two low-metallicity points along the minor axis could be the cause of the kriging map's 
		elongated shape. 
		This affects the radial profile in the outer regions, causing the presence of a metallicity bump between $35$ and $55~\rm{arcsec}$. 
		This feature is due to the adoption of the isophotal $b/a$ to measure the circular-equivalent distance of each map pixel 
		from the centre. 
		It is, however, remarkable that the metallicity isocurves are oriented similarly to the isophotes (see Figure \ref{fig:Q}). 
		
		\subsection{NGC~4278}
		NGC~4278 is an almost spherical elliptical galaxy. 
		According to \citet{Conroy12b}'s central measurements and \sauron\ radial profile, the central 
		metallicity for this galaxy is about solar. 
		However, \citet{Serra08} measure a much higher central metallicity. 
		Our CaT-derived metallicities match with \sauron\ radial profile in the overlapping radial region, 
		giving a high scatter in the outer profile. 
		The kriging extracted map shows a very flat 2D metallicity distribution. 
		
		\subsection{NGC~4365}
		This peculiar galaxy hosts a kinematically decoupled core \citep{Davies01}. 
		Several studies agree on a supersolar central metallicity for NGC~4365 \citep{Proctor02, 
		Howell05, Denicolo05b, Sanchez-Blazquez06, Idiart07}. 
		In general, our metallicity points in the outskirts show a scattered distribution about these central values 
		and the kriging extracted map presents a flat spatial metallicity distribution. 
		Because of the outermost higher-metallicity point, the extracted radial profile at $R>1~\rm{R_{e}}$ 
		has a slightly positive gradient. 
			
		\subsection{NGC~4374}
		The metallicity in the central regions of this galaxy has been measured by a number of studies, which found 
		a value consistent with a solar \citep{Trager00a, Sanchez-Blazquez06, Annibali07, Gallagher08} 
		or supersolar \citep{Proctor02, Howell05, Denicolo05b, Idiart07} metallicity. 
		Our points present a good agreement with respect to the \sauron\ profile. 
		This may be a consequence of an overestimated central velocity dispersion, which leads to an overcorrection 
		in metallicity.
		The kriging map for this galaxy is quite peculiar, showing patches of high and low metallicities in the 2D field. 
		Because of the relatively high number (35) and the good azimuthal coverage of the sampling 
		points, these apparent substructures may be real.

		\subsection{NGC~4473}
		NGC~4473 presents a kinematic transition from a disc-like rotation in the centre to 
		a triaxial `kinematically distinct halo' (KDC) outside $1.8~\rm{R_{e}}$ \citep{Foster13}. 
		For this galaxy, a weakly \citep{Idiart07, Conroy12b} and strongly \citep{Howell05} supersolar central 
		metallicities have been measured. 
		Our CaT-derived metallicities are compatible with the \sauron\ profile in the common radial range.   
		The global 2D metallicity field is quite flat, showing a slightly higher stellar metallicity along 
		the major axis. 
		
		\subsection{NGC~4494}
		In the literature we found for this galaxy only the central metallicity measurements by \citet{Denicolo05b}, which 
		shows a supersolar value. 
		With a sub-set of our data points, \citet{Foster09} measured a CaT index flat profile up to $1.4~\rm{R_{e}}$. 
		In addition, \citep{Foster11} found no evidence of CaT index azimuthal variation in a similar radial range. 	
		The metallicities we derived for this galaxy extend up to $1.5~\rm{R_{e}}$. 
		The metallicity profile we extract from the kriging map is consistent with being flat within $1~\rm{R_{e}}$ 
		and strongly steep outside this limit. 
		On the other hand, the map itself shows a higher and a lower metallicity patch, respectively, in the North and 
		South-East regions of our field. 
		While we do not have reasons to exclude the possibility of the first substructure being real, 
		the second seems driven by just two data points. 

		\subsection{NGC~4526}
		For NGC~4526 a supersolar central metallicity has been confirmed by \citet{Proctor02, Silchenko06, Gallagher08}. 
		The metallicities we measured in the outskirts closely match the \sauron\ radial 
		profile in the overlapping region. 
		The kriging metallicity map for this galaxy is centrally peaked and shows an elongation 
		almost aligned with the isophotal major axis. 

		\subsection{NGC~4649}
		For this massive galaxy, \citet{Gallagher08} found a central metallicity slightly below the solar value, while both the studies 
		by \citet{Trager00a} and \citet{Howell05} are above it. 
		Our point distribution is on average consistent with a supersolar metallicity also in the outskirts. 
		The kriging map presents a centrally peaked 2D metallicity distribution, with metallicity isocurves 
		following the galaxy isophotes. 
		
		\subsection{NGC~4697}
		For NGC~4697, many studies agree on a supersolar central metallicity and a decreasing profile with 
		radius \citep{Trager00a, Proctor02, Howell05, Sanchez-Blazquez06, Annibali07, Idiart07}.
		The 2D metallicity map is consistent with a centrally peaked metallicity and the isocurves of metallicity 
		resemble both the PA and the ellipticity of the isophotes. 		
		
		\subsection{NGC~5846}
		Over the years, many studies have measured NGC~5846's central metallicity, finding a range of values from weakly 
		undersolar to strongly supersolar \citep{Trager00a, Howell05, Denicolo05b, Sanchez-Blazquez06, 
		Annibali07, Idiart07, Gallagher08, Conroy12b}.
		The CaT-derived metallicities we measured are consistent with the \sauron\ profile in the overlapping 
		region. 
		NGC~5846 has a small companion (NGC~5846A) which we identified in both the SDSS image and the 
		CaT index kriging map (Figure \ref{fig:S}). 
		In order to retrieve the metallicity map for just the main galaxy, we excluded the slits within 15 arcsec 
		from the centre of NGC~5846A. 
		The kriging metallicity map we then retrieve shows a centrally peaked metallicity distribution. 
		However, there are only 2 points outside $1~\rm{R_{e}}$. 
		As a consequence, we do not measure the metallicity gradient in the outer region of this galaxy. 
			
		\subsection{NGC~5866}
		For NGC~5866 we retrieved in the literature only the weakly supersolar central metallicity value measured 
		by \citet{Silchenko06}. 
		Unfortunately, the poor azimuthal coverage of our metallicity measurements does not allow 
		to extract a reliable kriging map for the metallicity of this galaxy.

		\subsection{NGC~7457}
		The NGC~7457 central metallicity measurements by \citet{Silchenko06} and \citet{Serra08} agree on 
		a subsolar metal abundance. 
		On the contrary, the \sauron\ profile has a weak supersolar central value and a gentle 
		decline with radius. 
		The metallicity values we measure are mostly consistent with the \sauron\ radial 
		profile in our innermost probed regions. 
		In the outer regions, however, what we measure is more consistent with a metal-poor 
		stellar population, while \sauron\ profile seems still quite flat. 
		This difference is enhanced in the kriging extracted metallicity profile, which presents a steep 
		outer gradient. 
		The kriging map shows a centrally peaked metallicity 2D distribution, with both the orientation and the shape 
		of the metallicity isocurves roughly resembling the galaxy isophotes.

\end{document}